\documentclass[11pt]{article}

\pdfoutput=1
\usepackage[utf8]{inputenc}
\usepackage{amsmath}
\usepackage{amsthm}
\usepackage{amsfonts}          % if you want the fonts
\usepackage{amssymb}           % if you want extra symbols
\usepackage{color}
\usepackage{graphicx}
\usepackage{cite}
\usepackage{multirow,setspace,microtype}
\usepackage{tikz}
\usepackage{booktabs}
\usepackage[framemethod=tikz]{mdframed}
\AtBeginEnvironment{mdframed}{%
	\tikzset{every picture/.style={}}%
}
\mdfsetup{roundcorner=.5ex}
{\begin{mdframed}[backgroundcolor=black!5!white]}%
	{\end{mdframed}}

\allowdisplaybreaks
\PassOptionsToPackage{linecolor=blue,backgroundcolor=blue!25,bordercolor=blue,textsize=scriptsize}{todonotes}
\usepackage{todonotes}

\numberwithin{equation}{section}

\definecolor{darkblue}{rgb}{0.0,0.0,0.3} 		% Dark-blue links 

\usepackage[colorlinks]{hyperref}
\hypersetup{pdftitle={Explicit Soft Supersymmetry Breaking in the Heterotic M-Theory B-L MSSM},
	pdfauthor={Anthony Ashmore, Sebastian Dumitru and Burt Ovrut},
	linkcolor=darkblue, urlcolor=darkblue, citecolor=darkblue, filecolor=black, linktoc=page}

\usepackage[bf]{caption}
\let\originalleft\left
\let\originalright\right
\renewcommand{\left}{\mathopen{}\mathclose\bgroup\originalleft}
\renewcommand{\right}{\aftergroup\egroup\originalright}

\makeatletter
\g@addto@macro\bfseries{\boldmath}
\makeatother

\newlength{\xtrawidth}
\newlength{\xtraheight}
\newcommand{\Z}{\ensuremath{\mathbb{Z}}}

\newcommand{\Rep}[1]{\ensuremath{\boldsymbol{\underline{#1}}}}

\newcommand{\Rhat}{\ensuremath{{\widehat R}}}

\global\long\def\ii{\text{i}}%
\global\long\def\ee{\text{e}}%
\global\long\def\Uni#1{{U}(#1)}%
\global\long\def\Ex#1{{E}_{#1}}%
\global\long\def\eqspace{\mathrel{\phantom{{=}}{}}}%
\global\long\def\op#1{\operatorname{#1}}%

\DeclareMathOperator{\tr}{tr}

\DeclareMathOperator{\re}{Re}
\DeclareMathOperator{\imag}{Im}

\renewcommand{\thefootnote}{\fnsymbol{footnote}}

\usepackage{subcaption}
\usepackage{float}
\usepackage{caption}

\begin{document}

\begin{titlepage}
%\vspace{-4cm}
\title{\LARGE \bf{Explicit Soft Supersymmetry Breaking in\\[.15cm]the Heterotic M-Theory $B-L$ MSSM}\\[.3cm]}
                       
\author{{
   Anthony Ashmore,\textsuperscript{a,b}
   Sebastian Dumitru\textsuperscript{c}
   and Burt A.\,Ovrut\textsuperscript{c}}\\[0.7cm]
   {\it\small {}\textsuperscript{a}Enrico Fermi Institute \& Kadanoff Center for Theoretical Physics,} \\
   {\it\small  University of Chicago, Chicago, IL 60637, USA}\\[.1cm]
	{\it\small {}\textsuperscript{b}Sorbonne Universit\'e, Laboratoire de Physique Th\'eorique et Hautes Energies,} \\
	{\it\small F-75005 Paris, France} \\[.1cm]
	{\it\small {}\textsuperscript{c}Department of Physics, University of Pennsylvania,} \\
	{\it\small Philadelphia, PA 19104, USA}}         

\date{}

\maketitle

\begin{abstract}
\noindent
The strongly coupled heterotic M-theory vacuum for both the observable and hidden sectors of the $B-L$ MSSM theory is reviewed, including a discussion of the ``bundle'' constraints that both the observable sector $SU(4)$ vector bundle and the hidden sector bundle induced from a single line bundle must satisfy. Gaugino condensation is then introduced within this context, and the hidden sector bundles that exhibit gaugino condensation are presented. The condensation scale is computed, singling out one line bundle whose associated condensation scale is low enough to be compatible with the energy scales available at the LHC. The corresponding region of K\"ahler moduli space where all bundle constraints are satisfied is presented. The generic form of the moduli dependent $F$-terms due to a gaugino superpotential -- which spontaneously break $N=1$ supersymmetry in this sector -- is presented and then given explicitly for the unique line bundle associated with the low condensation scale. The moduli dependent coefficients for each of the gaugino and scalar field soft supersymmetry breaking terms are computed leading to a low-energy effective Lagrangian for the observable sector matter fields. We then show that at a large number of points in K\"ahler moduli space that satisfy all ``bundle'' constraints, these coefficients are initial conditions for the renormalization group equations which, at low energy, lead to completely realistic physics satisfying all phenomenological constraints. Finally, we show that a substantial number of these initial points also satisfy a final constraint arising from the quadratic Higgs-Higgs conjugate soft supersymmetry breaking term.

\noindent

{\let\thefootnote\relax\footnotetext{ashmore@uchicago.edu, dumitru@sas.upenn.edu, ovrut@elcapitan.hep.upenn.edu}}

%\vspace{.3in}
%\noindent
\end{abstract}

\thispagestyle{empty}
\end{titlepage}

\tableofcontents

\section{Introduction}

Hořava--Witten theory comes from eleven-dimensional M-theory compactified on an $S^{1}/\mathbb{Z}_{2}$ orbifold, with ten-dimensional planes at the ends of the interval, each supporting an $E_{8}$ gauge field~\cite{Horava:1995qa,Horava:1996ma}. This theory was further compactified to five-dimensional ``heterotic'' M-theory on a Calabi--Yau threefold in \cite{Lukas:1998yy, Lukas:1998tt} and then, by integrating out the $S^{1}/\mathbb{Z}_{2}$ modes, to a four-dimensional effective $E_{8} \times E_{8}$  heterotic string theory~\cite{Lukas:1997fg}. The resulting theory is strongly or weakly coupled for a large or small $S^{1}/\mathbb{Z}_{2}$ interval length respectively~\cite{Banks:1996ss}. One of the orbifold planes is referred to as the ``observable'' sector, while the other is labelled the ``hidden'' sector. By appropriately choosing a Calabi--Yau threefold admitting a holomorphic vector bundle, it has been shown that $N=1$ supersymmetric, phenomenologically realistic models of low-energy four-dimensional particle physics can be constructed on the observable sector. Many such models have been constructed, employing a variety of Calabi--Yau threefolds and a wide range of vector bundles -- see, for example, \cite{Donagi:2000vs,Braun:2005nv,Braun:2005bw,Braun:2005ux,Bouchard:2005ag,Anderson:2009mh,Braun:2011ni,Anderson:2011ns,Anderson:2012yf,Anderson:2013xka,Nibbelink:2015ixa,Nibbelink:2015vha,Braun:2017feb}.

One of these theories, named the $B-L$ MSSM or ``heterotic standard model'', is constructed by compactifying from eleven to five-dimensions on a specific class of Schoen Calabi--Yau threefolds with ${\mathbb{Z}}_{3} \times {\mathbb{Z}}_{3}$ fundamental group~\cite{MR923487,Braun:2004xv}. Together with a slope-stable holomorphic bundle with structure group $SU(4)$ \cite{Braun:2005zv} and two Wilson lines, the observable $E_{8}$ group is broken to the $SU(3)_{C}\times SU(2)_{L}\times U(1)_{Y}$ standard model gauge group with an additional $U(1)_{B-L}$ gauge group factor. This additional $U(1)_{B-L}$ gauge symmetry provides a ``natural'' explanation for the $R$-parity discrete symmetry imposed in an ad hoc manner in the MSSM. The matter spectrum of the $B-L$ MSSM was shown \cite{Braun:2005bw,Braun:2005nv,Braun:2005ux} to be precisely that of the MSSM; that is, three families of quark and lepton chiral superfields, each family with a right-handed neutrino chiral supermultiplet, as well as one Higgs-up and one Higgs-down doublet of chiral superfields. There are no exotic fields or vector-like pairs. It is important to note, however, that this theory, as with the MSSM, is first defined at a large mass scale -- usually just below the scale of the Calabi--Yau compactification. To analyze its predictions for low-energy phenomenology, it is necessary to break $N=1$ supersymmetry and then, via a complicated renormalization group analysis, compute its properties at the electroweak scale. 

This analysis has been carried out in a large number of papers. The exact form of the renormalization group equations, to the one-loop level, assuming ``soft'' supersymmetry breaking terms in the Lagrangian arising from spontaneous $N=1$ breaking in the ``hidden'' sector, was presented in \cite{Ovrut:2014rba,Ovrut:2015uea}. Using a statistical scan over the coefficients of the soft supersymmetry breaking terms, it was shown that for a large number of these coefficients -- referred to as ``black'' points in the space of possible couplings and masses -- all low-energy phenomenological requirements were satisfied; namely, the gauged $B-L$ symmetry is spontaneously broken consistent with experimental restrictions, electroweak symmetry is spontaneously broken with the exact $W^{\pm}$, $Z^{0}$ boson and Higgs masses, and that the masses of all sparticles exceed their present lower bounds \cite{Everett:2009vy,Ghosh:2010hy,Barger:2010iv,Mukhopadhyaya:1998xj,Chun:1998ub,Chun:1999bq,Hirsch:2000ef,FileviezPerez:2012mj,Perez:2013kla,Gamberini:1989jw,Agashe:2014kda}.\footnote{See also \cite{Dumitru:2019cgf} for a summary of these bounds.} Furthermore, these ``black'' points required no fine-tuning. Since $R$-parity is spontaneously broken, the various lightest supersymmetric particles (LSPs) associated with different initial conditions will decay at a relatively low mass scale to standard model particles. These $R$-parity violating (RPV) decay channels and production cross sections were computed in detail for various LSPs in \cite{Dumitru:2018jyb,Dumitru:2018nct} (see also \cite{FileviezPerez:2009gr,FileviezPerez:2012mj,Marshall:2014kea,Barger:2008wn}), and correlated to past and upcoming searches at the CERN LHC \cite{Dumitru:2018nct,Dumitru:2019cgf}.

Given these successes, it becomes important to provide one or more hidden sector gauge bundles that are consistent with the $B-L$ MSSM observable sector. That such a bundle is allowed was first discussed in \cite{Braun:2006ae}. Within the context of weakly coupled heterotic M-theory, this was first accomplished in \cite{Braun:2013wr}, where several explicit hidden sector line bundles, and Whitney sums of line bundles, were presented that satisfied all required ``bundle'' constraints. More recently \cite{Ashmore:2020ocb}, several explicit hidden sector line bundles, and Whitney sums of line bundles, were presented within the context of strongly coupled heterotic M-theory. These satisfy all required bundle constraints -- now including appropriate values for the unification mass and $SU(4)$ gauge coupling in the observable sector. In both cases the hidden sector bundles preserve $N=1$ supersymmetry at the perturbative level. However, it is clear from the previous discussion that one must somehow break supersymmetry spontaneously in the hidden sector. It is the purpose of this present paper, working within the context of the strongly coupled heterotic M-theory vacuum introduced in \cite{Ashmore:2020ocb}, to show how spontaneous supersymmetry breaking can occur non-perturbatively via gaugino condensation~\cite{Dine:1985rz,Horava:1996vs,Taylor:1990wr,Lukas:1997rb,Lukas:1999kt,Antoniadis:1997xk,Dudas:1997cd,Nilles:1998sx}.\footnote{See also \cite{Minasian:2017eur} for a discussion of gaugino condensation in the presence of $H$ flux.} Using an analysis of soft terms~\cite{Kaplunovsky:1993rd,Choi:1997cm}, we will then explicitly calculate the soft supersymmetry breaking parameters in the low-energy observable matter sector Lagrangian. Our conclusion will be that, of the soft supersymmetry breaking parameters explicitly calculated in this manner,  a large number will be ``black'' points; that is, they give good initial conditions for the renormalization group analysis that satisfy all low-energy phenomenological requirements. Furthermore, we will show that a substantial subset of these initial conditions -- referred to as ``red'' points -- also satisfy a new constraint imposed on the coefficient of the Higgs--Higgs conjugate soft supersymmetry breaking term.

The outline of this paper is as follows. In Section 2, we briefly review both the observable and hidden sectors of the heterotic M-theory $B-L$ MSSM vacuum presented in \cite{Ashmore:2020ocb}. We will discuss the gauge group and matter spectrum of the low-energy observable sector and the explicit region of K\"ahler moduli space where the associated $SU(4)$ holomorphic vector bundle is slope-stable. For the hidden sector, we present the generic properties a line bundle $L$ must possess to be equivariant on our specific Schoen Calabi--Yau, the explicit embedding of its $U(1)$ structure group into the $SU(2)$ factor of $SU(2) \times E_{7} \subset E_8$ via the ``induced'' Whitney sum bundle $L \oplus L^{-1}$, and the constraint imposed on $L$ so that this induced vector bundle is poly-stable. Having done this, we present three sets of ``bundle'' conditions -- referred to as the ``vacuum'', ``phenomenological'' and ``strong coupling'' constraints -- that any such heterotic vacuum must satisfy. Importantly, these are {\it new} constraints and not to be confused with the low-energy physical requirements discussed above. Section 3 introduces the non-perturbative moduli superpotential induced by the condensation of the $E_{7}$ gauginos at the scale where $E_{7}$ becomes strongly coupled. This condensation scale is then calculated explicitly in the region of K\"ahler moduli space satisfying the above constraints. This is accomplished using a topological analysis to find both the low-energy hidden sector matter spectrum and to compute the associated renormalization group coefficient for the unbroken $E_7$ gauge group. We give the line bundles that allow for gaugino condensation and then focus on an example where the condensation scale can be in the low-energy range studied at the CERN LHC. Section 4 gives explicit expressions for the $F$ auxiliary fields, assuming an $E_{7}$ gaugino condensate superpotential, associated with the complex dilaton, K\"ahler and five-brane location moduli. These are, generically, complicated functions of the various moduli. To be concrete, we give these for the line bundle whose condensation scale is lowest and, hence, most relevant to experiments at the LHC. In Section 5, using the formalism presented in \cite{Kaplunovsky:1993rd} and the results of Section 4, we compute the moduli dependent coefficients of the soft supersymmetry breaking terms of the matter fields in the $B-L$ MSSM observable sector. This is done for all soft breaking terms; that is, for the gaugino masses, the quadratic scalar masses, the holomorphic cubic scalar terms and the holomorphic quadratic scalar term. Again, we present the general formalism and then explicitly compute the soft supersymmetry breaking coefficients for the most promising example. These coefficients are calculated to the zeroth and first order in the strong coupling expansion parameter, using a relatively minor assumption about the form of the K\"ahler potential of the matter scalar superfields. These results are used in Section 6 to calculate the complete set of soft supersymmetry breaking coefficients at every point in the ``viable'' region\footnote{In \cite{Ashmore:2020ocb}, the ``viable'' region of K\"ahler moduli space for the specific line bundle $L={\cal{O}}_{X}(2,1,3)$ was often referred to as the ``magenta'' region. In this paper, since we deal with a more general set of line bundles, we will exclusively use the term ``viable'' region.} of K\"ahler moduli space associated with the specific line bundle $L$, each point of which satisfies all of the required ``bundle'' constraints introduced in Section 2. To explore these models in more detail, we arbitrarily choose one such ``viable'' point in K\"ahler moduli space and present two different sets of initial soft supersymmetry breaking coefficients at the unification scale. We then show, using a renormalization group analysis, that both of these correspond to ``black'' points and, hence, satisfy all low-energy phenomenological requirements. Using a new renormalization group analysis presented in Appendix A, we show that one of the two satisfies a further constraint on the Higgs--Higgs conjugate soft supersymmetry breaking coefficient that arises in gaugino condensation scenarios. Hence, it is a physically ``acceptable'' set of initial conditions which we refer to as a ``red'' point. Using a similar analysis, we show, however, that the second set of initial conditions does {\it not} satisfy  this Higgs-Higgs conjugate constraint. We then proceed to give a statistical analysis of the soft supersymmetry breaking terms at this specific point in K\"ahler moduli space, exhibit all ``black'' points, and determine which of these are, in addition, acceptable ``red'' points. We conclude by applying this analysis to every point in the ``viable'' region of K\"ahler moduli space and presenting all physically acceptable ``red'' points. \\

\section{\texorpdfstring{$B-L$}{B-L} MSSM Heterotic Vacuum}

\subsection{Observable Sector}

On the observable orbifold plane, the vector bundle is chosen to be a specific holomorphic bundle with structure group $SU(4)\subset E_8$~\cite{Braun:2005zv}. This $SU(4)$ bundle breaks the $E_8$ group down to 
\begin{equation}
  E_8 \to Spin(10) \ .
  \label{13}
\end{equation}
Hence, $Spin(10)$ is the ``grand unified'' group of the observable sector. This GUT group is then further broken at the scale $\langle M_{U}\rangle=3.15 \times 10^{16}$ GeV to the low energy gauge group of the $B-L$ MSSM by turning on two flat Wilson lines, each associated with a different $\Z_3$ factor of the $\Z_3 \times \Z_3$ symmetry of $X$. Doing this preserves the $N=1$ supersymmetry of the four-dimensional effective theory, but breaks the observable sector gauge group down to
\begin{equation}
  Spin(10) 
  \to 
  SU(3)_C \times SU(2)_L \times U(1)_Y \times U(1)_{B-L} \ .
  \label{17}
\end{equation}
That is, the gauge group of the $B-L$ MSSM is that of the MSSM with an additional $U(1)_{B-L}$ factor. The spectrum of the $B-L$ MSSM is determined from the exact structure of Schoen Calabi--Yau compactification threefold \cite{MR923487,Braun:2004xv}  and the specific choice of the holomorphic $SU(4)$ vector bundle \cite{Braun:2005zv,Braun:2006ae}. The spectrum was explicitly computed and extensively studied in \cite{Braun:2005nv,Braun:2005bw,Braun:2005ux}. It was found to be exactly the three quark and lepton families of the  
MSSM including three additional right-handed neutrino chiral multiplets, one per family. It also contains the conventional $H_{u}$ and $H_{d}$ Higgs doublet supermultiplets. There are no exotic fields or vector-like pairs.\\

\subsubsection*{Observable Sector Constraints}

In order to preserve $N=1$ supersymmetry in the four-dimensional effective theory, the $SU(4)$ bundle must be both slope-stable and have vanishing slope~\cite{Braun:2005zv,Braun:2006ae}. Since the first Chern class $c_1$ of any $SU(N)$ bundle is zero, it follows immediately that the slope of the observable $SU(4)$ bundle vanishes, as required. However, demonstrating that our chosen $SU(4)$ bundle is slope-stable is non-trivial. As proven in detail in~\cite{Braun:2006ae}, the $SU(4)$ vector bundle is slope-stable, and hence, satisfies the Hermitian Yang--Mills equations~\cite{UY,Donaldson}, in a restricted region of the positive K\"ahler cone specified by
\begin{equation}\label{51}
  \begin{gathered}
    \left(
      a^1
      < 
      a^2
      \leq 
      \sqrt{\tfrac{5}{2}} a^1
      \quad\text{and}\quad
      a^3
      <
      \frac{
        -(a^1)^2-3 a^1 a^2+ (a^2)^2
      }{
        6 a^1-6 a^2
      } 
    \right)
    \quad\text{or}\\
    \left(
      \sqrt{\tfrac{5}{2}} a^1
      <
      a^2
      <
      2 a^1
      \quad\text{and}\quad
      \frac{
        2(a^2)^2-5 (a^1)^2
      }{
        30 a^1-12 a^2
      }
      <
      a^3
      <
      \frac{
        -(a^1)^2-3 a^1 a^2+ (a^2)^2
      }{
        6 a^1-6 a^2
      }
    \right) \ ,
  \end{gathered}
\end{equation}
where the $a^{i}$, $i=1,2,3$, are the real parts of the three K\"ahler moduli. This first restriction on the K\"ahler moduli depends entirely on the non-Abelian observable sector $SU(4)$ bundle. We will show in the following that the physically realistic region  of K\"ahler moduli space is further constrained by the gauge bundle in the hidden sector.\\ 

\subsection{Hidden Sector}

Following our discussion in \cite{Ovrut:2018qog,Ashmore:2020ocb}, the hidden sector vector bundle is constructed from a single holomorphic line bundle $L$, with structure group $U(1) \subset E_8$. Any line bundle on a Calabi--Yau threefold is associated with a divisor of the manifold, and is conventionally expressed as 
\begin{equation}
L=\mathcal{O}_X(l^1,l^2,l^3)\ ,
\end{equation}
where $l^{i}$, $i=1,2,3$, are, generically, arbitrary integers. However, on the specific Schoen Calabi--Yau threefold $X$ we are using,one cannot choose the integers $l^i$ freely. Since $X$ has non-trivial $\Z_3 \times \Z_3$ holonomy, a line bundle will only be equivariant if one further constrains the integers $l^{i}$ so that
\begin{equation}
  (l^1+l^2) \op{mod} 3 = 0 \ .
  \label{22}
\end{equation}
We will impose this additional constraint on all line bundles from now on. Finally, as in \cite{Braun:2013wr,Ashmore:2020ocb}, we choose to embed the $U(1)$ structure group of $L$ into the $SU(2)$ gauge factor of $SU(2) \times E_{7} \subset E_8$ via the ``induced'' bundle $L \oplus L^{-1}$. It follows that the four-dimensional effective theory on the hidden sector has gauge symmetry
\begin{equation}
H= U(1) \times E_{7} \ .
\label{red7}
\end{equation}
The first factor is an ``anomalous''  $U(1)$. It is identical to the structure group of $L$ and arises in the low-energy theory since $U(1)$ commutes with itself. The particular embedding is characterised by a coefficient $a$ defined by
\begin{equation}
a=\frac{1}{4}\tr_{E_{8}}Q^{2} \ ,
\label{xf2}
\end{equation}
where $Q$ the generator of the $U(1)$ structure group of $L$ embedded into the $\Rep{248}$ representation of the hidden sector $E_{8}$. For us this takes the value
\begin{equation}
a=1 \ .
\label{white1}
\end{equation}
This simplifies many of the expressions for following constraints and will be used henceforth. \\

\subsubsection*{Hidden Sector Constraint}

We have chosen the $U(1)$ structure group of $L$ to embed into the $SU(2)$ subgroup of  $SU(2) \times E_{7} \subset E_8$ via the induced vector bundle $L \oplus L^{-1}$, ensuring that it defines an $E_8$ gauge connection. It is then necessary to show that the associated gauge connection satisfies the Hermitian Yang--Mills equations. This will be the case if $L \oplus L^{-1}$ is ``poly-stable''. Since the slope of $L \oplus L^{-1}$ vanishes by construction, it follows that the slope of $L$, denoted by $\mu (L)$, must  also be zero. Using previous work in \cite{Dine:1987xk,Lukas:1999nh,Blumenhagen:2005ga,Blumenhagen:2006ux}, the expression for the genus-one corrected slope of an arbitrary line bundle was presented in \cite{Braun:2013wr}. Using this expression, we now restrict our line bundle $L$ to satisfy, in addition to \eqref{22}, the constraint that
\begin{equation}
\mu(L)= d_{ijk} l^i a^j a^k - 
  \epsilon_S' \frac{\Rhat}{V^{1/3}} 
 \left(d_{ijk}l^i l^j l^k 
  + l^i(2,2,0)|_i
  -\bigl(\tfrac{1}{2}+\lambda\bigr)^2l^iW_i\right) =0 \ ,
\label{xf1}
\end{equation}
Here the $a^{i}$ are the real K\"ahler moduli, $\hat{R}$ is the modulus for the separation of the observable and hidden planes, $V$ is the ``average'' Calabi--Yau volume given by 
\begin{equation}
V=\frac{1}{6}d_{ijk}a^{i}a^{j}a^{k} \ ,
\label{wp1} 
\end{equation}
$\epsilon_{S}'$ is the strong coupling expansion parameter, and $d_{ijk}$ are the intersection numbers of the Schoen Calabi--Yau threefold (see \cite{Ashmore:2020ocb} for more details). In addition, $W_{i}$ is the effective class representing a single wrapped five-brane and $\lambda$ is the modulus specifying the location of the five-brane in the fifth dimension~\cite{Gray:2003vw,Brandle:2001ts}.\\

\subsection{Five-Dimensional Heterotic Vacuum}

So far we have specified the conditions that any line bundle on the hidden sector must satisfy in order for the $U(1)$ structure group of $L$ to embed properly into the hidden sector $SU(2) \times E_{7} \subset E_8$ group and for the induced bundle $L \oplus L^{-1}$ to be poly-stable, thus guaranteeing that the Hermitian Yang--Mills equations are satisfied and that the low-energy $d=4$ theory will be supersymmetric. However, there remain two further sets of constraints that any physically acceptable hidden sector line bundle must satisfy. These are: A) that the theory be anomaly free, $N=1$ supersymmetric with positive squares of the ``unified'' gauge couplings in both the observable and hidden sectors, and B) that gauge unification in the observable sector occurs for a realistic value of the gauge coupling and at a physically acceptable mass scale. These additional conditions were discussed in detail in \cite{Ashmore:2020ocb}. Here, we simply present the associated constraint equations. Let us begin with the conditions in A).\\

\subsubsection*{Vacuum Constraints}

The conditions for an anomaly-free heterotic M-theory vacuum containing a single five-brane between the observable and the hidden walls was presented in \cite{Lukas:1998hk}. For the compactification to four-dimensions on our specific Schoen threefold, for the precise $SU(4)$ gauge bundle on the observable sector discussed above, and for a single $U(1)$ line  bundle $L$ on the hidden sector orbifold plane embedded as described above, the anomaly condition can be expressed as
\begin{equation}
W_i= \bigl( \tfrac{4}{3},\tfrac{7}{3},-4\big)\big|_i
  +  d_{ijk} l^j l^k \ , \qquad i=1,2,3  \ .
\label{xf3}
\end{equation}
However, simply satisfying this constraint is not sufficient to guarantee that the low-energy theory is $N=1$ supersymmetric. For that to be the case, one must demand that $L$ is chosen so that the five-brane class $W_{i}$ is ``effective''; that is, that
\begin{equation}
W_i \geq 0 \ , \qquad i=1,2,3  \ .
\label{xf4}
\end{equation}
Conditions \eqref{xf3} and \eqref{xf4} each put additional constraints on the values one can choose for $(l^1,l^2,l^3)$. Finally, as stated above, it is essential that the squares of the ``unified'' gauge couplings in both the observable and hidden sectors be positive. The genus-one corrected expressions for the gauge couplings associated with the $Spin(10)$ and the $E_7\times U(1)$ symmetry groups on the observable and the hidden sector respectively, were presented and discussed in \cite{Braun:2013wr,Ashmore:2020ocb}. For the $U(1)$ embedding satisfying \eqref{white1}, the constraints that they be positive definite were found to be
\begin{equation}
\label{coup1}
  \begin{split}
%    {(g^{(1)})^{2}}\sim \text{Re}f_1=
d_{ijk}a^{i}a^{j}a^{k}-3 \epsilon_S^\prime \frac{\hat R}{V^{1/3}}\left(-\tfrac{2}{3}a^1+\tfrac{1}{3}a^2-4a^3-\bigl( \tfrac{1}{2}-\lambda \bigr)^2W_ia^i\right)>0\\
  \end{split}
\end{equation}
and
\begin{equation}
  \begin{split}
  \label{coup2}
%{(g^{(2)})^{2}}\sim \text{Re}f_2=
d_{ijk}a^{i}a^{j}a^{k}-3 \epsilon_S^\prime \frac{\hat R}{V^{1/3}}\left(d_{ijk}a^{i}l^{j}l^{k}+2(a^{1}+a^{2})-\bigl( \tfrac{1}{2}+\lambda \bigr)^2W_ia^i\right)>0 
  \end{split}
\end{equation}
respectively. Inequalities \eqref{coup1} and \eqref{coup2} complete the additional set of constraints labelled as conditions A) above. \\

\subsubsection*{Phenomenological Constraints}

So far, the constraints were mostly topological in nature. There is an additional set of ``phenomenological'' constraints, referred to as conditions B) above. They are that the 
$Spin(10)$ grand unification scale, $\langle M_{U} \rangle$, and the associated unified gauge coupling, $\langle \alpha_{u}\rangle=\langle g^{(1)}\rangle^{2}/4\pi$, in the observable sector be consistent with the phenomenologically viable values for these quantities~\cite{Deen:2016vyh,Ovrut:2015uea,Ovrut:2012wg}. We choose the unification scale to be
\begin{equation}
\langle M_{U}\rangle=3.15 \times 10^{16}~\text{GeV} \ .
\label{jack1}
\end{equation}
The gauge coupling associated with the $Spin(10)$ group can take several different values at the unification scale $\langle M_U\rangle$, depending on the unification scenario. In the observable sector, the $Spin(10)$ group is broken down by two flat Wilson lines to the $B-L$ MSSM group $SU(3)_C \times SU(2)_L \times U(1)_Y \times U(1)_{B-L}$. If the Wilson lines turn on at different scales (the split Wilson lines scenario), it was shown in \cite{Ashmore:2020ocb} that the value of the gauge coupling at the unification scale is
\begin{equation}
\langle \alpha_{u} \rangle = \frac{1}{20.08}  \ .
\label{bag3}
\end{equation}
On the other hand, if the Wilson lines turn on at the same energy scale (simultaneous Wilson lines scenario) we find \cite{Ashmore:2020ocb}
\begin{equation}
\langle \alpha_{u} \rangle = \frac{1}{26.46}  \ .
\label{bag31}
\end{equation}
Fixing these values at the unification scale allows us to compute the couplings and other, generically moduli dependent, parameters of our theory. A particularly useful parameter is the distance between the observable and hidden walls given by
$\pi \rho \hat R V^{-1/3}$ where
\begin{equation}
 \rho=\left( \frac{\hat{\alpha}_{GUT}}{8 \pi^{2}} \right)^{3/2}v^{1/2}M_{P}^{2} \ ,
\label{bag5}
\end{equation}
we define
\begin{equation}
\label{lala1}
\hat{\alpha}_\text{GUT}=\langle \alpha_{u} \rangle \re f_1, \qquad v=\frac{1}{VM_U^6}\ ,
\end{equation}
and $M_P$ is the Planck scale
\begin{equation}
M_P=1.22\times 10^{19}\text{GeV}\ .
\end{equation}
The coupling $\epsilon_S^\prime$ is given by
\begin{equation}
 \epsilon_S' =\frac{2\pi^{2}\rho^{4/3}}{v^{1/3}M_{P}^{2/3}} \ ,
\label{soc2}
\end{equation}
while $\hat R$ can be fixed after noticing that all of the above constraints are invariant under the scaling
\begin{equation}
a^{i} \to \frac{\epsilon_S^\prime \hat R}{V^{1/3}} a^{i} \ .
\label{wall1}
\end{equation}
It follows that the coefficient $\epsilon_S^\prime \hat R/V^{1/3}$ can be set to one. We refer to this choice as ``unity'' gauge
\begin{equation}
\epsilon_S'\frac{\Rhat}{V^{1/3}} = 1 \ .
\label{wall2}
\end{equation}
Finally, we note that the expression for $\re f_{1}$  in \eqref{lala1} was found in \cite{Ashmore:2020ocb} to be
\begin{equation}
\re f_{1}=V+\epsilon_S'\frac{\Rhat}{V^{1/3}} \left(\tfrac{1}{3} a^{1} -\tfrac{1}{6} a^{2} +2a^{3}+\tfrac{1}{2}\bigl(\tfrac{1}{2}-\lambda\bigr)^{2} W_{i}a^{i}\right) \ .
\label{white2A}
\end{equation}

\subsubsection*{Strong Coupling Constraint}

A final condition we impose is the ``strong coupling'' constraint. For our theory to be valid in a strongly coupled string regime, the observable and hidden walls must be sufficiently separated. We therefore require that the length of the $S^{1}/{\mathbb{Z}}_{2}$ orbifold interval is larger than the average Calabi--Yau radius. That is,
\begin{equation}
\frac{\pi \rho {\Rhat} V^{-1/3}}{(vV)^{1/6} } > 1 \ .
\label{seb3}
\end{equation}

A solution to the set of constraints discussed above, that is, the observable sector, hidden sector, vacuum, phenomenological and strong coupling constraints, is possible in the positive  K\"ahler cone.  As can be seen from \eqref{coup2}, the condition $(g^{(2)})^2>0$ is most easily satisfied when the value of $\lambda$ is as large as possible; that is, for the five-brane to be near the hidden wall. Note, however, that we will not simply set $\lambda=1/2$, so as to avoid unwanted ``small instanton'' transitions in the hidden sector \cite{Ovrut:2000qi}. Instead we will choose the value 
\begin{equation}
\lambda=0.49
\label{wall3}
\end{equation}
for the remainder of this paper, as was done in \cite{Ashmore:2020ocb}. Note also that equation \eqref{xf1} which sets $\mu(L)=0$ on the hidden wall can only have a solution for ample line bundles, that is, for $L=\mathcal{O}_X(l^1,l^2,l^3)$ where the $l^i$ all have the same sign. 

\begin{figure}[t]
   \centering
\includegraphics[width=0.6\textwidth]{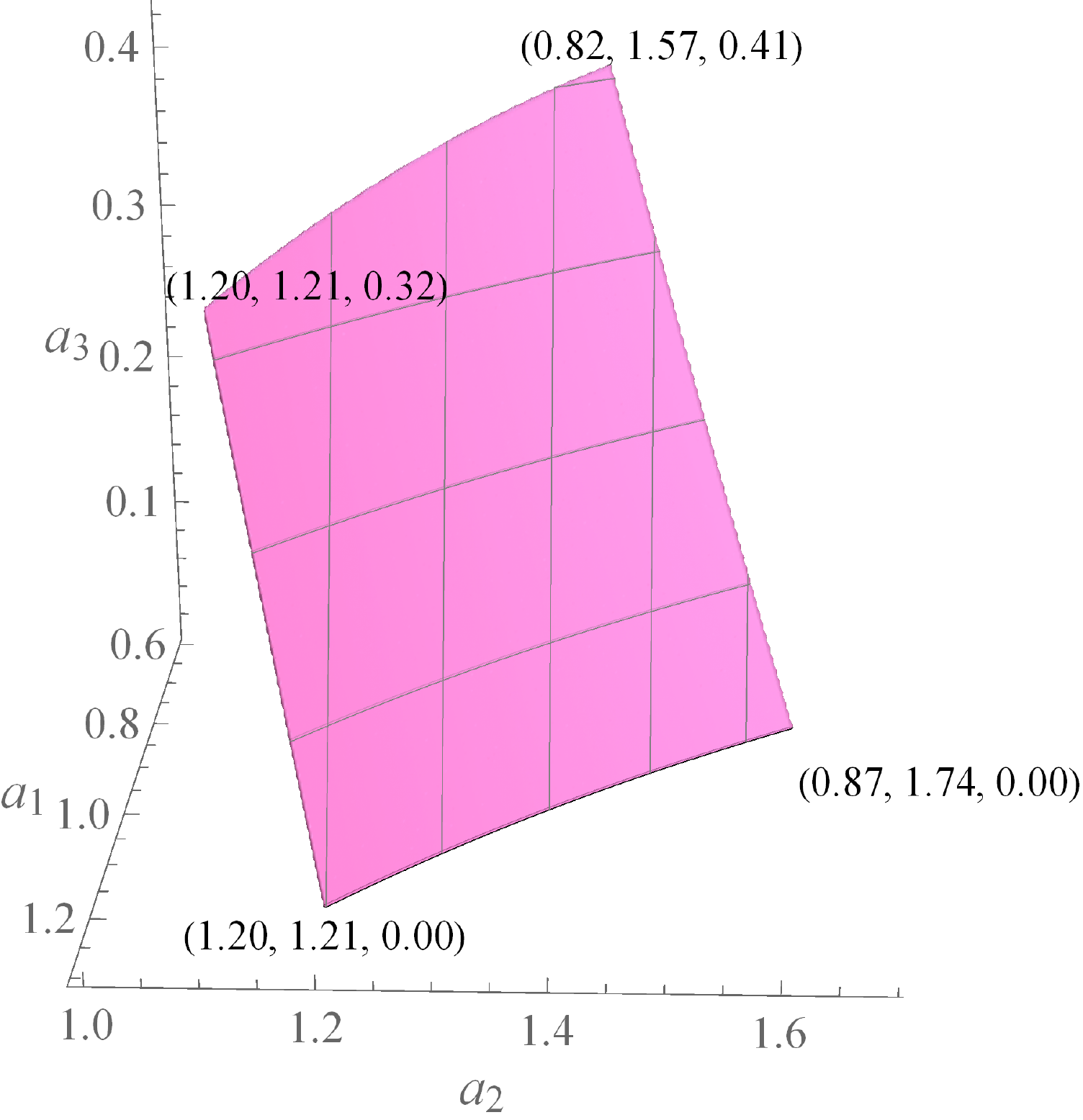}
\caption{ ``Viable'' region of K\"ahler moduli space space that satisfies all constraints in Section 2 for the line bundle $L= \mathcal{O}_X(2,1,3)$.}
\label{fig:KahlerViableRegion}
\end{figure}

It is straightforward to show that a large number of ample line bundles satisfy all of the above constraints. However, as mentioned in \cite{Ashmore:2020ocb}, and proven in detail in the following section, only a small subset of these, namely
\begin{gather}
\mathcal{O}_X(2,1,3)\ , \qquad \mathcal{O}_X(1,2,3)\ , \qquad \mathcal{O}_X(1,2,2)\ ,  \qquad\mathcal{O}_X(2,1,2)\ , \nonumber \\
\mathcal{O}_X(2,1,1)\ ,  \qquad\mathcal{O}_X(1,2,1)\ ,  \qquad\mathcal{O}_X(2,1,0) \ ,
\label{many1}
\end{gather}
can lead to $E_{7}$ gaugino condensation in the hidden sector that is compatible with realistic spontaneous breaking of $d=4$, $N=1$ supersymmetry. In \cite{Ashmore:2020ocb}, we solved the system of constraints and displayed the solution surface in K\"ahler moduli space for the specific hidden sector line bundle $\mathcal{O}_X(2,1,3)$ with the five-brane at $\lambda=0.49$. The reason for choosing $L=\mathcal{O}_X(2,1,3)$ over the other ample line bundles in \eqref{many1} will become clear after we discuss the gaugino condensation mechanism in the next section. We find that for $L=\mathcal{O}_X(2,1,3)$, the matter spectrum of the hidden sector is such as to maximally reduce the scale of gaugino condensation. Hence, this choice for $L$ provides the most promising model for spontaneous breaking of $N=1$ supersymmetry within the range of the LHC. A plot of the sub-region of K\"ahler moduli space in which all points satisfy all of the constraints listed in this section for line bundle $L=\mathcal{O}_X(2,1,3)$ was presented in \cite{Ashmore:2020ocb}. In the following section, we will use this ``viable'' region to discuss the full range of the supersymmetry breaking associated with this line bundle. Therefore, for clarity, we present the plot again in Figure \ref{fig:KahlerViableRegion}. Any point in this ``viable'' region of K\"ahler moduli space is consistent with {\it both} the split and simultaneous Wilson lines scenarios, but with somewhat different values for the fundamental parameters.

Before continuing, we point out that in \cite{Ashmore:2020ocb}--because of the ``color-coding'' employed in that paper--we referred to this ``viable'' region of K\"ahler moduli space as the ``magenta'' region. To be consistent with this previous terminology, we will henceforth in the present paper use the terms ``viable'' region and ``magenta'' region interchangeably.

\section{Gaugino Condensation}

In the previous section, we reviewed the constraints imposed on a heterotic M-theory vacuum whose hidden sector is defined by a single line bundle $L$ with its $U(1)$ structure group embedded into the $SU(2)$ subgroup of $SU(2) \times E_{7} \subset E_8$ via the induced vector bundle $L \oplus L^{-1}$. As in \cite{Ashmore:2020ocb}, we demanded that $d=4$,  $N=1$ supersymmetry be exactly preserved.  In this paper, however, we will analyze how spontaneous supersymmetry breaking in four dimensions can occur due to gaugino condensation of $E_{7}$ in the hidden sector.

As is well known \cite{Dine:1985rz,Lukas:1997rb,Nilles:1998sx}, if the $E_{7}$ gauge group of the hidden sector becomes strongly coupled below the compactification scale, then the associated gauginos condense and produce an effective superpotential for the relevant geometric moduli of the theory -- in our case, the dilaton $S$ and the complexified K\"ahler moduli $T^{i}$, $i=1,2,3$. The form of this gaugino condensate superpotential is given by~\cite{Dine:1985rz,Lukas:1997rb,Lukas:1999kt}
\begin{equation}
W=\langle M_{U} \rangle ^3 \exp\left({-\frac{6\pi}{b_L \hat{\alpha}_{\text{GUT}}}f_2}\right)\ ,
\label{sup1}
\end{equation}
where $\hat{\alpha}_{GUT}$ was defined in \eqref{lala1}, and $b_{L}$ is a real number associated with the beta-function of the $E_{7}$ gauge coupling $g^{(2)}$ for a choice of hidden sector line bundle $L$. Both $b_{L}$  and the associated ``condensation scale'' $\Lambda$ will be defined below. Furthermore, for any line bundle $L=\mathcal{O}_X(l^1,l^2,l^3)$  in the hidden sector, we find that
\begin{equation}
f_2=S+\frac{\epsilon_S^\prime}{2}\bigl( -(2,2,0)_{i}-d_{ijk}l^{j}l^{k}\bigr)T^{i} \ ,
\label{sup2}
\end{equation}
where $T^{i}=t^{i}+\ii 2\chi$,  $i=1,2,3$ and the strong coupling parameter $\epsilon_S^\prime$ was defined in \eqref{soc2}.
We want to emphasize that expressions \eqref{sup1} and \eqref{sup2} for the condensate superpotential and $f_{2}$ respectively are known only to linear order in $\epsilon_S^\prime$~\cite{Lukas:1997fg,Lukas:1998tt,Brandle:2001ts,Brandle:2003uya}.

An important point is that the potential induced by the gaugino superpotential is actually runaway and so does not fix the moduli fields at a minimum. Instead, we simply assume that the moduli can be fixed to some values (such as by picking a point in the viable region of K\"ahler moduli space) and then analyze the supersymmetry breaking due to the gaugino superpotential at those values of the moduli. In a complete model, one would be able to fix all of the moduli fields dynamically. However this is far outside the scope of the present paper.\footnote{One could use gaugino condensation in multiple gauge groups to generate a potential with a determined minimum and thus fix some of the moduli. However, this would still leave bundle moduli, complex structure moduli, and so on, to fix~\cite{Anderson:2010mh, Anderson:2011cza, Anderson:2011ty,Correia:2007sv}. To the knowledge of the authors, within the context of phenomenologically realistic models, there are \emph{no} models that fix all moduli.}

The observable sector $Spin(10)$ gauge coupling $g^{(1)}$ is fixed at the unification scale $\langle M_U \rangle$ based on phenomenological data. As discussed in \cite{Ashmore:2020ocb} and Section 2, we find that $\langle \alpha_{u}\rangle=\frac{1}{20.08}$ in the split Wilson lines scenario and $\langle \alpha_{u}\rangle=\frac{1}{26.64}$ in the simultaneous Wilson lines scenario (with $\alpha=g^{(1)2}/4\pi$). We cannot determine the value of the hidden sector gauge coupling $g^{(2)}$ at $\langle M_U \rangle$ based on direct observation. However, the two gauge couplings $g^{(1)}$ and $g^{(2)}$ both occur in the $d=10$ Hořava--Witten theory and, hence, it is possible to find an expression relating them to couplings in the $d=4$ effective theory after compactification. As shown in \eqref{lala1}, in the observable sector the $Spin(10)$ gauge coupling at the unification scale is given by~\cite{Banks:1996ss,Lukas:1997fg}
\begin{equation}
\langle \alpha_{u}\rangle=\frac{\hat \alpha_{\text{GUT}}}{\re f_1}\ ,
\label{need2}
\end{equation}
where $\re f_{1}$ is presented in \eqref{white2A}. Note that, for consistency of notation with previous work \cite{Ashmore:2020ocb}, throughout this paper we continue to denote $\frac{g^{(1)2}}{4\pi}$ in the observable sector simply as $\alpha$. Similarly, the $E_7$ hidden sector gauge coupling at the unification scale $\langle M_{U} \rangle$ is defined to satisfy
\begin{equation}
\langle \alpha^{(2)}_{u}\rangle=\frac{\hat \alpha_{\text{GUT}}}{\re f_2}\ ,
\label{need3A}
\end{equation}
where we denote the hidden sector gauge parameter by $\alpha^{(2)}=\frac{g^{(2)2}}{4\pi}$. The moduli dependent function $f_{2}$ was given in \eqref{sup2}. Using the relations 
\begin{equation}
\re S=V+\frac{\epsilon_{S}'}{2}\bigl(\tfrac{1}{2}+\lambda\bigr)^{2}W_{i}t^{i} 
\label{cl1}
\end{equation}
and
\begin{equation}
\re T^{i}=t^{i}=\frac{\hat{R}}{V^{1/3}} \ ,
\label{cl2}
\end{equation}
it follows from \eqref{sup2} that 
\begin{equation}
\re f_{2}=V+\epsilon_{S}'\frac{\hat{R}}{V^{1/3}}\left(\bigl(-(1,1,0)_{i}-\tfrac{1}{2}d_{ijk}l^{j}l^{k}\bigr)a^{i}+\bigl(\tfrac{1}{2}+\lambda\bigr)^{2}W_{i}a^{i}\right) \ .
\label{cl3}
\end{equation}
Finally, using \eqref{need2} and \eqref{need3A}, we find that
\begin{equation}
\langle \alpha^{(2)}_{u}\rangle=\frac{\re f_1}{\re f_2}\langle \alpha_{u}\rangle ,
\end{equation}
allowing one to solve for $\langle \alpha^{(2)}_{u}\rangle$ at any point in moduli space. Importantly, it is straightforward to show that any point in K\"ahler moduli space satisfying the constraints \eqref{coup1} and \eqref{coup2}, has positive definite values for ${\rm{Re}}f_{1}$ and ${\rm{Re}}f_{2}$ respectively. Hence, $\langle \alpha^{(2)}_{u}\rangle$ will be positive definite for any such moduli.

For an arbitrary momentum $p$ below the unification scale, the renormalization group equation for the hidden sector gauge parameter $\alpha^{(2)}$ is given by
\begin{equation}
\alpha^{(2)}(p)^{-1}=\langle \alpha_{u}^{(2)}\rangle^{-1}-\frac{b_L}{2\pi} \ln \left(\frac{ \langle M_U \rangle}{p}\right)\ .
\label{tea1}
\end{equation} 
Note that for $b_{L} < 0$, the value of $\alpha^{(2)}(p)^{-1}$ is identical to the perturbative gauge coupling $\langle \alpha_{u}^{(2)}\rangle^{-1}$ for $p= \langle M_U \rangle$ and only becomes more weakly coupled for $p< \langle M_U \rangle$. It follows that the $E_{7}$ gauge group never becomes strongly coupled in the effective theory and, therefore, gaugino condensation can never occur. Therefore, we will only consider line bundles $L$ for which 
\begin{equation}
b_{L}>0. 
\label{bl1}
\end{equation}
In this case,
roughly speaking, the hidden sector $E_{7}$ gauge theory becomes strongly coupled and, hence, its gauginos condense, at a momentum $p \approx \Lambda$ where $ \alpha^{(2)}(\Lambda)^{-1}$ can be  well approximated by $0$. It then follows from \eqref{tea1} that, at this scale, 
\begin{equation}
\langle \alpha_{u}^{(2)}\rangle^{-1}=\frac{b_L}{2\pi} \ln \left(\frac{\langle M_U\rangle}{\Lambda}\right)\ .
\label{tea2}
\end{equation}
The condensation scale $\Lambda$ can then be expressed as 
\begin{equation}
\label{eq:condesation_scale}
\Lambda=\langle M_U\rangle \ee^{\frac{-2\pi}{b_L} \langle \alpha^{(2)}_{u}\rangle^{-1}}=\langle M_U \rangle \ee^{\frac{-2\pi}{b_L}\frac{\re f_2}{\re f_1\langle \alpha_u \rangle}} \ .
\end{equation}
We reiterate that the functions $\re f_1$ and  $\re f_2$ are known, positive definite and presented in \eqref{white1} and \eqref{cl3} for any line bundle $L=\mathcal{O}_X(l^1,l^2,l^3)$ satisfying the constraints in Section 2. It follows that $\Lambda$, like $W$ defined in \eqref{sup1}, is a function of the K\"ahler moduli. Furthermore, $\langle \alpha_{u} \rangle$ is given in \eqref{bag3} and \eqref{bag31} for the split and simultaneous Wilson lines scenarios respectively. It remains to compute the coefficient $b_L$. 

For any line bundle $L=\mathcal{O}_X(l^1,l^2,l^3)$ in the hidden sector satisfying all constraints in Section 2, the beta-function coefficient $b_L$ for the $E_{7}$ gauge coupling is given by
\begin{equation}
b_L=3\,T(\Rep{133})-\sum_{\Rep{r}}n_{\Rep r}T(\Rep{r}) \ .
\label{hope1}
\end{equation}
Here, the sum is over the $E_{7}$ representations $\Rep{r}$ that arise in the decomposition of the of the adjoint representation $\Rep{248}$ of $E_{8}$ with respect to the low-energy hidden sector gauge group $U(1) \times E_{7}$, and the coefficients $n_{\Rep{r}}$ are the number of light chiral matter fields which transform as $\Rep{r}$.  $T(\Rep{r})$ denotes the Dynkin index of the representation $\Rep r$, defined by
\begin{equation}
T(\Rep{r})=\text{tr}_{\Rep{r}}{\mathrm T}_a^2 \ ,
\end{equation}
where ${\mathrm T}_a$ is an arbitrary Lie algebra generator of $E_{7}$ in the representation $\Rep{r}$. See, for example, \cite{Yamatsu:2015npn}. It follows that calculating the beta-function coefficient $b_{L}$ for a specific line bundle $L$ requires one to explicitly compute the particle content of its low-energy theory. To do this, we first recall that all line bundles under consideration have their $U(1)$ structure group embed into the $SU(2)$ subgroup of $SU(2) \times E_{7} \subset E_{8}$. It follows that for all such line bundles, the adjoint $\Rep{248}$ representation of the hidden sector $E_8$ decomposes under $U(1) \times\Ex 7$ as
\begin{equation}
\Rep{248} \to 
(0, \Rep{133}) \oplus 
\bigl( (1, \Rep{56}) \oplus (-1, \Rep{56})\bigr) \oplus 
\bigl( (2, \Rep{1}) \oplus (0, \Rep{1}) \oplus (-2, \Rep{1}) \bigr)\ .
\label{red55}
\end{equation}
The $(0,\Rep{133})$ corresponds to the adjoint representation of $\Ex 7$, while the $(\pm1,\Rep{56})$ give rise to chiral matter superfields with  $\pm1$ $U(1)$ charges transforming in the $\underline{\bf56}$ representation of $\Ex 7$ in four dimensions. 
The $(\pm2,\Rep 1)$ are $E_7$ singlet chiral superfields fields with charges $\pm 2$ under $U(1)$. Finally, the $(0,\Rep{1})$ gives the one dimensional adjoint representation of the $U(1)$ gauge group. The embedding of the line bundle is such that fields with $U(1)$ charge $-1$ are counted by $H^{*}(X,L)$, charge $-2$ fields are counted by $H^{*}(X,L^{2})$ and so on.

The low-energy massless spectrum can be determined by examining the chiral fermionic zero-modes of the Dirac operators for the various representations in the decomposition of the $\Rep{248}$. The Euler characteristic $\chi(\mathcal{F})$ counts $n_{\text{R}}-n_{\text{L}}$, where $n_{\text R}$ and $n_{\text L}$ are the number of right- and left-chiral zero-modes respectively transforming under the group representation associated with the bundle $\mathcal{F}$. With the notable exception of $\mathcal{F}=\mathcal{O}_{X}$, which corresponds the massless vector superfields of both the  $(0, \Rep{133})$ and $(0, \Rep{1})$ adjoint representations of $E_{7}$ and $U(1)$ respectively, right-chiral and left-chiral zero modes pair up and form massive fermion states, which can then be integrated out of the low-energy theory. However, the remaining unpaired zero modes of the $d=4$ effective theory remain massless and are precisely those counted by the Euler characteristic for each representation. On a Calabi--Yau threefold $X$,  $\chi(\mathcal{F})$ can be computed using the Atiyah--Singer index theorem and is given by
\begin{equation}
\chi(\mathcal{F})=\int_{X}{\rm {ch}}(\cal{F}) \wedge {\rm {Td}}({\rm X}) \ ,
\label{ni1}
\end{equation}
where ${\rm {ch}}(\cal{F})$ is the Chern character of $\mathcal{F}$ and ${\rm {Td}}({\rm X})$ is the Todd class of the tangent bundle of $X$.
It is useful to note that for a line bundle of the form ${\cal{F}}=\mathcal{O}_X(f^1,f^2,f^3)$, expression \eqref{ni1} simplifies to
\begin{equation}
\chi({\cal{F}})= \frac{1}{3}(f^{1}+f^{2})+\frac{1}{6}d_{ijk}f^{i}f^{j}f^{k} \ .
\label{st1}
\end{equation}
For the decomposition of the $\Rep{248}$ presented in \eqref{red55}, the bundles ${\cal{F}}$ corresponding to the $U(1) \times E_7$ representations $(0, \Rep{133})$, $(\pm1, \Rep{56})$, $(\pm2, \Rep{1})$ and 
$(0, \Rep{1})$ are $\mathcal{F}=\mathcal{O}_{X}, L^{\mp1}, L^{\mp2}$ and $\mathcal{O}_{X}$ respectively. Using the fact that the $E_{7}$ singlet states $(\pm2, \Rep{1})$ and $(0, \Rep{1})$ cannot contribute to $b_{L}$, that is, 

\begin{equation}
T({\Rep1})=0 \ ,
\label{no1}
\end{equation}
we need only consider $(0, \Rep{133})$ and$(\pm1, \Rep{56})$. Furthermore, since the $E_{7}$ adjoint representation can occur only once in the spectrum, it follows that we need to determine only the number of states $(\pm1, \Rep{56})$. Of these, the low-energy spectrum consists of the left chiral states only, which correspond to $(+1, \Rep{56})$. They are counted by the Euler characteristic of the line bundle $L^{-1}$. It follows from \eqref{st1} that for $L=\mathcal{O}_X(l^1,l^2,l^3)$, 
\begin{equation}
\chi(L^{-1})=-\frac{1}{6}\bigl(2l^1+2l^2+({l^1})^2l^2+l^{1}({l^2})^2+6l^{1}l^2l^3\bigr)\ .
\label{tr1}
\end{equation}

With this information about the low-energy spectrum, we can now compute the $b_L$ coefficient for any line bundle $L=\mathcal{O}_X(l^1,l^2,l^3)$ in the hidden sector satisfying all constraints in Section 2. 
Noting \cite{Yamatsu:2015npn} that the Dynkin indices satisfy \eqref{no1} and 
 \begin{equation}
  T(\Rep{133})=18 \quad T(\Rep{56})=6 \ ,
  \label{cut1}
 \end{equation}
we find, using \eqref{hope1}, that
 \begin{equation}
 \label{beta_expr}
 b_{L}=3\times T(\Rep{133})-|\chi(L^{-1})|\times T(\Rep{56})=54-6|\chi(L^{-1})| 
 \end{equation} 
 and, hence, from \eqref{tr1} that
 \begin{equation}
 b_{L}=54-| 2 l^{1} +2 l^{2}+(l^{1})^2 l^2+l^{1}(l^{2})^2+6l^{1}l^{2}l^{3}| \ .
 \label{cut2}
 \end{equation} 
 We can now explain why, out of all the line bundles satisfying the constraints in Section 2, only the seven ample line bundles listed in \eqref{many1} can exhibit $E_{7}$ gaugino condensation. Furthermore, we can now justify the preference for the line bundle $L=\mathcal{O}_X(2,1,3)$ in both our previous paper \cite{Ashmore:2020ocb} and in the rest of the present work.

In Section 2 we introduced a new constraint; that is, we demanded that the line bundle $L$ be such that $N=1$ supersymmetry is spontaneously broken by the condensation of the $E_{7}$ gauginos. However, as discussed above, for momenta below the unification scale $\langle M_{U} \rangle$, the $E_7$ gauge theory becomes strongly coupled if and only if condition \eqref{bl1} is satisfied. Therefore, for a line bundle $L=\mathcal{O}_X(l^1,l^2,l^3)$ satisfying all of the constraints in Section 2, we now impose, using \eqref{cut2},  the extra condition that
\begin{equation}
54-\left|2 l^{1} +2 l^{2}+(l^{1})^2 l^2+l^{1}(l^{2})^2+6l^{1}l^{2}l^{3}\right|>0\ .
\label{al1}
\end{equation}
We find that the only line bundles that satisfy all the constraints in Section 2, as well as this new constraint, are 
\begin{gather}
\mathcal{O}_X(2,1,3)\ , \qquad \mathcal{O}_X(1,2,3)\ , \qquad \mathcal{O}_X(1,2,2)\ ,  \qquad\mathcal{O}_X(2,1,2)\ , \nonumber \\
\mathcal{O}_X(2,1,1)\ ,  \qquad \mathcal{O}_X(1,2,1)\ ,  \qquad\mathcal{O}_X(2,1,0) \ ,
\label{many1A}
\end{gather}
which are exactly those presented in \eqref{many1} of Section 2. Calculating the $b_{L}$ coefficient for each of these bundles using \eqref{cut2}, we find that
\begin{itemize}
\item{$\mathcal{O}_X(2,1,3)$ and $\mathcal{O}_X(1,2,3)$: $b_{L}=6$;}
\item{$\mathcal{O}_X(2,1,2)$ and $\mathcal{O}_X(1,2,2)$: $b_{L}=18$;}
\item{$\mathcal{O}_X(2,1,1)$ and $\mathcal{O}_X(1,2,1)$: $b_{L}=30$;}
\item{ $\mathcal{O}_X(2,1,0)$: $b_{L}=42$.}
\end{itemize}
For any other line bundle satisfying all the constraints given in Section 2, one can show that $b_{L}<0$. For example, 
for $\mathcal{O}_X(2,1,4)$ and $\mathcal{O}_X(1,2,4)$, $b_{L}$ has already become negative; that is $b_{L}=-6$. 

So which of the above line bundles is, from the point of view of low-energy phenomenology, most interesting to study? It is clear from expression \eqref{eq:condesation_scale} that the gaugino condensation scale $\Lambda$, at any fixed point in K\"ahler moduli space, will be smallest for the line bundle $L$ with the lowest value of $b_{L}$. For this reason, we will focus in on the two line bundles $\mathcal{O}_X(2,1,3)$ and $\mathcal{O}_X(1,2,3)$, each of which was shown above to have $b_{L}=6$. Furthermore, a detailed study shows that the ``viable'' region of K\"ahler moduli space, each point of which satisfies all the constraints in Section 2, is considerably larger for the first of these two bundles. Therefore, we will focus on the line bundle $\mathcal{O}_X(2,1,3)$, as was done in \cite{Ashmore:2020ocb}.
For completeness, in Table \ref{tab:chiral_spectrum} we present the complete low energy spectrum, the Euler characteristics and the Dynkin coefficients for the line bundle $L=\mathcal{O}_{X}(2,1,3)$.
 
 \begin{table}
	\noindent \begin{centering}
		\begin{tabular}{rrrr}
			\toprule 
			$U(1) \times \Ex 7$ & Cohomology & Index $\chi$ &$T(\Rep{r})$\tabularnewline
			\midrule
			\midrule 
			$(0,\Rep{133})$ & $H^{*}(X,\mathcal{O}_{X})$ & $0$&18\tabularnewline
			\midrule 
			$(0,\Rep 1)$ & $H^{*}(X,\mathcal{O}_{X})$ & $0$&0\tabularnewline
			\midrule 
			$(-1,\Rep{56})$ & $H^{*}(X,L)$ & $8$&6\tabularnewline
			\midrule 
			$(1,\Rep{56})$ & $H^{*}(X,L^{-1})$ & $-8$&6\tabularnewline
			\midrule 
			$(-2,\Rep 1)$ & $H^{*}(X,L^{2})$ & $58$&0\tabularnewline
			\midrule 
			$(2,\Rep 1)$ & $H^{*}(X,L^{-2})$ & $-58$&0\tabularnewline
			\bottomrule
		\end{tabular}
		\par\end{centering}
	\caption{The chiral spectrum for the hidden sector $\protect\Uni 1\times\protect\Ex 7$ with a single line bundle $L=\mathcal{O}_{X}(2,1,3)$. The Euler characteristic (or index) $\chi$ gives the difference between the number of right- and left-chiral fermionic zero-modes transforming in the given representation. We denote the line bundle dual to $L$ by $L^{-1}$ and the trivial bundle $L^{0}$ by $\mathcal{O}_{X}$.\label{tab:chiral_spectrum}}
\end{table}

\begin{figure}[t]
   \centering
     \begin{subfigure}[b]{0.495\textwidth}
\includegraphics[width=1.0\textwidth]{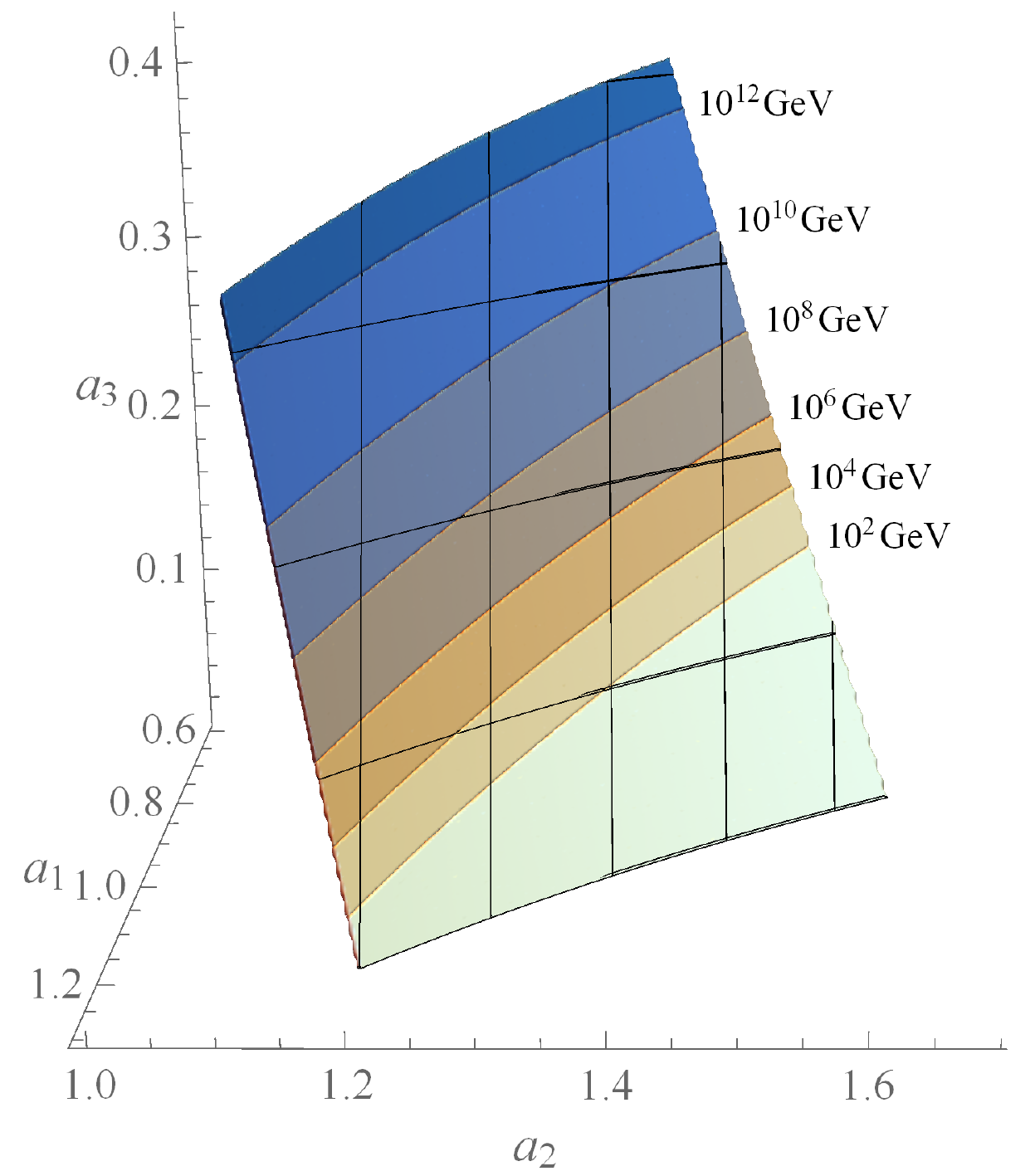}
\caption{$\langle \alpha_{u} \rangle=\frac{1}{20.08}$}
\end{subfigure}
     \begin{subfigure}[b]{0.495\textwidth}
\includegraphics[width=1.0\textwidth]{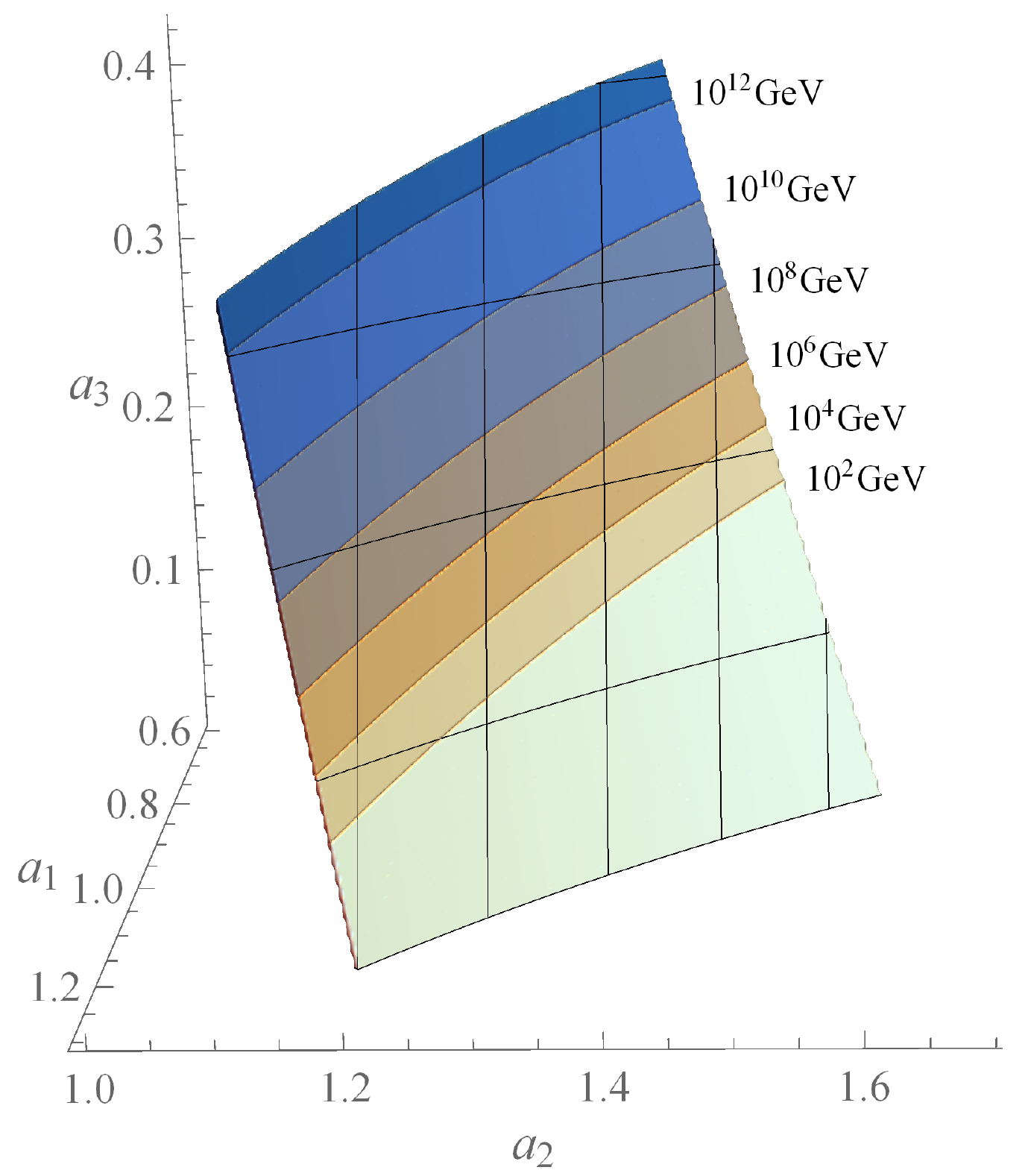}
\caption{$\langle \alpha_{u} \rangle=\frac{1}{26.46}$}
\end{subfigure}\\
\caption{Variation of the mass scale $m_{\text{susy}}\sim 8\pi\Lambda^3/M_p^2$ of the soft breaking terms across the ``viable'' region of K\"ahler moduli space space that satisfies all constraints in Section 2, for the line bundle $L= \mathcal{O}_X(2,1,3)$. The numbers indicate the $m_{\text{susy}}$ value corresponding to each contour.  Figure 2(a) and Figure 2(b) respectively, show the results for both the split and the simultaneous Wilson lines scenarios described in Section 2.  $m_{\text{susy}}$ scales below the EW scale $\sim10^2$ GeV become unphysically small and, therefore, are not displayed.}
\label{fig:susyScale_both}
\end{figure}

In a following section, we will explicitly calculate the soft supersymmetry breaking terms in the observable matter sector. Here, however, we can use what we have learned so far to predict the scale of those soft SUSY breaking terms. 
Supersymmetry breaking first occurs in the $S$ and $T^{i}$ moduli via the gaugino condensate superpotential $W$ given in \eqref{sup1}. Here, it is useful to note that writing $f_{2}=\re f_{2}+\ii \imag f_{2}$, and using \eqref{need3A} and \eqref{tea2}, it follows that 
\begin{equation}
W=\Lambda^{3}  \ee^{-\ii \frac{6 \pi}{b_{L}\hat{\alpha}_{\text GUT}}\imag f_{2}} \ .
\label{mk1}
\end{equation}
The supersymmetry breaking in the $S$ and $T^{i}$ moduli is then gravitationally mediated to the observable matter sector. The scale of SUSY breaking in the low-energy observable matter sector is then of order
\begin{equation}
m_{\rm{susy}} \sim \kappa_{4}^{2} \Lambda^3=8 \pi \frac{\Lambda^{3}}{M_{P}^{2}} \ ,
\label{ew1}
\end{equation}
where $\Lambda$ is the condensation scale given in \eqref{eq:condesation_scale} and we have used the fact that $\kappa_{4}^{2}=8\pi/M_{P}^{2}$. For $L=\mathcal{O}_{X}(2,1,3)$, it follows from \eqref{xf3} that $W_{i}=(9,17,0)_{i}$. Furthermore, in the unity gauge defined in \eqref{wall2}, $\re f_{1}$ and $\re f_{2}$ presented in \eqref{white2A} and \eqref{cl3} become
\begin{equation}
\re f_{1}=V+\frac{1}{3}a^{1}-\frac{1}{6}a^{2}+2a^{3}+\frac{1}{2}\bigl(\tfrac{1}{2}-\lambda\bigr)^{2}(9a^{1}+17a^{2})
\label{mw1}
\end{equation}
and
\begin{equation}
\re f_{2}=V-\frac{29}{6}a^{1}-\frac{25}{3}a^{2}-2a^{3}+\frac{1}{2}\bigl(\tfrac{1}{2}+\lambda\bigr)^{2}(9a^{1}+17a^{2})
\label{mw2}
\end{equation}
respectively, where from \eqref{wp1} we find that
\begin{equation}
V=\frac{1}{6}\bigl((a^{1})^{2}a^{2}+a^{1}(a^{2})^{2}+6a^{1}a^{2}a^{3}\bigr) \ .
\label{mw3}
\end{equation}
Using these results, as well as $\langle M_{U}\rangle$ given in \eqref{jack1}, $b_{L}=6$ and the values for $\langle \alpha_{u} \rangle$ presented in \eqref{bag3} and \eqref{bag31} for the split and unified Wilson lines scenarios respectively, one can compute $\Lambda$ in \eqref{eq:condesation_scale} and, hence, using \eqref{ew1} the value of $m_{\rm{susy}}$. Clearly, the result is a function of the point in   moduli space where $m_{\rm{susy}}$ is evaluated. Plots of $m_{\rm{susy}}$ evaluated over the ``viable'' region of K\"ahler moduli space for $L=\mathcal{O}_{X}(2,1,3)$ where all constraints in Section 2 are satisfied for both the split and unified Wilson lines scenarios are shown in Figure 2(a) and Figure 2(b) respectively.

\section{Spontaneous Supersymmetry Breaking in the Low-Energy Effective Theory}

Spontaneous $N=1$ supersymmetry breaking, induced by gaugino condensation in the hidden sector, appears in the low-energy effective theory as potentially non-vanishing $F$-terms in the chiral superfields of the dilaton $S$, the complexified K\"ahler moduli $T^{i}$ and the single five-brane modulus $Z$. Defining the index $a$ which runs over $(S, T^{1}, T^{2}, T^{3}, Z)$, the $F$-terms are given by
\begin{equation}
\bar F^{\bar b}=\kappa_4^2 \ee^{\hat K/2}\hat K^{\bar b a}(\partial_a W+W\partial_a \hat K) \ ,
\label{f1}
\end{equation}
where $\kappa_{4}^{2}=8\pi/M_{P}^{2}$ and $W$ is the gaugino condensate superpotential given in \eqref{sup1}. In addition, these $F$-terms depend strongly on the dimensionless K\"ahler potential $\hat{K}$. Before continuing, we note that since the scalar fields of all chiral multiplets have been chosen to be dimensionless, it follows that all $F$ auxiliary fields have dimension one.  Since $\kappa_{4}^{2}~ W \propto \Lambda^{3}/M_{P}^{2}$, it follows that the mass scale of each $F$-term is ${\cal{O}}(8\pi\Lambda^{3}/M_{P}^{2})$. Thus, although gaugino condensation and the superpotential $W$ occur in the hidden sector, their supersymmetry breaking effects on the $S$, $T^{i}$ and $Z$ moduli are mediated via gravitational interactions and thus are Planck mass suppressed. The  K\"ahler potential $\hat{K}$ was defined in \cite{Lukas:1999kt} and is given by
\begin{equation}
\hat K=\tilde K_S+K_T\ ,
\label{sum1}
\end{equation}
where
\begin{equation}
\begin{split}
\tilde K_S=&-\ln \left(  S+\bar S-\frac{\epsilon_S^\prime}{2}\frac{(Z+\bar Z)^2}{W_{i} (T+\bar T)^i} \right)\\
\simeq&-\ln(S+\bar S)+\frac{\epsilon_S^\prime}{2}\frac{(Z+\bar Z)^2}{(S+\bar S)W_i(T+\bar T)^i}\ ,
\label{split2}
\end{split}
\end{equation}
\begin{equation}
Z=W_it^iz+2iW_i(-\eta^i\nu+\chi^iz), \quad \text{with} \quad t^i=\tfrac{1}{2}(T+\bar T)^i,
\end{equation}
and 
\begin{equation}
K_T=-\ln \left( \tfrac{1}{48}d_{ijk}(T+\bar T)^i(T+\bar T)^j(T+\bar T)^k \right)
\label{tel1}
\end{equation}
Note that the second expression in \eqref{split2} is the expansion of $\hat{K}_S$ to linear order in $\epsilon_S^\prime$.

Given these K\"ahler potentials, it is tedious, but straightforward, to work out the expressions for $\ee^{\hat{K}/2}$, $\partial_{a}{\hat{K}}$ and $\hat{K}_{a\bar{b}}$. The results are the following. First of all, we find that the $e^{\hat K/2}$ factor is given by
\begin{equation}
\begin{split}
\ee^{\hat K/2}%&=\text{exp}\left[-\ln \left( S+\bar S-\frac{\epsilon_S^\prime}{2}\frac{Z+\bar Z}{W_i(T^i+\bar T^i)} \right)^{1/2}-\ln\left(\frac{1}{48}d_{ijk}(T+\bar T)^i(T+\bar T)^j(T+\bar T)^k\right)^{1/2}\right]\\
&=\left((S+\bar S)-\frac{\epsilon_S^\prime}{2}\frac{Z+\bar Z}{W_i(T+\bar T)^i}\right)^{-1/2}\left(\tfrac{1}{48}d_{ijk} (T+\bar T)^i(T+\bar T)^j(T+\bar T)^k \right)^{-1/2}\\
&=\left[ \frac{1}{\sqrt{2}V^{1/2}}-\frac{\left( \lambda+\frac{1}{2} \right)}{4\sqrt 2 V^{3/2}}\left( \left(\lambda+\tfrac{1}{2}\right)(9a^1+17a^2)-\frac{\epsilon_S^\prime}{2}\right) \right]\frac{1}{\hat R^{3/2}}
\end{split}
\label{exp}
\end{equation}
where, as always, we work to linear order in $\epsilon_S^\prime$ only. Second, it is straightforward to compute $\partial_{a}{\hat{K}}$ for $a=S,T^{1},T^{2},T^{3},Z$. The results are
\begin{equation}
\begin{split}
\frac{\partial \hat K}{\partial \bar S}&=-\frac{1}{S+\bar S}-\frac{\epsilon_S^\prime}{2}\frac{(Z+\bar Z)^2}{(S+\bar S)^2W_i(T+\bar T)^i}\ ,\\
\frac{\partial \hat K}{\partial \bar T^l}&=-\frac{\frac{1}{16}d_{lij}(T+\bar T)^i(T+\bar T)^j}{\frac{1}{48}d_{ijk}(T+\bar T)^i(T+T)^j(T+\bar T)^k}-\frac{\epsilon_S^\prime}{2}\frac{(Z+\bar Z)^2W_l}{(S+\bar S)(W_i(T+\bar T)^i))^2}\ ,\\
\frac{\partial \hat K}{\partial \bar Z}&=\frac{\epsilon_S^\prime(Z+\bar Z)}{(S+\bar S)W_i(T+\bar T)^i}\ .
\end{split}
\end{equation}
Using the expressions
\begin{gather}
Z+\bar Z=W_i(T+\bar T)^i\left( \lambda+\tfrac{1}{2}\right)\ ,\qquad \re T^i=t^i=\frac{\hat R a^i}{V^{1/3}}\ , \label{a}\\
\re  S=V+\frac{\epsilon_S^\prime}{2}\left( \lambda+\frac{1}{2}\right)^2W_it^i \ \label{b},
\end{gather}
one can rewrite the first derivatives as
\begin{equation}
\begin{split}
\frac{\partial \hat K}{\partial \bar S} &=-\frac{1}{2V}\ ,\\
\frac{\partial \hat K}{\partial \bar T^l} &=-\frac{\epsilon_S^\prime}{4V}\left(d_{lij}a^ia^j+\left(\lambda+\tfrac{1}{2}\right)^2W_l\right)\ , \\
\frac{\partial \hat K}{\partial \bar Z} &=\frac{\epsilon_S^\prime}{2}\frac{\left(\lambda+\frac{1}{2}\right)}{V} \ , 
\end{split}
\label{d}
\end{equation}
where we have dropped terms of order $\mathcal{O}({\epsilon_S^\prime}^2)$ and higher. Finally, one can compute the second derivative terms in the matrix $\hat{K}_{a\bar{b}}$. The components of $\hat{K}_{a\bar{b}}$ are given by
\begin{equation}
\begin{split}
\frac{\partial^2 \hat K}{\partial S\partial \bar S}&=\frac{1}{4V^2} \ ,\\
\frac{\partial^2 \hat K}{\partial T^m\partial \bar S}&=\frac{\epsilon_S^\prime}{8}\left(  \lambda+\tfrac{1}{2}\right)^2\frac{W_m}{V^2}\ ,\\
\frac{\partial^2 \hat K}{\partial Z\partial \bar S}&=-\frac{\epsilon_S^\prime}{4}\frac{\left(\lambda+\frac{1}{2}\right)}{V^2}\ ,\\
\frac{\partial^2 \hat K}{\partial T^m\partial \bar Z}&=-\frac{\epsilon_S^\prime}{4}\frac{\left(\lambda+\frac{1}{2}\right)W_m}{V^{2/3}\hat RW_ia^i}\ ,\\
\frac{\partial^2 \hat K}{\partial Z \partial \bar Z}&=\frac{\epsilon_S^\prime}{4}\frac{1}{V^{2/3}\hat RW_ia^i}\ ,\\
\frac{\partial^2 \hat K}{\partial T^m\partial \bar T^l}&=-\frac{d_{ilm}a^i}{4\hat R^2V^{1/3}}+\frac{d_{mij}a^ia^jd_{lpq}a^pa^q}{16\hat R^4V^{2/3}}+\frac{\epsilon_S^\prime}{4}\frac{\left(\lambda+\frac{1}{2}\right)^2W_mW_l}{V^{2/3}\hat RW_ia^i} \ ,
\label{K}
\end{split}
\end{equation}
where $l,m=1,2,3$ and, again, we have dropped all terms of $\mathcal{O}({\epsilon_S^\prime}^2)$ and higher. Of course, it is $K^{\bar{b} a}$, the inverse of $\hat{K}_{a\bar{b}}$, that appears in the expression \eqref{f1} for the $F$-terms. However, an analytic expression for $K^{\bar{b} a}$ is very complicated and unenlightening. Hence, in the following we will compute it only numerically from $\hat{K}_{a\bar{b}}$ for given values of the moduli. In Section 2, we concluded that the line bundles ${\mathcal{O}}_{X} (2, 1, 3)$ and ${\mathcal{O}}_{X} (1, 2, 3)$  are the most promising for building a model in which supersymmetry is broken at a lower scale.
%, although above the energy range already explored experimentally. 
In the end, we settled for ${\mathcal{O}}_{X} (2, 1, 3)$ because it produced a larger ``viable'' surface in the $(a^1, a^2, a^3)$ parameter space where all the conditions presented in Section 2 are met. Therefore, it is useful to evaluate \eqref{K} for the line bundle $L={\mathcal{O}}_{X} (2, 1, 3)$. Recalling that for this choice of line bundle we found $W_i=(9,17,0)_i$, it follows that the matrix $\hat{K}_{a\bar{b}}$ becomes
\begin{equation}
\hat K_{a\bar b}=
\left(
\begin{matrix}
\frac{1}{4V^2}&\frac{9\epsilon_S^\prime \left(\lambda +\frac{1}{2} \right)}{8V^2}&\frac{17\epsilon_S^\prime \left(\lambda +\frac{1}{2} \right)}{8V^2}&0&-\frac{\epsilon_S^\prime\left(\lambda +\frac{1}{2} \right)}{4V^2}\\
\frac{9\epsilon_S^\prime \left(\lambda +\frac{1}{2} \right)}{8V^2}&\frac{\partial^2\hat K}{\partial T^1 \partial \bar T^1}&\frac{\partial^2\hat K}{\partial T^1 \partial \bar T^2}&\frac{\partial^2\hat K}{\partial T^1 \partial \bar T^3}
&\frac{-9\epsilon_S^\prime\left(\lambda+\frac{1}{2}\right)}{4V^{2/3}\hat R(9a^1+17a^2)}\\
\frac{17\epsilon_S^\prime \left(\lambda +\frac{1}{2} \right)}{8V^2}&\frac{\partial^2\hat K}{\partial T^2 \partial \bar T^1}&\frac{\partial^2\hat K}{\partial T^2 \partial \bar T^2}&\frac{\partial^2\hat K}{\partial T^2 \partial \bar T^3}&\frac{-17\epsilon_S^\prime\left(\lambda+\frac{1}{2}\right)} {4V^{2/3}\hat R(9a^1+17a^2)}\\
0&\frac{\partial^2\hat K}{\partial T^3 \partial \bar T^1}&\frac{\partial^2\hat K}{\partial T^3 \partial \bar T^2}&\frac{\partial^2\hat K}{\partial T^3 \partial \bar T^3}&0\\
-\frac{\epsilon_S^\prime\left(\lambda +\frac{1}{2} \right)}{4V^2}&\frac{-9\epsilon_S^\prime\left(\lambda+\frac{1}{2}\right)}{4V^{2/3}\hat R(9a^1+17a^2)}&\frac{-17\epsilon_S^\prime\left(\lambda+\frac{1}{2}\right)}{4V^{2/3}\hat R(9a^1+17a^2)}&0&\frac{\epsilon_S^\prime}{4V^{2/3}\hat R(9a^1+17a^2)}
\end{matrix}
\right)
\end{equation}
Inside the matrix we have the submatrix
\begin{equation}
\frac{\partial^2 \hat K}{\partial T^m \partial \bar T^n}=-\frac{d_{imn}a^i}{4\hat R^2V^{1/3}}+d_{mij}a^ia^jd_{npq}a^pa^q+\frac{\epsilon_S^\prime}{4}\frac{\left(\lambda+\frac{1}{2}\right)^2W_mW_n}{V^{2/3}\hat R(9a^1+17a^2)} \ .
\end{equation}

Having discussed the K\"ahler potential components, $\ee^{\hat{K}/2}$, $\partial_{a}{\hat{K}}$ and $\hat{K}_{a\bar{b}}$,  we now turn to the superpotential contributions to the $F$-terms; that is, the K\"ahler covariant derivatives of $W$ given by
\begin{equation}
{\mathcal{D}}_{a}W=\partial_a W+W\partial_a \hat K\ ,  \quad a=S, T^{1}, T^{2}, T^{3}, Z\ .
\label{w1}
\end{equation}
We have already computed in \eqref{d} the first derivatives of the K\"ahler potential. We are left to compute the first derivatives of the superpotential $W$. It follows from  \eqref{sup1} and \eqref{sup2} that for the line bundle $L={\mathcal{O}}_{X} (2, 1, 3)$
\begin{equation}
\begin{split}
\partial_S W&=W\left( -\frac{6\pi}{b_L\hat \alpha_\text{GUT}} \right)\ ,\\
\partial_{T_1} W&=W\left( -\frac{6\pi}{b_L\hat \alpha_\text{GUT}} \right)\left(-\frac{29}{6}\epsilon_S^\prime \right)=W\left( \epsilon_S^\prime\frac{29\pi}{b_L\hat \alpha_\text{GUT}} \right)\ ,\\
\partial_{T_2} W&=W\left( -\frac{6\pi}{b_L\hat \alpha_\text{GUT}} \right)\left(-\frac{25}{3}\epsilon_S^\prime \right)=W\left( \epsilon_S^\prime\frac{50\pi}{b_L\hat \alpha_\text{GUT}} \right)\ ,\\
\partial_{T_3} W&=W\left( -\frac{6\pi}{b_L\hat \alpha_\text{GUT}} \right)\left(-2\epsilon_S^\prime \right)=W\left( \epsilon_S^\prime\frac{12\pi}{b_L\hat \alpha_\text{GUT}} \right)\ ,\\
\partial_{Z}W&=0\ ,
\end{split}
\label{w2}
\end{equation}
where, as shown in Section 2, $b_{L}=6$.
It then follows from \eqref{w1}, using \eqref{d} and \eqref{w2}, that the covariant derivatives of the superpotential are
\begin{equation}
\begin{split}
\partial_S W+W\partial_S \hat K&=W\left[ -\frac{6\pi}{b_L\hat \alpha_\text{GUT}}-\frac{1}{2V} \right]\ ,\\
\partial_{T_1} W+W\partial_{T^1}\hat K&=W\epsilon_S^\prime\left[ \frac{29\pi}{b_L\hat \alpha_\text{GUT}}-\frac{1}{4V}d_{1ij}a^ia^j-9\frac{\left(\lambda+\frac{1}{2}\right)^2}{4V}\right]\ ,\\
\partial_{T_2} W+W\partial_{T^2}\hat K&=W\epsilon_S^\prime\left[ \frac{50\pi}{b_L\hat \alpha_\text{GUT}}-\frac{1}{4V}d_{2ij}a^ia^j-17\frac{\left(\lambda+\frac{1}{2}\right)^2}{4V}\right]\ ,\\
\partial_{T_3} W+W\partial_{T^3}\hat K&=W\epsilon_S^\prime\left[ \frac{12\pi}{b_L\hat \alpha_\text{GUT}}-\frac{1}{4V}d_{3ij}a^ia^j\right]\ ,\\
\partial_{Z}W+W\partial_Z \hat K&=W\epsilon_S^\prime\left[ \frac{1}{2}\frac{\left( \lambda+\frac{1}{2}\right) }{V} \right]\ ,
\end{split}
\label{D}
\end{equation}
respectively.

Using the expressions for $\ee^{\hat{K}/2}$, $\partial_{a}{\hat{K}}$, $\hat{K}_{a\bar{b}}$ and ${\mathcal{D}}_{a}$ given in \eqref{exp}, \eqref{d}, \eqref{K} and \eqref{D} respectively, one can compute the value of any $F$-term $F^{\bar{b}}$ given in \eqref{f1} explicitly at any point in moduli space. Note that in what follows, we always set the \emph{imaginary} components of the moduli to zero. We now want to discuss the explicit form of the  ``soft'' supersymmetry breaking terms induced in the observable sector of the theory by these $F$-terms and, more specifically, the exact form of their coefficients. These will be presented in the following section.

\section{Soft Supersymmetry Breaking Terms in the Observable Sector}

As discussed in detail in \cite{Ovrut:2012wg}, in the $B-L$ MSSM, the $Spin(10)$ group of the observable sector is broken near the unification scale $\langle M_{U} \rangle$ to the low energy gauge group $SU(3)_{C}\times SU(2)_{L}\times U(1)_{3R}\times U(1)_{B-L}$ by two independent Wilson lines associated with the $\mathbb{Z}_{3} \times \mathbb{Z}_{3}$ homotopy group of the Calabi--Yau threefold. 
The particle content of the resulting effective theory
is precisely that of the MSSM, with three right-handed neutrino chiral multiplets and
a single Higgs-Higgs conjugate pair, with no exotic fields.  

The superpotential of the observable sector evaluated at $\langle M_{U} \rangle$ is given by
\begin{equation}
W=\mu H_{u} H_{d}+Y_{u}QH_{u}u^{c}-Y_{d}QH_{d}d^{c}-Y_{e}QH_{d}\ee^{c}+Y_{\nu}QH_{u}\nu^{c} \ ,
\label{con1}
\end{equation}
where flavor and gauge indices have been suppressed and the Yukawa couplings are three-by-three matrices in flavor space. The observed smallness of the three CKM mixing angles and the CP-violating phase dictate that the quark and lepton Yukawa matrices should be nearly diagonal and real. Furthermore, the smallness of the first and second family fermion masses implies that all components of the up/down quark and lepton Yukawa couplings, with the exception of the top and bottom quarks, and the tau lepton, can be neglected for the purposes of this paper. Similarly, the very light neutrino masses imply that the neutrino Yukawa couplings can also be neglected in our analysis. The $\mu$-parameter can be chosen to be real, but not necessarily positive, without loss of generality. We emphasize that we make no attempt in this paper -- nor was there any attempt in the previous $B-L$ MSSM papers -- to solve the ``$\mu$-problem''. Hence, we allow the dimension-one parameter $\mu$ to take any value required to obtain the correct $Z$-boson mass.

To align with the formalism presented in \cite{Kaplunovsky:1993rd} and \cite{Lukas:1999kt}, where the general form of the soft supersymmetry breaking terms in the Lagrangian which we use in this analysis were computed, we rewrite the superpotential \eqref{con1} in the form
\begin{equation}
W= \frac{1}{2}\hat{\mu}_{IJ}C^{I}C^{J}+\frac{1}{3}\hat{Y}^{IJK}C^{I}C^{J}C^{K} \ ,
\label{con2}
\end{equation}
where $C^{I}$ are the chiral superfields associated with the top and bottom quarks, the tau lepton and the up- and down-Higgs particles. Comparing \eqref{con1} and \eqref{con2}, it follows that  $\hat{\mu}_{IJ}$ is symmetric and all components vanish with the exception of $\hat{\mu}_{H_{u}H_{d}}=\mu$. Similarly, the coefficients $\hat{Y}_{IJK}$ are completely symmetric and, following the formalism in \cite{Lukas:1999kt}, are given by
\begin{equation}
\hat{Y}_{IJK}=2\sqrt{2\pi\hat{\alpha}_{GUT}} \>y_{IJK} \ ,
\label{con3}
\end{equation}
where $y_{IJK}$ are the physical Yukawa parameters for the top, bottom and tau particles scaled up to $\langle M_{U} \rangle$.

The soft supersymmetry breaking terms associated with the observable sector of the $B-L$ MSSM were presented in \cite{Ovrut:2014rba,Ovrut:2015uea}. At the unification scale $\langle M_{U} \rangle$, they were found to be of the form
\begin{align}
 \label{eq:6}
\begin{split}
	-\mathcal L_{\mbox{\scriptsize soft}}  &= 
	\left(
		\frac{1}{2} M_3 \tilde g^2+ \frac{1}{2} M_2 \tilde W^2+ \frac{1}{2} M_R \tilde W_R^2+\frac{1}{2} M_{BL} \tilde {B^\prime}^2
	\right.
		\\
	& \eqspace\left.
		\hspace{0.4cm} +a_u \tilde Q H_u \tilde u^c - a_d \tilde Q H_d \tilde d^c - a_e \tilde L H_d \tilde e^c
		+ a_\nu \tilde L H_u \tilde \nu^c + b H_u H_d + \text{h.c.}
	\right)
	\\
	&\eqspace + m_{\tilde Q}^2|\tilde Q|^2+m_{\tilde u^c}^2|\tilde u^c|^2+m_{\tilde d^c}^2|\tilde d^c|^2+m_{\tilde L}^2|\tilde L|^2
	+m_{\tilde \nu^c}^2|\tilde \nu^c|^2+m_{\tilde e^c}^2|\tilde e^c|^2  \\  
	&\eqspace+m_{H_u}^2|H_u|^2+m_{H_d}^2|H_d|^2 \ .
\end{split}
\end{align}
The $b$ parameter can be chosen to be real and positive without loss of generality. The gaugino soft masses can, in principle, be complex. This, however, could lead to CP-violating effects that are not observed. Therefore, we proceed by assuming they all are real. The $a$-parameters and soft scalar masses can, in general, be Hermitian matrices in family space. Again, however, this could lead to unobserved flavor and CP violation. Therefore, we will assume they all are diagonal and real. For more explanation of these assumptions, see \cite{Ovrut:2015uea}.

As we did for the superpotential, to compare with the formalism presented in \cite{Kaplunovsky:1993rd,Lukas:1999kt}, we will rewrite the soft supersymmetry breaking terms in \eqref{eq:6} as
\begin{equation}
-\mathcal L_{\mbox{\scriptsize soft}}= \left( \frac{1}{2}M_{i}(\lambda^{i})^{2}+\frac{1}{3}a_{IJK}\tilde{C}^{I}\tilde{C}^{J}\tilde{C}^{K} +\frac{1}{2}B_{IJ}\tilde{C}^{I}\tilde{C}^{J}+\text{h.c.} \right)+m_{I\bar{J}}^{2}\tilde{C}^{I}\bar{\tilde{C}}^{\bar{J}} \ ,
\label{con4}
\end{equation}
where $\lambda^{i}$ are the gauginos for $i=3,2,3R,BL$, and $\tilde{C}^{I}$ are the scalar components of the chiral superfields associated with the top and bottom quarks, the tau lepton and up- and down-Higgs particles. Comparing \eqref{con4} with \eqref{eq:6}, it follows that a) the matrix $m_{I\bar{J}}^{2}$ is diagonal with its $I,\bar{J}$ indices running over stop, sbottom, stau, and Higgs-up and Higgs-down scalars, b) $B_{IJ}$  is symmetric, all of whose terms vanish except $B_{H_{u}H_{d}}=b$ and c) $a_{IJK}$ is totally symmetric where, as with $m_{I\bar{J}}^{2}$, its indices $I,J,K$ run over stop, sbottom, stau and Higgs-up and Higgs-down scalars only, and are associated with the $a$-parameters in \eqref{eq:6} accordingly.

The parameters for each of these soft supersymmetry breaking terms -- including the induced gravitino mass which enters some of the soft breaking coefficients -- can be explicitly computed for any spontaneous SUSY breaking mechanism. Using the notation of \cite{Kaplunovsky:1993rd}, the generic expressions for these parameters are the following:
\begin {enumerate}
\item The gravitino mass:

$m_{3/2}=\kappa_{4}^{2}\ee^{\hat{K}/2}|W|. $  

\item The gaugino masses:

$M_{i}=\frac{1}{2}F^{a}\partial_{a}\ln g^{-2}_{i}.$    

Note that the gaugino mass is, in general not ``universal'' -- that is, it is not necessarily the same for all gauginos. However, for the present analysis of the $B-L$ MSSM, the gaugino masses will turn out to be identical. We explain why this is the case in the discussion to follow.

\item The quadratic scalar masses:

$m_{I\bar{J}}^{2}=m^{2}_{3/2}Z_{I\bar{J}}-F^{a}\bar{F}^{\bar{b}}R_{a \bar{b} I\bar{J}}\ .$   

Here $Z_{I\bar{J}}$ is defined by $K_{\text {matter}}=Z_{I\bar{J}}\tilde{C}^{I}{\bar{\tilde{C}}}{}^{\bar{J}}$ for generic observable-sector scalar fields $\tilde{C}^{I}$.
The explicit forms for $Z_{I\bar{J}}$ and $R_{a \bar{b} I\bar{J}}$ are presented in the discussion below.

\item The cubic scalar coefficients:

$a_{IJK}=F^{a}\big( \partial_{a}Y_{IJK}+\frac{1}{2}\hat{K}_{a}Y_{IJK}-3\Gamma^{N}_{a(I}Y_{JK)N} \big)\ .$ 

The parameters $Y_{IJK}$ and $\Gamma^{N}_{aI}$ will be given in the following analysis. 

\item The holomorphic quadratic coefficient:

$ B_{IJ}=F^{a}\bigl(\partial_{a}\mu_{IJ}+\frac{1}{2}(\partial_{a}\hat{K})-3\Gamma^{N}_{a(I}\mu_{J)N} \bigr)- m_{3/2}\mu_{IJ} \ .$

The parameter $\mu_{IJ}$ will be discussed below.

\end{enumerate}

\noindent We will now discuss each of these five parameters in detail, plotting their values over the physically acceptable ``viable'' region of K\"ahler moduli space introduced in Section 2 for the hidden sector line bundle $L={\mathcal{O}}_{X} (2, 1, 3)$.

The above expressions for the soft supersymmetry breaking terms are generic; that is, they can arise from any vacuum state which spontaneously breaks SUSY via non-vanishing $F$-terms. However, for the remainder of this paper, we will consider supersymmetry breaking to occur explicitly from a ``gaugino condensate'' in the hidden sector of the $B-L$ MSSM theory.

\subsection{Gravitino Mass}

The gravitino mass is simply defined to be
\begin{equation}
m_{3/2}=\kappa_{4}^{2}\ee^{\hat{K}/2}|W| \ ,
\label{van1}
\end{equation}
where $\kappa_{4}^{2}=8 \pi/M_{P}^{2}$, $\hat{K}$ is defined in \eqref{sum1} and $W$ is the gaugino condensate superpotential presented in \eqref{sup1}.

\subsection*{Gaugino Mass}

The generic expression for the gaugino mass  associated with the $i$-th factor of an observable sector gauge group of the form $G=\Pi_{i} G_{i}$ is given by
\begin{equation}
M_i=\frac{1}{2}F^a\partial_a \ln g_i^{-2}\qquad a=S,T^1,T^2,T^3,Z \ .
\label{g1}
\end{equation}
In our case, the index $i$ spans the factors in the $d=4$ low-energy gauge group $G=SU(3)_{C}\times SU(2)_{L}\times U(1)_{3R}\times U(1)_{B-L}$ of the  $B-L$ MSSM in the observable sector. That is,
\begin{equation}
i=3,\>2,\>3R,\> B-L \ .
\end{equation}
As discussed in \cite{Ashmore:2020ocb}, in the simultaneous Wilson lines scenario each gauge coupling $g_{i}^{2}$ is related to its average value $\langle g_{u}^{2}\rangle$  at the unification scale $\langle M_{U} \rangle=3.15 \times 10^{16}$ GeV by
\begin{equation}
g_i^2=c_i\langle g_u^2\rangle
\end{equation}
for some constant coefficient $c_{i}$.
Then
\begin{equation}
M_i=\frac{1}{2}F^a\partial_a \ln g_i^{-2}=\frac{1}{2}F^a\frac{1}{g_i^{-2}}\partial_a g_i^{-2}\\
=\frac{1}{2}F^{a} c_{i} \langle g_u^{2} \rangle \partial_a \frac{1}{c_{i} \langle g_u^{2} \rangle}
 \end{equation}
Now  \eqref{need2} implies that 
\begin{equation}
\langle g_u^2 \rangle=\frac{4\pi\hat \alpha_{\text{GUT}}}{\re  f_1} \ ,
\end{equation}
where $ \hat{\alpha}_{\text{GUT}}$ is a constant parameter. It follows that the constants $c_{i}$ and ${ \hat{\alpha}}_{\text{GUT}}$ drop out
and, hence, the gaugino masses defined \eqref{g1} are all identical. Defining this unique parameter to be $M_{1/2}$, we find that
\begin{equation}
M_{1/2}=\frac{1}{2\re f_1}F^a\partial_a \re  f_1, 
\end{equation}
as presented previously. A similar argument can be made for the split Wilson lines scenario. 

The expression for the real part of $f_{1}$, prior to imposing unity gauge,  for the line bundle $L=\mathcal{O}_{X}(2,1,3)$ with $W_{i}=(9,17,0)$ can be written as
\begin{equation}
\re f_{1}=\frac{S+\bar{S}}{2}+\epsilon_{S} ' \left( \tfrac{29}{12}(T^{1}+\bar{T}^{1}) +\tfrac{25}{6}(T^{2}+\bar{T}^{2}) +(T^{3}+\bar{T}^{3}) -\tfrac{1}{2}(Z+\bar{Z}) \right) \ .
\label{need0}
\end{equation}
It follows that 
\begin{gather}
\partial_{S}\re f_{1}=\frac{1}{2}, \quad \partial_{T^{1}}\re f_{1}=\frac{29}{12} \epsilon_{S} ' ,\quad\partial_{T^{2}}\re f_{1}=\frac{25}{6} \epsilon_{S} ' ,  \\
\qquad \qquad \partial_{T^{3}}\re f_{1}=\epsilon_{S} ' ,\quad \partial_{Z}\re f_{1}=-\frac{1}{2}\epsilon_{S}'  \ .
\label{need00}
\end{gather}
In unity gauge, the real part of $f_{1}$ for the line bundle $L=\mathcal{O}_{X}(2,1,3)$  is presented in \eqref{white2A}. It was found to be
\begin{equation}
{\rm {Re}} f_1=V+\frac{1}{3}a^1-\frac{1}{6}a^2+2a^3+\frac{1}{2}\left( \tfrac{1}{2}-\lambda \right)^2W_ia^i \ ,
\label{need3}
\end{equation}
where 
\begin{equation}
W_{i}=(9,17,0) \ .
\end{equation}
Putting everything together, the universal gaugino soft supersymmetry breaking coefficient is given, in unity gauge, by
\begin{equation}
\begin{split}
M_{1/2}&=\frac{1}{2\big( V+\frac{1}{3}a^1-\frac{1}{6}a^2+2a^3+\frac{1}{2}\left( \frac{1}{2}-\lambda \right)(9a^{1}+17a^{2}) \big)}  \\
& \times \left[ \frac{1}{2}F^{S}+ \epsilon_{S} '  \frac{29}{12}F^{T^{1}} +\epsilon_{S} '  \frac{25}{6}F^{T^{2}} +\epsilon_{S} ' F^{T^{3}} - \epsilon_{S} ' \frac{1}{2}F^{Z}  \right] \ .
\end{split}
\label{need5}
\end{equation}
Taking $\lambda=0.49$, as done in \cite{Ashmore:2020ocb} and above, one can compute the value of $m_{1/2}$ at any point in the physical ``viable'' subspace of K\"ahler moduli space for $L=\mathcal{O}_{X}(2,1,3)$, as we did above for both $m_{\rm{susy}}$ and $m_{3/2}$. 

\subsection{Quadratic Scalar Masses}

The generic form for the quadratic scalar mass coefficients is given by~\cite{Kaplunovsky:1993rd,Brignole:1998dxa,Lukas:1997rb,Lukas:1999kt}
\begin{equation}
m_{I\bar{J}}^{2}=m^{2}_{3/2}Z_{I\bar{J}}-F^{a}\bar{F}^{\bar{b}}R_{a \bar{b} I\bar{J}} \ .
\label{q1}
\end{equation}
%
%Begin by considering $Z_{I\bar{J}}$. 
To linear order in $\epsilon_S^\prime$, it was shown in \cite{Lukas:1999kt} that, for a {\it single} five-brane located at $z \in [0,1]$,
\begin{equation}
Z_{I\bar{J}}=\ee^{K_{T}/3}\left[K_{BI\bar{J}}-\frac{\epsilon_S^\prime}{2(S+\bar{S})}\tilde{\Gamma}^{i}_{BI\bar{J}}\big(\beta_{i}^{(0)}+(1-z)^{2}W_{i}  \big)   \right] \ ,
\label{q2} 
\end{equation}
where $\beta_{i}^{(0)}$ is the ``charge'' on the observable wall and $W_{i}$ is the five-brane class. 
Note that for the $B-L$ MSSM vacuum, $\beta_{i}^{(0)}$ is given by \cite{Ashmore:2020ocb}
\begin{equation}
\beta_{i}^{(0)}=\left(\frac{2}{3},-\frac{1}{3},4 \right)_{i} \ .
\label{l1}
\end{equation}
The $T^i$-dependent K\"ahler potential was presented in \eqref{tel1} and $K_{BI\bar{J}}$ is defined by
\begin{equation}
K_{BI\bar{J}}=G_{I\bar{J}} \ ,
\label{kn1}
\end{equation}
where $G_{I\bar{J}}$ is a positive-definite Hermitian metric on the $H^{1}$ cohomologies associated with the $\tilde{C}^{I}$ matter scalars in the observable sector \cite{Lukas:1997rb,Lukas:1999kt}. Generically, $G_{I\bar{J}}$ is moduli dependent.  The quantity  $\tilde{\Gamma}^{i}_{BI\bar{J}}$ is given by
\begin{equation}
\tilde{\Gamma}^{i}_{BI\bar{J}}={\Gamma}^{i}_{BI\bar{J}}-(T^{i}+\bar{T}^{i})K_{BI\bar{J}}-\tfrac{2}{3}(T^{i}+\bar{T}^{i})(T^{k}+\bar{T}^{k})K_{Tkj}{\Gamma}^{j}_{BI\bar{J}} \ ,
\label{kn2}
\end{equation}
with
\begin{equation}
\Gamma^{i}_{BI\bar{J}}=K_{T}^{ij} \frac{\partial K_{BI\bar{J}}}{\partial T^{j}}\ ,
\label{kn3}
\end{equation}
and $K_{T}^{ij}$ is the inverse of the matrix 
%$K_{Tij}=\frac{\partial^{2}K_{T}}{\partial T^{i} \partial T^{j}}$.
%
\begin{equation}
K_{Tij}=\frac{\partial^{2}K_{T}}{\partial T^{i} \partial T^{j}} =-\frac{d_{lmi}a^{i}}{4\hat{R}^{2}V^{1/3}} +\frac{d_{mij}a^{i}a^{j}d_{lpq}a^{p}a^{q}}{16\hat{R}^{4}V^{2/3}}\ .
\label{kn4}
\end{equation}

Next, we consider the tensor $R_{a \bar{b} I\bar{J}}$ in \eqref{q1}. It is defined to be
\begin{equation}
R_{a \bar{b} I\bar{J}}=\partial_{a}\partial_{\bar{b}}Z_{I\bar{J}}-\Gamma^{N}_{aI}Z_{N\bar{L}}{\bar{\Gamma}}^{\bar{L}}_{\bar{b}\bar{J}}\ ,
\label{r1}
\end{equation}
where
\begin{equation}
\Gamma^{N}_{aI}=Z^{N\bar{J}}\partial_{a}Z_{\bar{J}I} \ .
\label{r2}
\end{equation}
Writing the generic expression for $Z_{I\bar{J}}$ in \eqref{q2} as
\begin{equation}
Z_{I\bar{J}}=Z_{I\bar{J}}^{(0)}+Z_{I\bar{J}}^{(\epsilon_{S}')} \ ,
\label{r3}
\end{equation}
it follows that
\begin{equation}
\Gamma^{N}_{aI}=\Gamma^{(0)N}_{aI}+\Gamma^{(\epsilon_{S}')N}_{aI}\ ,
\end{equation}
where
\begin{equation}
\Gamma^{(0)N}_{aI}=Z^{(0)N\bar{J}}\partial_{a}Z^{(0)}_{\bar{J}I}, \qquad \Gamma^{(\epsilon_{S}')N}_{aI}=Z^{(0)N\bar{J}}\partial_{a}Z^{(\epsilon_{S}')}_{\bar{J}I}+Z^{(\epsilon_{S}')N\bar{J}}\partial_{a}Z^{(0)}_{\bar{J}I} \ .
\end{equation}
Inserting these expressions into \eqref{r1}, we find that
\begin{equation}
R_{a\bar{b}I\bar{J}}= R^{(0)}_{a\bar{b}I\bar{J}}+R^{(\epsilon_{S}')}_{a\bar{b}I\bar{J}}\ ,
\label{bc2}
\end{equation}
with
\begin{equation}
R^{(0)}_{a\bar{b}I\bar{J}}=\partial_{a}\partial_{\bar{b}}Z^{(0)}_{I\bar{J}} - \Gamma^{(0)N}_{aI}Z^{(0)}_{N\bar{L}}\bar{\Gamma}^{(0)\bar{L}}_{\bar{b}\bar{J}} \ ,
\label{bc3}
\end{equation}
and
\begin{equation}
R^{(\epsilon_{S}')}_{a\bar{b}I\bar{J}}=\partial_{a}\partial_{\bar{b}}Z^{(\epsilon_{S}')}_{I\bar{J}} - \Gamma^{(\epsilon_{S}')N}_{aI}Z^{(0)}_{N\bar{L}}\bar{\Gamma}^{(0)\bar{L}}_{\bar{b}\bar{J}} - \Gamma^{(0)N}_{aI}Z^{(\epsilon_{S}')}_{N\bar{L}}\bar{\Gamma}^{(0)\bar{L}}_{\bar{b}\bar{J}} - \Gamma^{(0)N}_{aI}Z^{(0)}_{N\bar{L}}\bar{\Gamma}^{(\epsilon_{S}')\bar{L}}_{\bar{b}\bar{J}} \ .
\label{bc4}
\end{equation}

Putting everything together, one can express the quadratic scalar soft coefficients in \eqref{q1} as 
\begin{equation}
m_{I\bar{J}}^{2} = m^{(0)2}_{I\bar{J}} +m^{(\epsilon_{S}')2}_{I \bar{J} } \ ,
\label{bc5}
\end{equation}
where
\begin{equation}
m^{(0)2}_{I\bar{J}}=m_{3/2}^{2}Z^{(0)}_{I\bar{J}}-F^{a}\bar{F}^{\bar{b}}R^{(0)}_{a\bar{b}I\bar{J}}\ ,
\label{bc6}
\end{equation}
and
\begin{equation}
m^{(\epsilon_{S}')2}_{I\bar{J}}=m_{3/2}^{2}Z^{(\epsilon_{S}')}_{I\bar{J}}-F^{a}\bar{F}^{\bar{b}}R^{(\epsilon_{S}')}_{a\bar{b}I\bar{J}} \ ,
\label{bc7}
\end{equation}
with $Z^{(0)}_{I\bar{J}}$ and  $Z^{(\epsilon_{S}')}_{I\bar{J}}$ defined in \eqref{r3}, and  $R^{(0)}_{a\bar{b}I\bar{J}}$ and $R^{(\epsilon_{S}')}_{a\bar{b}I\bar{J}}$ given in \eqref{bc3} and \eqref{bc4} respectively.

Let us now compute these quantities explicitly. First of all, we note that there is currently no known method to explicitly compute $G_{I\bar J}$.\footnote{This \emph{should} be computable using numeric metrics on Calabi--Yau threefolds~\cite{Headrick:2005ch,Douglas:2006rr,Braun:2007sn,Headrick:2009jz,Douglas:2006hz,Anderson:2011ed,Anderson:2010ke,Cui:2019uhy,Anderson:2020hux,Douglas:2020hpv} and their moduli spaces~\cite{Keller:2009vj}, and the corresponding eigenmodes of the Laplacian~\cite{Braun:2008jp,Ashmore:2020ujw}.} With this in mind, for the rest of this work we shall assume that $K_{BI\bar{J}}=G_{I\bar{J}}$ is {\it moduli independent}, that is, simply an Hermitian matrix of numbers. We will denote this choice by
\begin{equation}
G_{I\bar{J}}={\cal{G}}_{I\bar{J}} \ .
\label{s1}
\end{equation}
We assume this to be the case henceforth. It then follows from \eqref{kn3} that $\Gamma^{i}_{BI\bar{J}}=0$ and, hence,
\begin{equation}
\tilde{\Gamma}^{i}_{BI\bar{J}}=-(T+\bar{T} )^{i}{\cal{G}}_{I\bar{J}} \ .
\label{s2}
\end{equation}
Then, using the metric \eqref{s1}, expression \eqref{q2} for $Z_{I\bar{J}}$ simplifies to 
\begin{equation}
Z_{I\bar{J}}=Z_{I\bar{J}}^{(0)}+Z_{I\bar{J}}^{(\epsilon_{S}')}
\label{s3}
\end{equation}
where
\begin{equation}
Z_{I\bar{J}}^{(0)}=\ee^{K_{T}/3} {\cal{G}}_{I\bar{J}}
\label{s4}
\end{equation}
and
\begin{equation}
Z_{I\bar{J}}^{(\epsilon_{S}')}=\frac{\epsilon_S^\prime}{2}\ee^{K_{T}/3}  \frac{(T+\bar{T})^{i}}{S+\bar{S}} \left[ \left( \frac{2}{3}, -\frac{1}{3},4 \right)_{i}+\left(1-\frac{Z+\bar{Z}}{ W_{l}(T+\bar{T})^{l}} \right) ^{2} W_{i} \right] {\cal{G}}_{I\bar{J}} \ .
\label{s5}
\end{equation}
Note that we have rewritten the last term using \eqref{a} so as to be able to differentiate this expression with respect to $Z$. This will be necessary in order to compute $R^{(\epsilon_{S}')}_{a\bar{b}I\bar{J}}$ below. It is also useful to rewrite the expressions for $Z_{I\bar{J}}^{(0)}$ and $Z_{I\bar{J}}^{(\epsilon_{S}')}$ in terms of the $\hat{R}$, $a^{i}$, $V$ and $\lambda$ variables using \eqref{a} and \eqref{b}, and the fact that $V=\frac{1}{6}d_{ijk}a^{i}a^{j}a^{k}$. Doing this, we find
\begin{equation}
Z_{I\bar{J}}^{(0)}=\frac{1}{\hat{R}}{\cal{G}}_{I\bar{J}}
\label{s4A}
\end{equation}
and
\begin{equation}
Z_{I\bar{J}}^{(\epsilon_{S}')}=\epsilon_S^\prime \frac{a^{i}}{2V^{4/3}} \left[ \left( \frac{2}{3}, -\frac{1}{3},4 \right)_{i}+ (\frac{1}{2}-\lambda)^{2} W_{i} \right] {\cal{G}}_{I\bar{J}} \ .
\label{s5A}
\end{equation}
Recall from \eqref{wall1} that unity gauge is defined by setting
\begin{equation}
\epsilon_S'\frac{\Rhat}{V^{1/3}} = 1 \ .
\label{wall2A}
\end{equation}
It follows that in unity gauge the expression for $Z_{I\bar{J}}^{(\epsilon_{S}')}$ becomes
\begin{equation}
Z_{I\bar{J}}^{(\epsilon_{S}')}=\frac{a^{i}}{2\hat{R}V} \left[ \left( \frac{2}{3}, -\frac{1}{3},4 \right)_{i}+ (\frac{1}{2}-\lambda)^{2} W_{i} \right] {\cal{G}}_{I\bar{J}} \ .
\label{s5AA}
\end{equation}
Finally, evaluating this expression for the specific line bundle $L=\mathcal{O}_{X}(2,1,3)$ with $W_{i}=(9,17,0)$ discussed above, we find that
\begin{equation}
Z_{I\bar{J}}^{(\epsilon_{S}')}=\frac{1}{2\hat{R}V} \left[ \left( \frac{2}{3}+9 (\frac{1}{2}-\lambda)^{2} \right) a^{1}+ \left(-\frac{1}{3}+17(\frac{1}{2}-\lambda)^{2} \right) a^{2} +4a^{3} \right]{\cal{G}}_{I\bar{J}} \ .
\label{s5AAA}
\end{equation}

Let us now compute $R^{(0)}_{a\bar{b}I\bar{J}}$ and $R^{(\epsilon_{S}')}_{a\bar{b}I\bar{J}}$ using expressions \eqref{bc3} and \eqref{bc4} respectively. We begin with $R^{(0)}_{a\bar{b}I\bar{J}}$. Using the fact that
\begin{equation}
Z^{(0)N\bar{J}}=\ee^{-K_{T}/3}{ \cal{G}}^{N\bar{J}} \ ,
\label{s6}
\end{equation}
it follows that 
\begin{equation}
\Gamma^{(0)N}_{aI}=\frac{1}{3}(\partial_{a}K_{T})\delta^{N}_{I}
\label{s7}
\end{equation}
and, hence
\begin{equation}
-\Gamma^{(0)N}_{aI}Z^{(0)}_{N\bar{L}}\bar{\Gamma}^{(0)\bar{L}}_{\bar{b}\bar{J}} = \frac{1}{9}\ee^{K_{T}/3} (\partial_{a}K_{T})(\partial_{\bar{b}}K_{T}){\cal{G}}_{I\bar{J}} \ .
\label{s8}
\end{equation}
It is straightforward to show from \eqref{s4} that
\begin{equation}
\partial_{a}\partial_{\bar{b}}Z^{(0)}_{I\bar{J}}= \ee^{K_{T}/3}\left(\frac{1}{9}(\partial_{a}K_{T})(\partial_{\bar{b}}K_{T})+\frac{1}{3}(\partial_{a}\partial_{\bar{b}}K_{T}) \right){\cal{G}}_{I\bar{J}} \ .
\label{s9}
\end{equation}
Using \eqref{s8} and \eqref{s9}, expression \eqref{bc3} becomes
\begin{equation}
R^{(0)}_{a\bar{b}I\bar{J}}=\frac{\ee^{K_{T}/3}}{3}(\partial_{a}\partial_{\bar{b}}K_{T}){\cal{G}}_{I\bar{J}} \ .
\label{s10}
\end{equation}
Note that this vanishes if index $a$ and/or ${b}$ is $S, Z$. For  $a=i$, ${b}=j$ for $i,j=1,2,3$, $\partial_{i}\partial_{{j}}K_{T}$ is given by \eqref{kn4}.
Let us now compute $R^{(\epsilon_{S}')}_{a\bar{b}I\bar{J}}$. It follows from \eqref{bc4} that, in addition to  the inverse of $Z^{(0)}_{I\bar{J}}$ given in \eqref{s6}, one also needs to know the inverse $Z^{(\epsilon_{S}'))}_{I\bar{J}}$ in \eqref{s5}. Calculating this to linear order in $\epsilon_{S}'$ is straightforward. It is found to be
\begin{equation}
Z^{(\epsilon_{S}')N\bar{J}}=-\frac{\epsilon_S^\prime}{2}\ee^{-K_{T}/3}  \frac{(T+\bar{T})^{i}}{S+\bar{S}} \left[ \left( \frac{2}{3}, -\frac{1}{3},4 \right)_{i}+\left(1-\frac{Z+\bar{Z}}{ W_{l}(T+\bar{T})^{l}} \right) ^{2} W_{i} \right]{ \cal{G}}^{N\bar{J}} \ .
\label{ep1}
\end{equation}
To continue, recall from \eqref{s7} that
\begin{equation}
\Gamma^{(0)N}_{aI}=\frac{1}{3}(\partial_{a}K_{T})\delta^{N}_{I} \ .
\label{ep2}
\end{equation}
Furthermore, using \eqref{s5} and the inverse \eqref{ep1} one can show that
\begin{equation}
\Gamma^{(\epsilon_S^\prime)N}_{aI}=\frac{\epsilon_S^\prime}{2} \partial_{a}\left( \frac{(T+\bar{T})^{i}}{S+\bar{S}}[X]_{i}\right)    \delta^{N}_{I} \ ,
\label{ep3}
\end{equation}
where we have introduced 
\begin{equation}
[X]_{i}=\left( \frac{2}{3}, -\frac{1}{3},4 \right)_{i}+\left(1-\frac{Z+\bar{Z}}{W_{l}(T+\bar{T})^{l}} \right) ^{2} W_{i}
\label{ep4}
\end{equation}
to simplify the notation. Using \eqref{ep3} and \eqref{ep4} it is straightforward to show that the last three terms in \eqref{bc4} are given by
\begin{align}
& - \Gamma^{(\epsilon_{S}')N}_{aI}Z^{(0)}_{N\bar{L}}\bar{\Gamma}^{(0)\bar{L}}_{\bar{b}\bar{J}} - \Gamma^{(0)N}_{aI}Z^{(\epsilon_{S}')}_{N\bar{L}}\bar{\Gamma}^{(0)\bar{L}}_{\bar{b}\bar{J}} - \Gamma^{(0)N}_{aI}Z^{(0)}_{N\bar{L}}\bar{\Gamma}^{(\epsilon_{S}')\bar{L}}_{\bar{b}\bar{J}}\nonumber \\
&=-\frac{\epsilon_{S}'}{6}\ee^{K_{T}/3}  \Big(  \frac{1}{3} \frac{(T+\bar{T})^{i}}{S+\bar{S}} [X]_{i}(\partial_{a}K_{T})(\partial_{\bar{b}}K_{T})  \label{ep5}     \\
&\eqspace +\partial_{a} ( \frac{(T+\bar{T})^{i}}{S+\bar{S}} [X]_{i} ) (\partial_{\bar{b}}K_{T})  + (\partial_{a}K_{T})\partial_{\bar{b}} ( \frac{(T+\bar{T})^{i}}{S+\bar{S}} [X]_{i} ) \Big){ \cal{G}}_{I\bar{J}} \ . \nonumber
\end{align}
Similarly, using $Z_{I\bar{J}}^{(\epsilon_{S}')}$ in \eqref{s5}, it is tedious but straightforward to show that
\begin{align}
\partial_{a}\partial_{\bar{b}}Z_{I\bar{J}}^{(\epsilon_{S}')}&= \frac{\epsilon_{S}'}{6}\ee^{K_{T}/3}  \Big( \frac{1}{3} \frac{(T+\bar{T})^{i}}{S+\bar{S}} [X]_{i}(\partial_{a}K_{T})(\partial_{\bar{b}}K_{T}) \label{ep6} \\ 
& \eqspace+\partial_{a} ( \frac{(T+\bar{T})^{i}}{S+\bar{S}} [X]_{i} ) (\partial_{\bar{b}}K_{T})  + (\partial_{a}K_{T})\partial_{\bar{b}} ( \frac{(T+\bar{T})^{i}}{S+\bar{S}} [X]_{i} ) \nonumber \\
&\eqspace+(\partial_{a}\partial_{\bar{b}}K_{T}) \frac{(T+\bar{T})^{i}}{S+\bar{S}} [X]_{i} +3\partial_{a}\partial_{b} (\frac{(T+\bar{T})^{i}}{S+\bar{S}} [X]_{i}) \Big) {\cal{G}}_{I\bar{J}} \ . \nonumber
%\label{ep6}
\end{align}
Adding \eqref{ep5} and \eqref{ep6}, we see that the first three terms in each expression exactly cancel and, hence, it follows from \eqref{bc4} that
\begin{equation}
R^{(\epsilon_{S}')}_{a\bar{b}I\bar{J}}=\frac{\epsilon_{S}'}{6}\ee^{K_{T}/3}  
\Big((\partial_{a}\partial_{\bar{b}}K_{T}) \frac{(T+\bar{T})^{i}}{S+\bar{S}} [X]_{i} +3\partial_{a}\partial_{b} (\frac{(T+\bar{T})^{i}}{S+\bar{S}} [X]_{i}) \Big) {\cal{G}}_{I\bar{J}} \ .
\label{ep7}
\end{equation}
Having presented the generic expression for $R^{(0)}_{a\bar{b}I\bar{J}}$ and $R^{(\epsilon_{S}')}_{a\bar{b}I\bar{J}}$ in \eqref{s10} and \eqref{ep7} respectively, it is again useful to rewrite them  in terms of the $\hat{R}$, $a^{i}$, $V$. For $R^{(0)}_{a\bar{b}I\bar{J}}$ we find that
\begin{equation}
R^{(0)}_{a\bar{b}I\bar{J}}=\frac{1}{3\hat{R}}(\partial_{a}\partial_{\bar{b}}K_{T}){\cal{G}}_{I\bar{J}} \ ,
\label{s10A}
\end{equation}
where $\partial_{a}\partial_{\bar{b}}K_{T}$ is given in \eqref{kn4}. The expression for $R^{(\epsilon_{S}')}_{a\bar{b}I\bar{J}}$, however, is considerably more complicated. In this paper, in order to simplify a long calculation, we will present the components of this quantity, not only in terms of the variables $\hat{R}$, $a^{i}$, $V$ and $\lambda$, but will further restrict the result to the line bundle $L=\mathcal{O}_{X}(2,1,3)$ with $W_{i}=(9,17,0)$ discussed above only. Moreover, we will present the results in the unity gauge defined in \eqref{wall2A}. For this specific case, defining the indices $a=S,T^{1},T^{2},T^{3},Z$ and $\bar{b}=\bar{S},\bar{T}^{1},\bar{T}^{2},\bar{T}^{3},\bar{Z}$, we find that
\begin{equation}
R^{(\epsilon_{S}')}_{a\bar{b}I\bar{J}}= M^{(\epsilon_{S}')}_{a\bar{b}}{\cal{G}}_{I\bar{J}}   \ ,
\label{fi1}
\end{equation}
where $M^{(\epsilon_{S}')}_{a\bar{b}}$ is the real, symmetric matrix specified by
\begin{align}
\noindent \bar{S}~ {\rm Terms}:~~ M_{S\bar{S}}&=\frac{1}{4\hat{R}V^{3}} \left( \frac{2a^{1}}{3}-\frac{a^{2}}{3}+4a^{3}+\left(\tfrac{1}{2}-\lambda\right)^{2}(9a^{1}+17a^{2}) \right) \label{help1}\ , \\
M_{T^{1}\bar{S}}&=-\frac{1}{8\hat{R}^{2}V^{5/3}} \left( \frac{2}{3}+9\left(\tfrac{1}{2}-\lambda\right)^{2} \right)\ , \\
M_{T^{2}\bar{S}}&=-\frac{1}{8\hat{R}^{2}V^{5/3}} \left( -\frac{1}{3}+17\left(\tfrac{1}{2}-\lambda\right)^{2} \right) \ ,\\
M_{T^{3}\bar{S}}&=-\frac{1}{2\hat{R}^{2}V^{5/3}} \ ,\\
M_{Z\bar{S}}&=\frac{1}{\hat{R}V}\left(\tfrac{1}{2}-\lambda\right)\left(1-\frac{1}{2V}\left(\lambda+\tfrac{1}{2}\right)(9a^{1}+17a^{2})  \right)\ ,
\label{res1}\\
\bar{T}^{1}~ {\rm Terms}:~~ M_{T^{1}\bar{T}^{1}}&=\frac{1}{6 \hat{R} V} (\partial_{T^{1}}\partial_{\bar{T}^{1}}K_{T})\biggl(  \frac{2a^{1}}{3}-\frac{a^{2}}{3}+4a^{3}  \\
&\eqspace+\left(\tfrac{1}{2}-\lambda\right)^{2}(9a^{1}+17a^{2}) \biggr)\ ,  \nonumber\\
M_{T^{2}\bar{T}^{1}}&=\frac{1}{6 \hat{R} V} (\partial_{T^{2}}\partial_{\bar{T}^{1}}K_{T})\biggl(  \frac{2a^{1}}{3}-\frac{a^{2}}{3}+4a^{3} \\
&\eqspace+\left(\tfrac{1}{2}-\lambda\right)^{2}(9a^{1}+17a^{2}) \biggr)\ ,  \nonumber\\
M_{T^{3}\bar{T}^{1}}&=\frac{1}{6 \hat{R} V} (\partial_{T^{3}}\partial_{\bar{T}^{1}}K_{T})\biggl(  \frac{2a^{1}}{3}-\frac{a^{2}}{3}+4a^{3} \\
&\eqspace+\left(\tfrac{1}{2}-\lambda\right)^{2}(9a^{1}+17a^{2}) \biggr)\ ,  \nonumber \\
M_{Z\bar{T}^{1}}&=-\frac{9}{4\hat{R^{3}} V^{1/3}}  \frac{\left(\frac{1}{2}-\lambda\right)}{(9a^{1}+17a^{2})}\ ,
\label{res2}\\
\bar{T}^{2}~ {\rm Terms}:~~ M_{T^{2}\bar{T}^{2}}&=\frac{1}{6 \hat{R} V} (\partial_{T^{2}}\partial_{\bar{T}^{2}}K_{T})\biggl(  \frac{2a^{1}}{3}-\frac{a^{2}}{3}+4a^{3} \\
&\eqspace+\left(\tfrac{1}{2}-\lambda\right)^{2}(9a^{1}+17a^{2}) \biggr)\ ,  \nonumber\\
M_{T^{3}\bar{T}^{2}}&=\frac{1}{6 \hat{R} V} (\partial_{T^{3}}\partial_{\bar{T}^{2}}K_{T})\biggl(  \frac{2a^{1}}{3}-\frac{a^{2}}{3}+4a^{3} \\
&\eqspace+\left(\tfrac{1}{2}-\lambda\right)^{2}(9a^{1}+17a^{2}) \biggr)\ , \nonumber\\
M_{Z\bar{T}^{2}}&=-\frac{17}{4\hat{R}^{3} V^{1/3}}  \frac{\left(\tfrac{1}{2}-\lambda\right)}{(9a^{1}+17a^{2})} \ ,
\label{res3}\\
\bar{T}^{3}~ {\rm Terms}:~~ M_{T^{3}\bar{T}^{3}}&=\frac{1}{6 \hat{R} V} (\partial_{T^{3}}\partial_{\bar{T}^{3}}K_{T})\biggl(  \frac{2a^{1}}{3}-\frac{a^{2}}{3}+4a^{3} \\
&\eqspace+\left(\tfrac{1}{2}-\lambda\right)^{2}(9a^{1}+17a^{2}) \biggr)\ ,  \nonumber\\
M_{Z\bar{T}^{2}}&=0\ ,
\label{res4}\\
\bar{Z}~ {\rm Terms}:~~ M_{Z\bar{Z}}&= \frac{1}{8\hat{R}^{3}V^{1/3}}  \frac{1}{(9a^{1}+17a^{2})}  \ . 
\label{res5}
\end{align}
For completeness, we restate that
\begin{equation}
\partial_{T^{m}}\partial_{\bar{T}^{l}}K_{T}=\frac{-d_{lmi}a^{i}}{4\hat{R}^{2}V^{1/3}} +\frac{d_{mij}a^{i}a^{j}d_{lpq}a^{p}a^{q}}{16\hat{R}^{4}V^{2/3}} \ ,
\label{kn4A}
\end{equation}
where
\begin{align}
d_{1jk}a^{j}a^{k}&=\frac{2}{3}a^{1}a^{2}+\frac{1}{3}(a^{2})^{2}+2a^{2}a^{3}\ , \\
d_{2jk}a^{j}a^{k}&=\frac{(a^{1})^{2}}{3}+\frac{2}{3}a^{1}a^{2}+2a^{1}a^{3}\ , \\
d_{3jk}a^{j}a^{k}&=2a^{1}a^{2} \ .
\label{ing1}
\end{align}

Using the above results, one can now calculate the coefficients of the quadratic scalar soft supersymmetry breaking terms to linear order in $\epsilon_{S}'$. Recall from \eqref{bc5} that
\begin{equation}
m_{I\bar{J}}^{2} = m^{(0)2}_{I\bar{J}} +m^{(\epsilon_{S}')2}_{I \bar{J} } \ .
\label{bc5A}
\end{equation}
Then it follows from \eqref{bc6}, \eqref{s4} and \eqref{s10A} that 
\begin{equation}
m^{(0)2}_{I\bar{J}}=\frac{1}{\hat{R}} \left( m_{3/2}^{2}-\frac{1}{3}F^{a}\bar{F}^{\bar{b}}(\partial_{a}\partial_{\bar{b}}K_{T})  \right) {\cal{G}}_{I\bar{J}} \ ,
\label{bc6AA}
\end{equation}
and from \eqref{bc7}, \eqref{s5AAA} and \eqref{fi1} that, in unity gauge for $L=\mathcal{O}_{X}(2,1,3)$ with $W_{i}=(9,17,0)$,
\begin{align}
\label{eq:squared_scalar_mass}
m^{(\epsilon_{S}')2}_{I\bar{J}}=&\biggl(\frac{m_{3/2}^{2}}{2\hat{R}V} \left[ \left( \frac{2}{3}+9 \left(\tfrac{1}{2}-\lambda\right)^{2} \right) a^{1}+ \left(-\frac{1}{3}+17\left(\tfrac{1}{2}-\lambda\right)^{2} \right) a^{2} +4a^{3} \right] \\
& \eqspace -F^{a}\bar{F}^{\bar{b}}M^{(\epsilon_{S}')}_{a\bar{b}} \biggr) {\cal{G}}_{I\bar{J}}\ , \nonumber
\end{align}
with the coefficients of $M^{(\epsilon_{S}')}_{a\bar{b}}$ given in \eqref{help1} -- \eqref{res5}.
Adding \eqref{bc6AA} and \eqref{eq:squared_scalar_mass}, the scalar masses squared $m_{I\bar{J}}^{2}$ can be put in the simple form
\begin{equation}
m_{I\bar{J}}^{2} =m^2_s(a^1,a^2,a^3)\mathcal{G}_{I\bar J}\ ,
\end{equation}
 where $m_s^2$ is a moduli-dependent function which is independent of the $I,\bar{J}$ indices. This function can be computed at any point inside the ``viable'' region of K\"ahler moduli space associated with $L=\mathcal{O}_{X}(2,1,3)$. Recall that we have assumed that $\mathcal{G}_{I\bar J}$ is a moduli-independent matrix, with numerical entries.

\subsection{Cubic Scalar Coefficients}

The generic form for the mass-dimension-one coefficients of the cubic scalar soft supersymmetry breaking terms were shown in \cite{Soni:1983rm,Kaplunovsky:1993rd,Louis:1994ht,Brignole:1998dxa} to be 
\begin{equation}
a_{IJK}=F^{a}\left( \partial_{a}Y_{IJK}+\tfrac{1}{2}(\partial_{a}\hat{K})Y_{IJK}-3\Gamma^{N}_{a(I}Y_{JK)N} \right) \ ,
\label{night1}
\end{equation}
where $\hat{K}$ is given in \eqref{sum1}, \eqref{split2} and \eqref{tel1}, and $Y_{IJK}$ is
\begin{equation}
Y_{IJK}=\ee^{\hat{K}/2}\hat{Y}_{IJK}=\ee^{\hat{K}/2} 2\sqrt{2\pi\hat{\alpha}_{GUT}} ~y_{IJK}\ ,
\label{night2}
\end{equation}
with $\hat{\alpha}_{GUT}$ defined in \eqref{lala1}, $y_{IJK}$ the Yukawa couplings at mass scale $\langle M_{U} \rangle$ and, as defined in \eqref{r2},
\begin{equation}
\Gamma^{N}_{aI}=Z^{N\bar{J}}\partial_{a}Z_{\bar{J}I} \ .
\label{r2A}
\end{equation}
Noting that \eqref{night2} implies 
\begin{equation}
\partial_{a}Y_{IJK}=\frac{1}{2}(\partial_{a}\hat{K})Y_{IJK} \ ,
\label{night3}
\end{equation}
it follows that expression \eqref{night1} can be simplified to
\begin{equation}
a_{IJK}=F^{a}\left( (\partial_{a}\hat{K})Y_{IJK}-3\Gamma^{N}_{a(I}Y_{JK)N} \right) \ .
\label{night1A}
\end{equation}
As discussed previously, subject to the assumption that
\begin{equation}
Z_{I\bar{J}}^{(0)}=\ee^{K_{T}/3}{ \cal{G}}_{I\bar{J}} \ ,
\label{doc1}
\end{equation}
it follows from \eqref{ep2} and \eqref{ep3} that
\begin{equation}
\Gamma^{N}_{aI}=\Gamma^{(0)N}_{aI} +\Gamma^{(\epsilon_{S}')N}_{aI}=\left( \frac{1}{3}(\partial_{a}K_{T}) +\frac{\epsilon_{S}'}{2} \partial_{a}\big( \frac{(T+\bar{T})^{i}}{(S+\bar{S})}[X]_i \big)  \right)\delta^{N}_{I} \ ,
\label{doc2}
\end{equation}
where $[X]_i$ is defined in \eqref{ep4}. Furthermore, the symmetry of the cubic couplings $Y_{IJK}$ implies that
\begin{equation}
\Gamma^{N}_{a(I}Y_{JK)N}=\left[ \frac{1}{3}(\partial_{a}K_{T}) +\frac{\epsilon_{S}'}{2} \partial_{a}\left( \frac{(T+\bar{T})^{i}}{(S+\bar{S})} [X]_i \right) \right] Y_{IJK} \ .
\label{doc3}
\end{equation}
Inserting this into \eqref{night1A} and using \eqref{sum1}, it follows that
\begin{equation}
a_{IJK}=F^{a}\partial_{a} \left( \tilde{K}_{S}-\frac{3}{2} \epsilon_{S}' \frac{(T+\bar{T})^{i}}{(S+\bar{S})} [X]_i \right) Y_{IJK}\ .
\label{doc4}
\end{equation}

To compare this result to the formalism for the soft terms in the $B-L$ MSSM presented in \cite{Ovrut:2012wg}, it is convenient to write \eqref{doc4} in the form
\begin{equation}
\begin{split}
a_{IJK}&= \mathcal{A}(S,T^{i},Z) ~y_{IJK} \ ,\\
\end{split}
\label{doc5}
\end{equation}
where $\mathcal{A}$ is the specific function of the moduli given by
\begin{equation}
\mathcal{A}(S,T^{i},Z)=2 \sqrt{2\pi\hat{\alpha}_{GUT}}\ee^{\hat{K}/2}F^{a}\partial_{a} \left( \tilde{K}_{S}-\frac{3}{2} \epsilon_{S}' \frac{(T+\bar{T})^{i}}{(S+\bar{S})} [X]_i \right)\ .
\label{doc6}
\end{equation}
As discussed in detail in \cite{Ovrut:2014rba,Ovrut:2015uea}, in the renormalization analysis of the $B-L$ MSSM the experimental values of the quark and lepton Yukawa parameters $y_{IJK}$ are entered into the theory at the electroweak scale. These parameters are then run up using the RGEs to give precise values for the Yukawa couplings $ y_{IJK}$ at the unification scale $\langle M_{U} \rangle$. Hence, the $y_{IJK}$ parameters in the above analysis and in \eqref{doc5} are completely specified. Hence, the only unknown part of the soft supersymmetry breaking cubic parameters is the {\it universal} moduli function $\cal{A}$ defined in \eqref{doc6}. Its exact value will depend on where it is evaluated in moduli space. We note, for completeness, that the renormalization group equation used in \cite{Ovrut:2014rba,Ovrut:2015uea} sets all Yukawa parameters to zero except for the top and bottom quarks and for the tau lepton, including in the soft supersymmetry breaking terms. As shown in earlier work, the remaining Yukawa parameters are too small to lead to significant effects and are hence ignored. 

The $\partial_{a} \big( \tilde{K}_{S}-\frac{3}{2} \epsilon_{S}' \frac{(T+\bar{T})^{i}}{(S+\bar{S})} [X]_i \big)$ factors in the universal soft supersymmetry breaking cubic coefficient \eqref{doc6} can be explicitly calculated as functions of the moduli using \eqref{d} and the expression for $[X]_i$ in \eqref{ep4}. However, the generic results are not particularly enlightening. As we did in previous sections, we will present each of these quantities written in terms of the variables $\hat{R}$, $a^{i}$, $V$ and $\lambda$, and restricted to the case of the line bundle $L=\mathcal{O}_{X}(2,1,3)$ discussed above. Furthermore, we will present the results in unity gauge defined in \eqref{wall2A}. For this specific case, defining the indices $a=S,T^{1},T^{2},T^{3},Z$, we find that 
\begin{align}
\partial_{S} \left( \tilde{K}_{S}-\frac{3}{2} \epsilon_{S}' \frac{(T+\bar{T})^{i}}{(S+\bar{S})} [X]_i \right)=&-\frac{1}{2V}\biggl(1-\frac{1}{2V}(2a^{1}-a^{2}+12a^{3}) \label{toi1}\\
&\eqspace-\frac{3}{2V}\left(\lambda-\tfrac{1}{2}\right)^{2} (9a^{1}+17a^{2}) \biggr)\ , \nonumber \\
\partial_{T^{i}} \left( \tilde{K}_{S}-\frac{3}{2} \epsilon_{S}' \frac{(T+\bar{T})^{l}}{(S+\bar{S})} [X]_l \right)=& -\frac{1}{4\hat{R}V^{2/3}} \biggl[(2,-1,12)_{i} \label{toi2} \\
& \eqspace +\left(3-2\left(\lambda+\tfrac{1}{2}\right)^{2}\right)(9,17,0)_{i} \biggr]\ ,  \nonumber \\
\partial_{Z} \left( \tilde{K}_{S}-\frac{3}{2} \epsilon_{S}' \frac{(T+\bar{T})^{i}}{(S+\bar{S})} [X]_i \right)=&\frac{1}{2\hat{R}V^{2/3}}\left(3-2\left(\lambda-\tfrac{1}{2} \right) \right) \label{toi3} \ .
\end{align}
Putting these results into \eqref{doc6} and computing the associated F-term fields $F^{a}$ using the results of Section 3, it follows from \eqref{doc6} that one can compute the universal $\mathcal{A}$ coefficient for any given point in the ``viable'' region of K\"ahler moduli space associated with the line bundle $L=\mathcal{O}_{X}(2,1,3)$.

\subsection{The Holomorphic Quadratic Term}

The generic form for the mass-dimension-two coefficient of the quadratic scalar soft supersymmetry breaking terms $C_{I}C_{J}+\text{h.c.}$ was shown in \cite{Soni:1983rm,Kaplunovsky:1993rd,Louis:1994ht} to be
\begin{equation}
B_{IJ}=F^{a}\left(\partial_{a}\mu_{IJ}+\tfrac{1}{2}(\partial_{a}\hat{K})-3\Gamma^{N}_{a(I}\mu_{J)N} \right)- m_{3/2}\mu_{IJ}
\label{naan1}
\end{equation}
where $\hat{K}$ is given in \eqref{sum1}, \eqref{split2} and \eqref{tel1}, $m_{3/2}$ is defined in \eqref{van1} and 
\begin{equation}
\mu_{IJ}=\ee^{\hat{K}/2}\hat{\mu}_{IJ}\ ,
\label{naan2}
\end{equation}
with $\hat{\mu}_{IJ}$ the dimension-one parameter of the $C^{I}C^{J}$ holomorphic term in the superpotential.

Note that the three terms in the brackets in \eqref{naan1} are exactly of the same form as the expression for $a_{IJK}$ given in \eqref{night1} in the previous subsection, with $Y_{IJK}$ now replaced by $\mu_{IJ}$. Using the fact that $\mu_{IJ}$ is symmetric in $IJ$, and following the same procedure as was used to evaluate $a_{IJK}$ above, we find that 
\begin{equation}
F^{a}\left(\partial_{a}\mu_{IJ}+\tfrac{1}{2}(\partial_{a}\hat{K})-3\Gamma^{N}_{a(I}\mu_{J)N} \right)=F^{a}\partial_{a} \left( \tilde{K}_{S}-\frac{3}{2} \epsilon_{S}' \frac{(T+\bar{T})^{i}}{(S+\bar{S})} [X]_i \right) \mu_{IJ} \ ,
\label{naan3}
\end{equation}
where $\tilde{K}_{S}$ is defined in \eqref{split2} and $[X]_i$ is presented in \eqref{ep4}. It follows that 
\begin{equation}
B_{IJ}=\ee^{\hat{K}/2}\left[F^{a}\partial_{a} \left( \tilde{K}_{S}-\frac{3}{2} \epsilon_{S}' \frac{(T+\bar{T})^{i}}{(S+\bar{S})} [X]_i \right)-m_{3/2}   \right]\hat  \mu_{IJ} \ .
\label{naan4}
\end{equation}
Restricting these expression to the line bundle $L=\mathcal{O}_{X}(2,1,3)$ discussed above and expressing the results in the variables $\hat{R}$, $a^{i}$, $V$ and $\lambda$ in unity gauge, $B_{IJ}$ can then be evaluated using expressions in \eqref{toi1}, \eqref{toi2}, \eqref{toi3} and \eqref{van1}. As was the case for the $a_{IJK}$ coefficients above, it is useful to write $B_{IJ}$ in the form
\begin{equation}
B_{IJ}={\mathcal{B}}(S, T^{i}, Z) {\hat{\mu}}_{IJ}\ ,
\label{pa1}
\end{equation}
where $\mathcal{B}$ is a function of the moduli given by
\begin{equation}
{\mathcal{B}}(S, T^{i}, Z) =\ee^{\hat{K}/2}\left[F^{a}\partial_{a} \left( \tilde{K}_{S}-\frac{3}{2} \epsilon_{S}' \frac{(T+\bar{T})^{i}}{(S+\bar{S})} [X]_i \right)-m_{3/2}   \right] \ .
\label{pa2}
\end{equation}
To compare this result to the superpotential and the soft terms in the $B-L$ MSSM presented in \eqref{con2} and \eqref{eq:6} respectively, we note that 
\begin{equation}
	{\hat{\mu}}_{IJ}=   \begin{cases}
		\mu & \text{for }  I=H_{u}, J=H_{d} \text{ or } I=H_{d}, J=H_{u} \ , \\
		0 & \text{for other choices of }I,J \ ,
	\end{cases}                               
	\label{pa3}  
\end{equation}
and, therefore, that
\begin{equation}
	B_{IJ}=   \begin{cases}
		b & \text{for }  I=H_{u}, J=H_{d} \text{ or } I=H_{d}, J=H_{u} \ , \\
		0 & \text{for other choices of }I,J \ ,
	\end{cases}                               
	\label{pa4}  
\end{equation}
where 
\begin{equation}
b={\mathcal{B}}(S, T^{i}, Z)\, \mu \ .
\label{pa5}
\end{equation}
We note that in this paper, as with previous work \cite{Ovrut:2014rba,Deen:2016zfr,Dumitru:2018jyb,Dumitru:2018nct,Dumitru:2019cgf}, we make no attempt to solve the ``${\mu}$ problem" \cite{Martin:1997ns}. Therefore, the value of the parameter $\mu$ at the unification scale is unconstrained. However, unlike our previous work, where the $b$ parameter was also unconstrained, in the present paper the ratio
\begin{equation}
\frac{b}{\mu}={\mathcal{B}}(S, T^{i}, Z)
\label{pa6}
\end{equation}
{\it is} constrained at the unification scale for any given viable point in moduli space. Hence, constraint \eqref{pa6} must be satisfied for any initial black point to be acceptable.

\section{Realistic Soft Supersymmetry Breaking Terms}

Prior to the present paper, the 24 independent coefficients of the soft supersymmetry breaking terms for the $B-L$ MSSM were unknown and, with the exception of a small number of constraints, completely arbitrary. Hence, to compute the low energy phenomenological predictions of the theory, in previous work \cite{Ovrut:2014rba,Ovrut:2015uea} the RGEs were run down from the unification scale to the electroweak scale using arbitrary ``initial'' soft breaking coefficients, chosen {\it statistically} by scattering their values arbitrarily over a large range. To achieve phenomenologically realistic results, it was necessary that a) the gauged $B-L$ symmetry be spontaneously broken at a sufficiently high scale, b) electroweak symmetry be broken with the correct $W^{\pm}$ and $Z^{0}$ boson masses, c) that the Higgs mass have its experimental value and, finally, d) that all sparticle masses be above their present experimental lower bounds. Not all such initial ``points'' in the space of coefficients satisfied all of these criteria and, hence, these specific coefficients were not acceptable. However, it was shown in a number of papers \cite{Ovrut:2014rba,Ovrut:2015uea,Deen:2016zfr} that a surprisingly large number of such arbitrarily chosen, non fine-tuned initial points did, indeed, satisfy all experimental criteria. These were named ``black'' points. The physical predictions associated with different ``black'' points were studied in detail in  \cite{Ovrut:2014rba,Dumitru:2018jyb,Dumitru:2018nct,Dumitru:2019cgf}.

However, in Sections 3, 4 and 5 of this paper, we have presented {\it explicit} calculations of all of the soft supersymmetry breaking parameters in the $B-L$ MSSM assuming gaugino condensation in the hidden sector of the theory. Since these parameters are dependent on the form of the hidden sector vector bundle, as well as on the various moduli 
of the vacuum, we restricted our generic results to the specific case of the hidden sector associated with the line bundle $L=\mathcal{O}_{X}(2,1,3)$. In this Section, we will compute the values of these soft supersymmetry breaking parameters at all points in the ``viable'' moduli space, discussed in Section 2, associated with $L=\mathcal{O}_{X}(2,1,3)$. Importantly, however, as noted at the end of Section 2, each point in the ``viable'' moduli space is associated with {\it both} the split Wilson lines and the simultaneous Wilson lines scenarios, albeit with somewhat different values of the fundamental parameters. A full RG analysis would require that one state which of these two scenarios will be chosen. The split Wilson lines scenario is technically more complicated since it requires the addition of an  ``intermediate'' scaling regime between the mass scales of the two Wilson lines. However,  the length of this additional mass interval turns out to be much smaller than an order of magnitude and, in the end, the values of the fundamental parameters are only slightly different from those of the simultaneous Wilson lines scenario. It follows that it is considerably simpler, and numerically almost identical to the accuracy we are working, to use the simultaneous scenario when doing a RG analysis in the context of the soft supersymmetry breaking parameters. For these reasons, {\it we will use the simultaneous Wilson lines scenario for the remainder of this paper}. Doing this, we will determine, at any given ``viable'' point in K\"ahler moduli space, whether or not any of these initial coefficients are phenomenologically acceptable ``black'' points. Furthermore, we will present the complete K\"ahler moduli subspace for which each point allows such ``black'' points.

We begin our analysis with the soft supersymmetry breaking gaugino mass terms presented in
\eqref{need5} of Section 5. There we found that
\begin{equation}
 \label{eq:ReGauginoMass}
\begin{split}
M_{1/2}&=\frac{1}{2\re f_1}F^a\partial_a \re  f_1\\
&=\frac{ \frac{1}{2}F^{S}+ \epsilon_{S} '  \frac{29}{12}F^{T^{1}} +\epsilon_{S} '  \frac{25}{6}F^{T^{2}} +\epsilon_{S} ' F^{T^{3}} - \epsilon_{S} ' \frac{1}{2}F^{Z}  }{2\big( V+\frac{1}{3}a^1-\frac{1}{6}a^2+2a^3+\frac{1}{2}\left( \frac{1}{2}-\lambda \right)(9a^{1}+17a^{2}) \big)}\ .  \\
\end{split}
\end{equation}
It is important to note that, although dependent on the K\"ahler moduli, it has the {\it same value} for each of the gauginos associated with $SU(3)_{C}, SU(2)_{L}, U(1)_{3R}$ and $U(1)_{B-L}$. That is, it does not depend on the gauge group factor. Hence, the first prediction of our theory is that the soft supersymmetry breaking 
gaugino masses are {\it universal} at any fixed point in K\"ahler moduli space at the unification scale. That is,
\begin{equation}
M_3=M_2=M_{3R}=M_{BL}=M_{1/2}\ .
\end{equation}
Computing the expression in  \eqref{eq:ReGauginoMass} at different points $(a^1,a^2,a^3)$ across the ``viable'' region of K\"ahler moduli space associated with $L=\mathcal{O}_{X}(2,1,3)$ -- shown 
in Figure \ref{fig:KahlerViableRegion} -- will further fix the scale of these mass terms and their sign.

Next, consider the soft supersymmetry breaking cubic scalar couplings. These were computed in \eqref{doc5} of the previous section and found to be
\begin{equation}
\begin{split}
a_{IJK}&= \mathcal{A}(S,T^{i},Z) ~y_{IJK} \ ,\\
\end{split}
\label{doc5X}
\end{equation}
where $y_{IJK}$ are the Yukawa couplings at the unification scale and $\mathcal{A}$ is the specific function of the $S$, $T^{i}$ and $Z$ moduli given by
\begin{equation}
\mathcal{A}(S,T^{i},Z)=2 \sqrt{2\pi\hat{\alpha}_{GUT}}\ee^{\hat{K}/2}F^{a}\partial_{a} \left( \tilde{K}_{S}-\frac{3}{2} \epsilon_{S}' \frac{(T+\bar{T})^{i}}{(S+\bar{S})} [X]_i \right) \ .
\label{doc6X}
\end{equation}
The Yukawa couplings at the unification scale are computed from their experimental values at the electroweak scale using the RG. That is, near the unification scale, all cubic couplings are 
proportional to the Yukawa couplings with the same moduli-dependent proportionality function $\cal{A}$.

Finally, it follows from \eqref{bc5A}, \eqref{bc6AA} and \eqref{eq:squared_scalar_mass} that the expression for the squared scalar masses $m_{I\bar{J}}^{2}$ can be put in the simple form
\begin{equation}
m_{I\bar{J}}^{2} =m^2_s(a^1,a^2,a^3)\mathcal{G}_{I\bar J}\ ,
\end{equation}
 where $m_s^2$ is a moduli-dependent function, independent of the $I,\bar{J}$ indices, given by
\begin{align}
m^2_s(a^1,a^2,a^3)&=\frac{1}{\hat{R}} \left( m_{3/2}^{2}-\tfrac{1}{3}F^{a}\bar{F}^{\bar{b}}(\partial_{a}\partial_{\bar{b}}K_{T})  \right) 
+\frac{m_{3/2}^{2}}{2\hat{R}V} \Bigl[ \left( \tfrac{2}{3}+9 \left(\tfrac{1}{2}-\lambda\right)^{2} \right) a^{1} \nonumber  \\
&\eqspace+ \left(-\tfrac{1}{3}+17\left(\tfrac{1}{2}-\lambda\right)^{2} \right) a^{2} +4a^{3} \Bigr]  -F^{a}\bar{F}^{\bar{b}}M^{(\epsilon_{S}')}_{a\bar{b}} \ . \label{eq:DefineMsA}
\end{align}
Unlike  the gaugino mass terms and the cubic couplings which are determined for fixed K\"ahler moduli, 
there is a level of arbitrariness in the moduli-independent matrix $\mathcal{G}_{I\bar J}$ that one cannot eliminate in the case of the scalar squared masses. As we mentioned earlier, currently there 
is no way to compute $\mathcal{G}_{I\bar J}$ explicitly. However, the absence of flavor changing neutral currents in low-energy experiments restricts this matrix to be diagonal~\cite{Llacer:2019owe,Aad:2019pxo,ATLAS:2019pcn,Peixoto:2018apt,Aaboud:2018nyl}. Considering this, the simplest choice one can make for 
 $\mathcal{G}_{I\bar J}$ is to assume that it has equal entries; that is, it is proportional, with an arbitrary non-vanishing constant, to $\delta_{I\bar{J}}$. This is a typical case considered in unification scenarios, where all soft squared scalar masses are 
 taken to be universal. We find, however, that we can never break $B-L$ symmetry in this simple universal case. 
 The reason is that in the $B-L$ MSSM, the third generation right-handed sneutrino mass squared must turn
 negative at low energy in order to  trigger $B-L$ symmetry breaking.  However, when the scalar squared masses are 
 universal at the input scale $\langle M_{U} \rangle$ this is never possible, as we now show.
 
 As presented in \cite{Ambroso:2009jd,Ambroso:2009sc,Ovrut:2012wg}, the right-handed sneutrino mass RGE is
 \begin{equation}
 16\pi^2 \frac{d}{dt}m^2_{\tilde \nu_3^c}=-3g_{BL}^2 M_{BL}^2-2g_{R}^2 M_{3R}^2+\frac{3}{4}g_{BL}^2S_{BL}-g_R^2S_R\ ,
 \end{equation}
 where
 \begin{equation}
  \label{eq:SBL_SR}
 \begin{split}
 S_{BL}&={\bf{Tr}}(2m^2_{\tilde Q}-m^2_{\tilde u^c}-m^2_{\tilde d^c}-2m^2_{\tilde L}+m^2_{\tilde \nu^c}+m^2_{\tilde e^c})\ ,\\
 S_R&=m_{H_u}^2-m_{H_d}^2+{\bf{Tr}}\left( -\frac{3}{2}m^2_{\tilde u^c}+ \frac{3}{2}m^2_{\tilde d^c}
 - \frac{1}{2}m^2_{\tilde \nu^c}+ \frac{1}{2}m^2_{\tilde e^c} \right)\ .
 \end{split}
 \end{equation}
 The analytic solution for the right-handed sneutrino mass at the $B-L$ scale is then given by
 \begin{align}
& m^2_{\tilde \nu_3^c}(M_{B-L})\nonumber\\
&=m^2_{\tilde \nu_3^c}(\langle M_U \rangle)\nonumber  \\
 &+\frac{1}{14}\frac{g^4_R(\langle M_U \rangle)-g_R^4(M_{B-L})}{g_U^4}M_{3R}(\langle M_{U }\rangle)^2+ \frac{1}{8}\frac{g^4_{BL}(\langle M_{U}\rangle)-g_{BL}^4(M_{B-L})}{g_U^4}M_{BL}(\langle M_U \rangle)^2 \nonumber \\
  &+\frac{1}{14}\frac{g^2_R(\langle M_{U} \rangle)-g_R^2(M_{B-L})}{g_R^2(\langle M_{U} \rangle)}S_R(\langle M_{U} \rangle)- \frac{1}{16}\frac{g^2_{BL}(\langle M_{U} \rangle)-g_{BL}^2(M_{B-L})}{g_U^4}S_{BL}(\langle M_{U} \rangle) \ . \label{eq:sneutrinoRGE}  
\end{align}
Since Abelian gauge couplings grow larger at higher scales, a tachyonic sneutrino is possible only when $S_R$ 
is negative and/or $S_{BL}$ is positive at the unification scale. For universal soft scalar masses, however, both $S$ terms 
exactly vanish at the unification scale. It follows that universal scalar soft masses are physically unacceptable in the $B-L$ MSSM.

A more general, and more suitable, choice for $\mathcal{G}_{I\bar J}$ is a diagonal matrix with distinct values 
for different particle types. The matrix elements of $\mathcal{G}_{I\bar J}$ are not expected to depend on the particle 
families. Hence, we set diagonal entries to be equal for different generations of the same sparticle representations--differentiating only between squarks, sleptons and up- and down-Higgs scalars.  
Specifically, we set
\begin{equation}
 \label{eq:GIJ_elements}
\begin{split}
\mathcal{G}_{\tilde Q_1}=&\mathcal{G}_{\tilde Q_2}=\mathcal{G}_{\tilde Q_3} \ ,\\
\mathcal{G}_{\tilde u^c_1}=\mathcal{G}_{\tilde u^c_2}=\mathcal{G}_{\tilde u^c_3} \ ,&\qquad \mathcal{G}_{\tilde d^c_1}=\mathcal{G}_{\tilde d^c_2}=\mathcal{G}_{\tilde d^c_3} \ ,\\
\mathcal{G}_{\tilde L_1}=&\mathcal{G}_{\tilde L_2}=\mathcal{G}_{\tilde L_3}\ ,\\ 
\mathcal{G}_{\tilde \nu^c_1}=\mathcal{G}_{\tilde \nu^c_2}=\mathcal{G}_{\tilde \nu^c_3}  \ , &\qquad \mathcal{G}_{\tilde e^c_1}=\mathcal{G}_{\tilde e^c_2}=\mathcal{G}_{\tilde e^c_3}\ ,\\
 \mathcal{G}_{\tilde H_u}&\neq\mathcal{G}_{\tilde H_d}\ .
\end{split}
\end{equation}
That is, $\mathcal{G}_{I\bar J}$ will be a diagonal matrix composed of the eight independent parameters in \eqref{eq:GIJ_elements}.
Having done this, we must state how to choose the numerical values of these quantities. Since, at present, it is unknown how to calculate them, one could either a) simply assign them fixed values or b) as was done for {\it all} soft supersymmetry breaking coefficients in previous work~\cite{Ovrut:2012wg,Ovrut:2014rba,Dumitru:2018nct,Dumitru:2019cgf} -- including the gaugino masses and cubic scalars -- choose them statistically within some suitable size range. In this paper, we will follow the statistical approach. That is, at any given point in moduli space, we will calculate, using \eqref{eq:DefineMsA}, the universal coefficient $m^2_s(a^1,a^2,a^3)$. The $\mathcal{G}_{I\bar J}$ matrix will be taken to be of the generic form given in \eqref{eq:GIJ_elements}. However, the values of its eight independent diagonal coefficients will be chosen {\it statistically}. That is, we will treat each such coefficient as uncorrelated from the others, and assign to any given coefficient random values, varying between $\frac{1}{{10}}$ and ${10}$, according to a log-normal 
distribution. That is, 
\begin{equation}
 \label{eq:GIJ_range}
\mathcal{G}_{\tilde Q_1,\tilde u^c_1,\tilde d^c_1,\tilde L_1,\tilde \nu^c_1,\tilde e^c_1,\tilde H_u,\tilde H_d}\in \left[\frac{1}{{10}},{10}\right] \ .
\end{equation}
This choice of distribution for the matrix elements ensures that no scalar mass squared differs by more than an order of magnitude 
from the universal coefficient value $m_s^2$ at any given point in moduli space. Furthermore, within this ensemble, all scalar masses are within the same order of magnitude of 
each other. Therefore, this statistical distribution can be considered a mild deviation from the universal case where $\mathcal G_{I\bar J}=\delta_{I\bar J}$, 
but one that, as we show below, allows for $B-L$ symmetry breaking. Before continuing, we want to emphasize that we are using a statistical approach for $m^{2}_{I\bar{J}}$ only. The soft gaugino and cubic scalar coefficients are fixed by \eqref{eq:ReGauginoMass} and \eqref{doc6X} respectively and, therefore, are not statistical. This should be compared with previous work \cite{Ovrut:2014rba,Deen:2016zfr,Dumitru:2018nct,Dumitru:2019cgf} where, having no theory for calculating any of the soft supersymmetry breaking terms (the gauginos, cubic scalars and scalar mass terms) one was forced to chose the values of each of these 24 independent coefficients statistically. There is one final parameter in the initial data of the RG simulation that is thrown at random; that is
\begin{equation}
\tan\beta \in [1.2,65] \ .
\label{eq:tanBrange}
\end{equation}
The upper and lower bounds for $\tan \beta$ are taken from \cite{Martin:1997ns} and are consistent with present bounds that 
ensure perturbative Yukawa couplings. 

Importantly, in this paper, there is now an {\it additional} constraint imposed by the $b$ parameter in the $B-L$ MSSM soft SUSY breaking Lagrangian. 
Although we can now compute the value of this parameter at the unification scale (see \eqref{pa5} and \eqref{pa6}), we have so far omitted it from our discussion. 
The reason we have, thus far, neglected $b$ is that it is not among the initial parameters that are thrown in the RG simulation in \cite{Ovrut:2014rba,Ovrut:2015uea}. In the original RG formulation, both parameters $\mu$ and $b$, 
from the $B-L$ MSSM superpotential  \eqref{con1} and the soft supersymmetry breaking terms \eqref{eq:6} respectively, are computed separately {\it at the electroweak scale} to ensure the correct value of the $Z^{0}$ mass and to guarantee that the low energy vacuum is stable. Specifically, the RG simulations calculate the values of $\mu$ and $b$ at the {\it EW scale} using the mathematical constraints 
\begin{equation}
\label{eq:mu_expression2}
\mu^2=\frac{m_{H_u}^2\tan^2 \beta-m_{H_d}^2}{1-\tan^2 \beta}-\frac{1}{2}M_Z^2
\end{equation}
and 
\begin{equation}
\label{eq:b_expression}
\frac{2b}{\sin 2\beta}=2\mu^2+m_{H_u}^2+m_{H_d}^2
\end{equation}
respectively. The RG simulation that we use is hard-coded this way. 
Modifying it so as to include the $b$ parameter in the initial data at high energy scale would pose a significant challenge, that we can 
easily avoid. We will do so by running the initial RG simulation using, as previously, the initial data for the soft gaugino masses, as well as the soft cubic scalar couplings, all computed at fixed points in the moduli space. The remaining free-parameters, that we will throw at random, are the matrix 
elements $\mathcal{G}_{I\bar J}$ and ${\rm tan}\beta$. If the full RGE analysis produces ``black'' points, each will have different values of $\mu$ and $b$ 
computed at low energy, using \eqref{eq:mu_expression2} and \eqref{eq:b_expression}, for the different values of $\tan \beta$ and matrix elements that were thrown. However, not all such
``black'' points are truly acceptable. For each of them, we need to run the values of $b$ and $\mu$ up to the {\it unification scale} using their own RGEs, and check to see if the ratio $\frac{b}{\mu}$ satisfies \eqref{pa6}.
If the ratio is correct, then this is called an {\it acceptable} ``black'' point. However, if \eqref{pa6} is not satisfied, then this ``black'' point is not physically acceptable. The RGEs for both $b$ and $\mu$ are presented in the Appendix, and discussed there in detail.

\begin{figure}[t]
	\centering
	\begin{subfigure}[b]{0.59\textwidth}
		\includegraphics[width=1.0\textwidth]{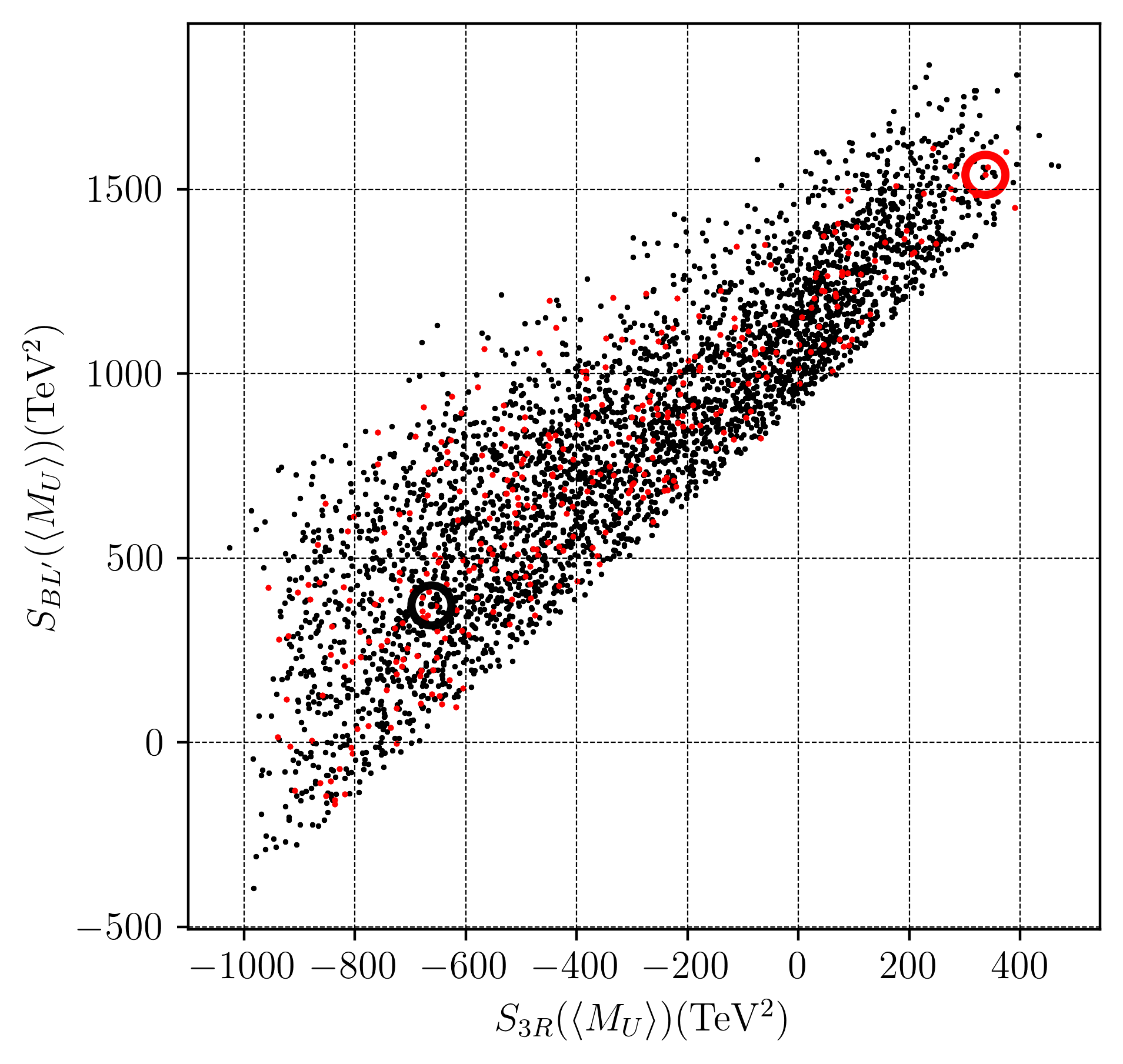}
		\centering
	\end{subfigure}
	\caption{ Plot of the ``black'' points obtained after running the RGE simulation at the point $(a^1,a^2,a^3)=(0.910, 1.401, 0.163)$. We statistically throw 10 million sets of random initial data. The 3,330 ``black'' points we obtain are further divided into two sets by color. With red we display the {\it acceptable} ``black'' points'' that, in addition, have the correct $\frac{b}{\mu}$ value at the unification scale (eq.~\eqref{cil3}). The remaining 2,974 {\it unacceptable} ``black'' points that do not satisfy this constraint are shown in black. Two points are circled in this plot. The point circled in red is an acceptable ``black'' point which has the initial data given in \eqref{eq:ExamplePoint2ss}. The point circled in black is an unacceptable ``black'' point which has the initial data shown in \eqref{eq:ExamplePoint2ssA}.}
	\label{fig:redScatterPlot}
\end{figure}

We will now give an example of this process. To begin,  we choose a point $(a^1,a^2,a^3)=(0.910, 1.401,0.163)$ inside the ``viable'' region of K\"ahler moduli space 
shown in Figure 1. Restricting our analysis to this point, we use \eqref{eq:ReGauginoMass}, \eqref{doc6X} and \eqref{eq:DefineMsA} to compute
\begin{equation}
\label{eq:ExamplePoint1}
M_{1/2}=-13{,}290~\text{GeV}\ , \quad \mathcal{A}=2{,}170~\text{GeV}\ , \quad m_s=4{,}079~\text{GeV} \ .
\end{equation}
These values can be used as initial data for our RG simulation. However, there remain unfixed initial data -- specifically, the eight scalar matrix elements of 
$\mathcal{G}_{I\bar J}$ and $\tan \beta \in [1.2,65]$, both of which are thrown randomly.
Using the so-created initial data sets, we run the RG simulation presented above.
Some of these initial data sets pass all the physical requirements discussed above. We call these initial throws ``black'' points.  We can plot these as ``black'' points on a two-dimensional graph of $S_{BL}$ versus $S_R$, which are defined in \eqref{eq:SBL_SR}. These two parameters are of great importance in our model as they set the scale of $B-L$ breaking. To put some numbers on this, at our chosen point we statistically throw 10 million sets of random initial data. We find that 3,330 of these sets are ``black'' points.
However, as discussed above, not all of these ``black'' points lead to the correct value of  $b/\mu$ at the unification scale. To analyze this, one must take each choice of initial conditions, run them down to the EW scale using the RGEs, calculate both parameters $b$ and $\mu$ at that scale and then, using the RG formalism given in Appendix A, run both parameters up to the unification scale and compute $b/\mu$ at the scale $\langle M_{U} \rangle$. If the result satisfies constraint \eqref{pa6}, then that ``black'' point is an {\it acceptable} ``black'' point. We find that out of the 3,330 ``black'' points, only 356 additionally satisfy the constraint on $b/\mu$. We present these results in Figure 3. All 3,330 points are shown and colored either red or black. The ``red'' points are ``black'' points that, in addition, satisfy the $b/\mu$ constraint, whereas the remaining 2,974 ``black'' points that do not satisfy this constraint are simply shown in black.

To make this process more concrete, let us consider a set of initial parameters given by
\begin{equation}
\begin{gathered}
m_{\tilde Q_1}=m_{\tilde Q_2}=m_{\tilde Q_3}=12{,}688 \>\text{GeV}\ ,\qquad
m_{\tilde u^c_1}=m_{\tilde u^c_2}=m_{\tilde u^c_3}=2{,}232 \>\text{GeV}\ ,\\
m_{\tilde d^c_1}=m_{\tilde d^c_2}=m_{\tilde d^c_3}=3{,}247\> \text{GeV}\ ,\qquad
m_{\tilde L_1}=m_{\tilde L_2}=m_{\tilde L_3}=	2{,}168\> \text{GeV}\ ,\\
m_{\tilde \nu^c_1}=m_{\tilde \nu^c_2}=m_{\tilde \nu^c_3}=2{,}495 \>\text{GeV}\ ,\qquad
m_{\tilde e^c_1}=m_{\tilde e^c_2}=m_{\tilde e^c_3}=14{,}480\> \text{GeV}\ ,\\
m_{H_u}=3879\> \text{GeV}\ ,\qquad m_{H_d}=2{,}783\> \text{GeV}\ ,\\
M_{1/2}~ (=M_3=M_2=M_{3R}=M_{BL})=-13{,}290\>\text{GeV}\ ,\\
\mathcal{A}~ (=\mathcal{A}_t=\mathcal{A}_b=\mathcal{A}_\tau)=2{,}170\>\text{GeV}\ ,\qquad
\tan \beta=4.79\ .
\end{gathered}
\label{eq:ExamplePoint2ss}
\end{equation}
Running these initial conditions down to the EW scale with our RG formalism, we find that it is indeed a ``black'' point. But is it an acceptable ``black'' point, that is, can it be promoted to a ``red'' point? To determine this, we calculate the associated values of $\mu$ and $b$  at the EW scale using the fact that, at the EW scale,
\begin{equation}
m_{H_u}^2=-(12{,}643\>\text{GeV})^2\>,\qquad m_{H_d}^2=(8{,}702\>\text{GeV})^2\ .
\end{equation}
It then follows from \eqref{eq:mu_expression2} and \eqref{eq:b_expression} that at this scale
\begin{equation}
\label{eq:B_low1}
\mu=13{,}060~ {\rm GeV}\ , \qquad b=(7{,}169~ {\rm GeV} )^{2} \ .
\end{equation}
Before proceeding, it is important to note that we found the low energy value of $\mu$ using eq.~\eqref{eq:mu_expression2}, 
which also allows for the negative solution $\mu=-13{,}060$ GeV. Although negative values  of $\mu$ are allowed in the $B-L$ MSSM, we find that the low energy ratio $b/\mu$ can never be RG scaled to the value of this ratio predicted by \eqref{pa6}.
Hence, one need only consider the positive roots of $\mu$ in eq.~\eqref{eq:mu_expression2}.
Using \eqref{eq:B_low1}, we find that
\begin{equation}
\frac{b}{\mu}(M_{Z}) = 3{,}936~{\rm GeV} .
\end{equation}
However, as discussed above, one must now run this ratio back up to the unification scale using the RG analysis given in Appendix A. 

\begin{figure}[t]
   \centering
     \begin{subfigure}[b]{0.49\textwidth}
\includegraphics[width=1.0\textwidth]{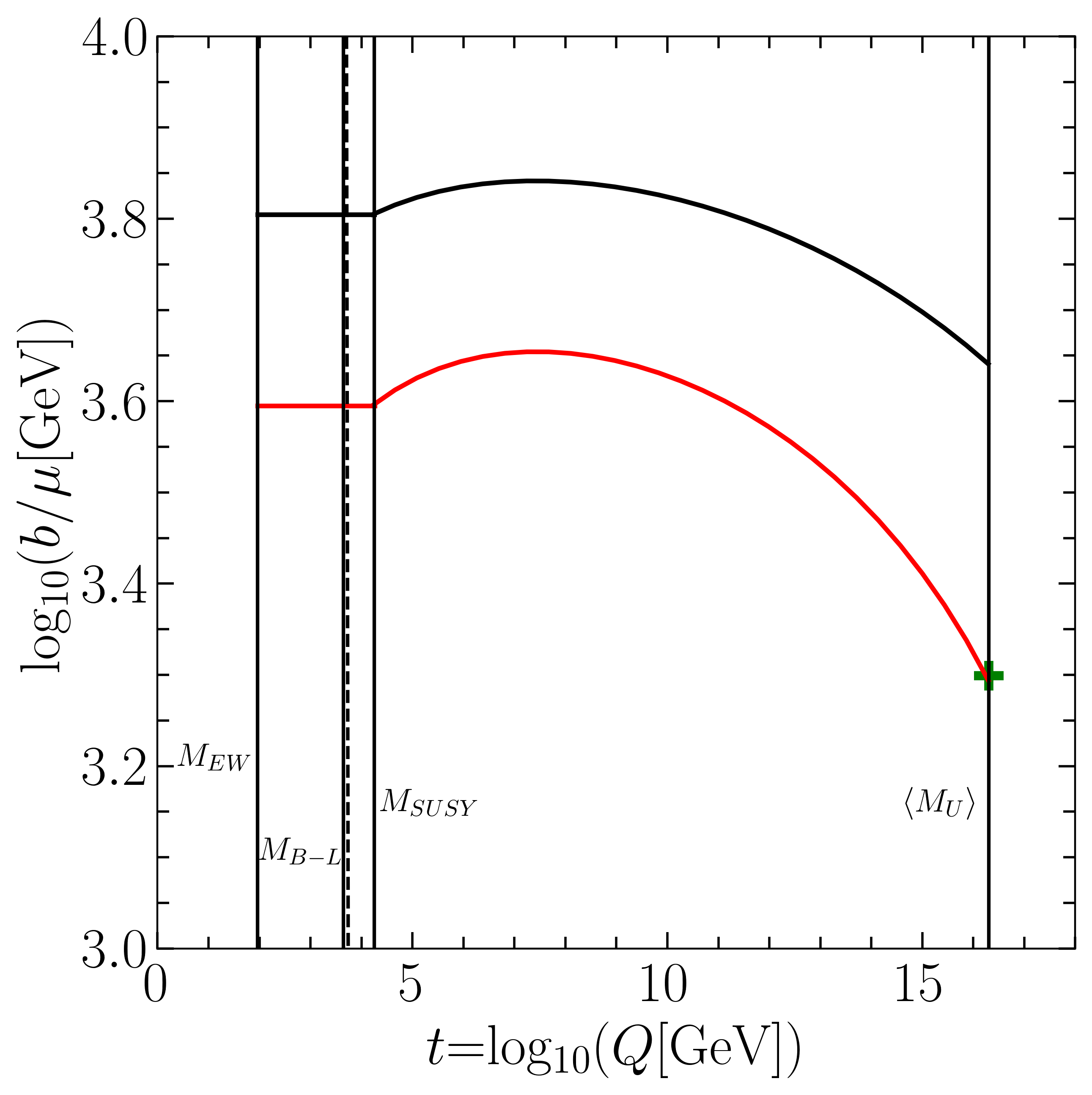}
\label{fig:RGE_b}
\centering
\end{subfigure}
\caption{We show how the ratio $b/\mu$ runs between the EW scale and the unification scale for two different points in our simulation. The red line corresponds to the acceptable ``black'' point circled in red in Figure \ref{fig:redScatterPlot} and with initial data shown in \eqref{eq:ExamplePoint2ss}. This point corresponds to an ``upside-down'' hierarchy with $M_{BL}=4{,}257$~GeV, $M_{\text{SUSY}}=18{,}101$~GeV  (these scales are indicated with vertical solid lines). The ratio $b/\mu$ runs from $3{,}936$~GeV at the EW scale, to $2{,}008$~GeV near the unification scale. This matches the high energy value predicted in \eqref{cil3}, and is marked with a green cross. The black line corresponds to the unacceptable ``black'' point circled with black in Figure \ref{fig:redScatterPlot} and which has initial data shown in  \eqref{eq:ExamplePoint2ssA}. This point corresponds to an ``upside-down'' hierarchy with $M_{BL}=6{,}099$~GeV, $M_{\text{SUSY}}=18{,}845$~GeV (these scales are indicated with vertical dashed lines, though they are very close to the solid lines in the ``acceptable'' case).  The ratio $b/\mu$ runs from $6{,}433$~GeV at the EW scale, to $4{,}436$~GeV near the unification scale, {\it far} from the correct high energy value.}
\label{fig:RGE_both}
\end{figure}

To do this it is necessary to determine the scaling ``regimes'' appropriate for the chosen set of initial conditions. Define
\begin{equation}
M_{SUSY}=\sqrt{m_{{\tilde{t}}_{1}}{m_{{\tilde{t}}_{2}}}}
\label{cil1}
\end{equation}
to be the approximate scale of the sparticle masses\footnote{$M_{SUSY}$ is  a low energy supersymmetry breaking scale and should not be confused with $m_{\rm susy}$.}, and $M_{B-L}$ the scale of spontaneous $B-L$ breaking.  Then our RG analysis tells us that, for our chosen initial conditions \eqref{eq:ExamplePoint2ss} at the point $(a^1,a^2,a^3)=(0.910, 1.401, 0.163)$ in K\"ahler moduli space,
\begin{equation}
M_{\text{SUSY}}=18{,}101~{\rm GeV}, \qquad M_{B-L}=4{,}257~{\rm GeV}.
\end{equation}
Since $M_{B-L} < M_{SUSY}$, this is known as an ``upside-down'' mass hierarchy. It follows that the RGEs for $\mu$ and $b$ pass through three scaling regimes; namely, $M_Z \rightarrow M_{B-L}$, then $M_{B-L} \rightarrow M_{SUSY}$ and finally $M_{SUSY} \rightarrow \langle M_{U} \rangle$. Using the RG analysis given in the appendix, we plot the value of $b/\mu$ from $M_{Z}$, through $M_{B-L}$, through $M_{SUSY}$ up to the unification scale. The result is shown as the red line in Figure 4. In particular, we find that
\begin{equation}
\frac{b}{\mu}(\langle M_{U} \rangle) = 2{,}008~{\rm GeV} \ .
\label{cil3}
\end{equation}
This is {\it exactly} the value predicted by expression \eqref{pa6} at this point in moduli space. Thus, the initial conditions we gave in \eqref{eq:ExamplePoint2ss} actually define a ``red'' point. We emphasize this specific ``red'' point in Figure 3 by surrounding it with a red circle.
However, at the same point in K\"ahler moduli space, $(a^1,a^2,a^3)=(0.910, 1.401, 0.163)$, not all ``black'' points satisfy the constraint on $b/\mu$ at the unification scale. For example, consider a different set of random initial data given by
 \begin{equation}
\begin{gathered}
m_{\tilde Q_1}=m_{\tilde Q_2}=m_{\tilde Q_3}=11{,}490 \>\text{GeV}\ ,\qquad
m_{\tilde u^c_1}=m_{\tilde u^c_2}=m_{\tilde u^c_3}=14{,}042 \>\text{GeV}\ ,\\
m_{\tilde d^c_1}=m_{\tilde d^c_2}=m_{\tilde d^c_3}=3{,}529\> \text{GeV}\ ,\qquad
m_{\tilde L_1}=m_{\tilde L_2}=m_{\tilde L_3}=	5{,}974\> \text{GeV}\ ,\\
m_{\tilde \nu^c_1}=m_{\tilde \nu^c_2}=m_{\tilde \nu^c_3}=1{,}176 \>\text{GeV}\ ,\qquad
m_{\tilde e^c_1}=m_{\tilde e^c_2}=m_{\tilde e^c_3}=11{,}809\> \text{GeV}\ ,\\
m_{H_u}=3{,}557\> \text{GeV}\ ,\qquad m_{H_d}=7{,}095\> \text{GeV}\ ,\\
M_{1/2}~ (=M_3=M_2=M_{3R}=M_{BL})=-13{,}290\>\text{GeV}\ ,\\
\mathcal{A}~ (=\mathcal{A}_t=\mathcal{A}_b=\mathcal{A}_\tau)=2{,}170\>\text{GeV}\ ,\qquad
\tan \beta=3.16\ .
\end{gathered}
\label{eq:ExamplePoint2ssA}
\end{equation}
Running these initial conditions down to the EW scale with our RG formalism, we find that it is indeed a ``black'' point. But, is it an acceptable ``black'' point; that is, a ``red'' point? To determine this, we calculate the associated values of $\mu$ and $b$ at the EW scale using the fact that, at the EW scale,
\begin{equation}
m_{H_u}^2=-(13{,}555 \>\text{GeV})^2\>,\qquad m_{H_d}^2=(9{,}162 \>\text{GeV})^2\ .
\end{equation}
It then follows from \eqref{eq:mu_expression2} and \eqref{eq:b_expression} that at this scale
\begin{equation}
\label{eq:B_low2}
\mu=14{,}610~ {\rm GeV} , \qquad b=(9{,}695~ {\rm GeV} )^{2} \ .
\end{equation}
Using \eqref{eq:B_low2}, we find that the ratio is
\begin{equation}
\frac{b}{\mu}(M_{Z}) =6{,}433~{\rm GeV}\ .
\end{equation}
However, as discussed above, one must now run this ratio back up to the unification scale using the RG analysis given in the appendix. 
Again, we need to determine the scaling ``regimes'' appropriate for this black point. For the initial coefficients given in \eqref{eq:ExamplePoint2ssA}, our RG analysis tells us that
\begin{equation}
M_{\text{SUSY}}=18{,}845~{\rm GeV}, \qquad M_{B-L}=6{,}099 ~{\rm GeV}.
\end{equation}
As in the previous example, since $M_{B-L} < M_{SUSY}$, this is an ``upside-down'' mass hierarchy. It follows that the RGEs for $\mu$ and $b$ pass through the same three scaling regimes as before; namely, $M_Z \rightarrow M_{B-L}$, then $M_{B-L} \rightarrow M_{SUSY}$ and finally $M_{SUSY} \rightarrow \langle M_{U} \rangle$. Using the RG analysis given in the appendix, we plot the value of $b/\mu$ from $M_{Z}$, through $M_{B-L}$, through $M_{SUSY}$ up to the unification scale. The result is shown as the black line in Figure 4. In particular, we find that
\begin{equation}
\frac{b}{\mu}(\langle M_{U} \rangle) = 4{,}436~{\rm GeV} \ .
\label{cil3A}
\end{equation}
This result is {\it quite far} from the required value $b/\mu=2,008~{\rm GeV}$ at the unification scale given in \eqref{cil3}. That is, the initial conditions given in \eqref{eq:ExamplePoint2ssA} define a ``black'' point, but they cannot be promoted to a ``red'' point since they do not satisfy the constraint on $\frac{b}{\mu}$ at the unification scale. This specific initial data is plotted as a ``black'' point in Figure 3 surrounded by a black circle.

\begin{figure}[t]
   \centering
     \begin{subfigure}[b]{0.4958\textwidth}
\includegraphics[width=1.0\textwidth]{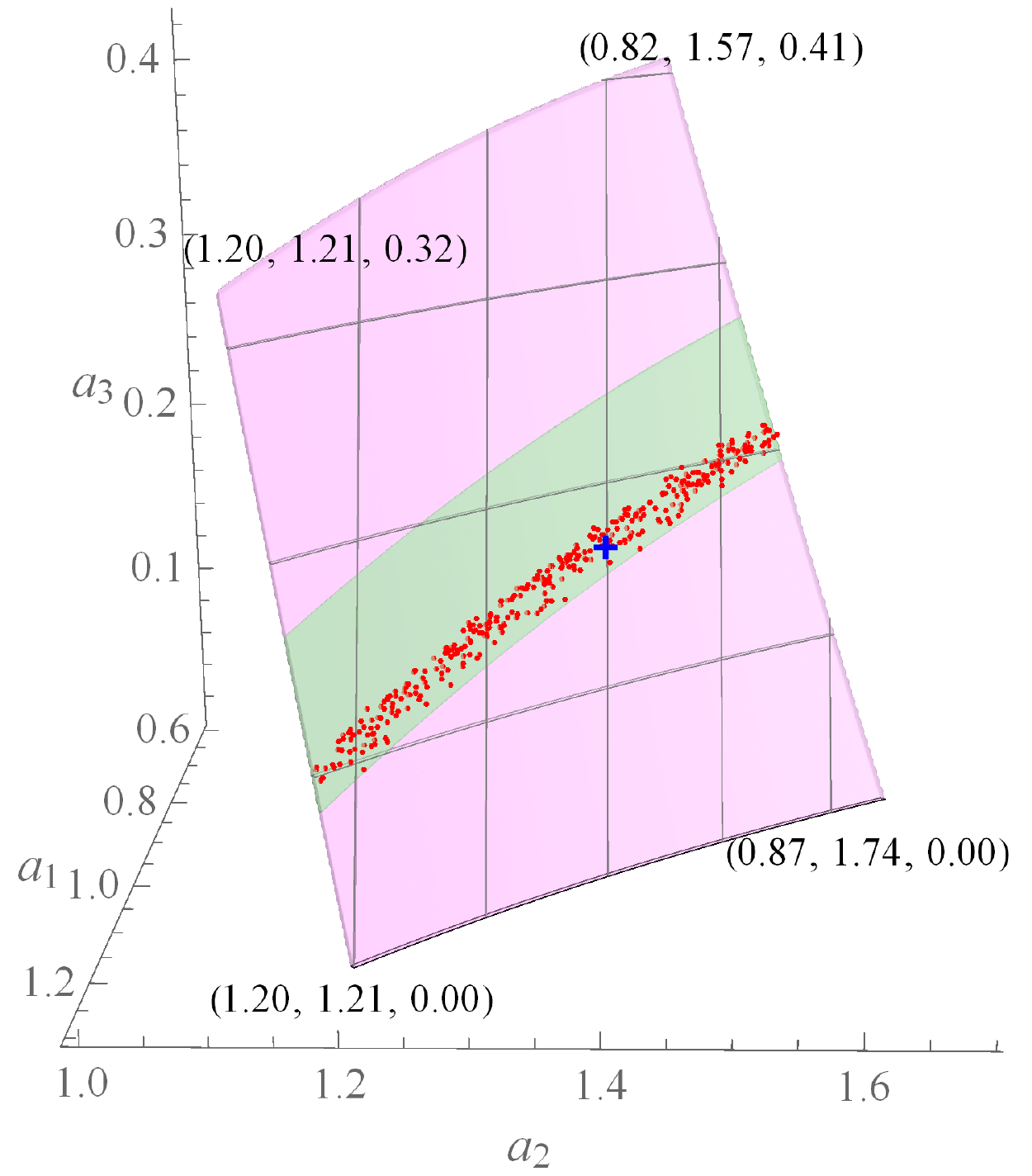}
\caption{}
\label{fig:AllRedPointsA}
\end{subfigure}
     \begin{subfigure}[b]{0.496\textwidth}
\includegraphics[width=1.0\textwidth]{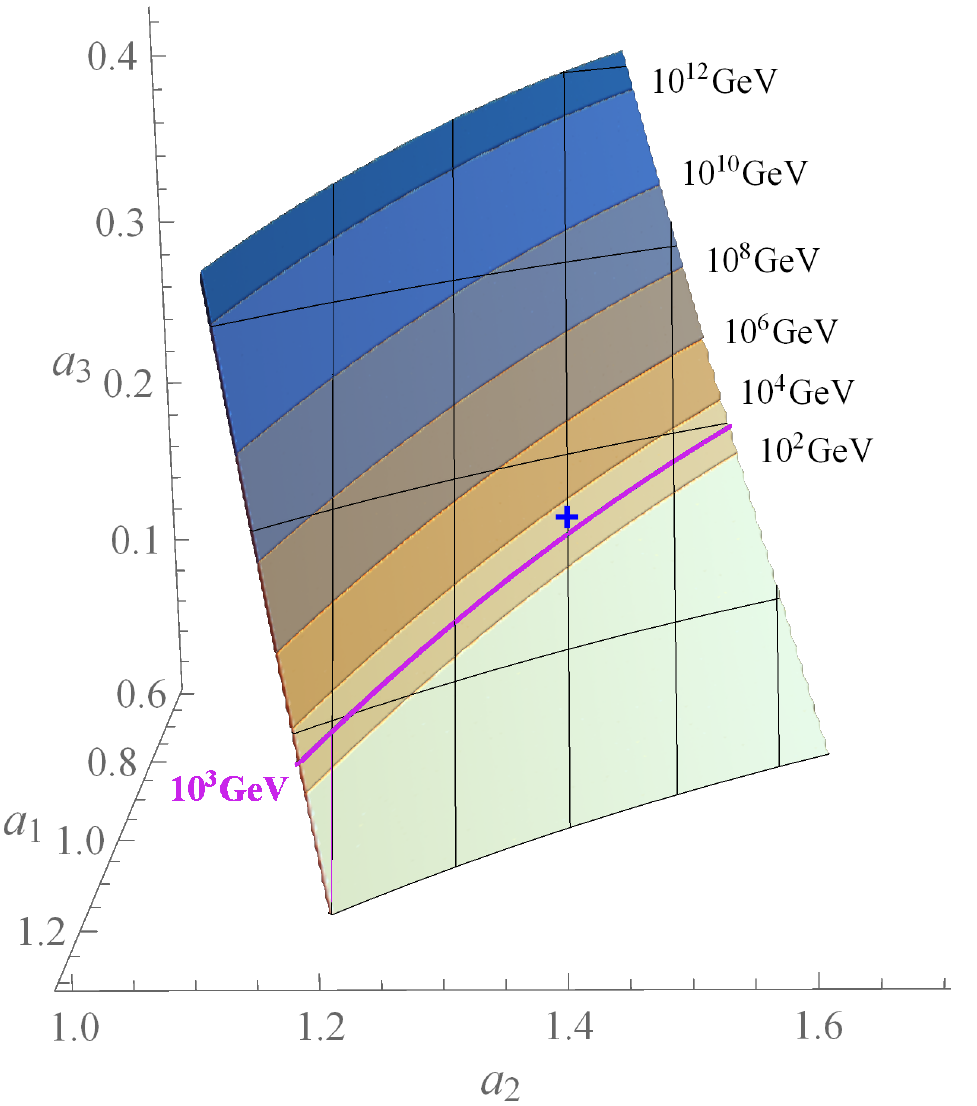}
\caption{}
\label{fig:AllRedPointsB}
\end{subfigure}
\caption{(a) In magenta we show the ``viable'' region of K\"ahler moduli space that satisfies all constraints in Section 2 for the line bundle $L= \mathcal{O}_X(2,1,3)$. The RGE simulation produces ``black'' points when $(a^1,a^2,a^3)$ are sampled within the green subregion. With red we show points in this green subspace where we find at least one {\it acceptable} ``black'' point--that is, a ``red'' point.  The region of possible  {\it acceptable} ``black'' points forms a strip. The blue cross indicates the point $(a^1,a^2,a^3)=(0.910, 1.401, 0.163)$ we used for the previous examples where we found the ``black'' and ``red'' points shown in Figure \ref{fig:redScatterPlot}. (b) Plot of the variation of the soft SUSY breaking parameter $m_{\rm susy}$ across the ``viable'' region of K\"ahler moduli space in the simultaneous Wilson lines scenario with $\langle \alpha_{u} \rangle^{-1}=26.46$. This was also shown in Figure \ref{fig:susyScale_both}(b), but we reproduce it here for ease of comparison. The blue cross indicates the point $(a^1,a^2,a^3)=(0.910, 1.401, 0.163)$. The purple line is the $m_{\text{susy}}=1~\text{TeV}$ contour.}
\label{fig:AllRedPoints}
\end{figure}

In the rest of this section, we will extend our analysis from the single sample point $(a^1,a^2,a^3)=(0.910, 1.401, 0.163)$ inside the ``viable'' region of K\"ahler moduli space in Figure 1, to the entire ``viable''  region. We do this using a statistical analysis, exactly as was done for the specific point $(a^1,a^2,a^3)=(0.910, 1.401, 0.163)$ above, but now applied to every point in the viable region of moduli space. We sample 10,000 values of $(a^1,a^2,a^3)$ across the ``viable'' region of the K\"ahler moduli space. At each point, we compute $m_s$, $M_{1/2}$ and $\mathcal{A}$, and then run the RGE simulation with 1,000 initial ``throws''. The results are shown in Figure \ref{fig:AllRedPoints}. 

The structure of Figure 5(a) is the following. To begin, we a find subregion, highlighted in green, in which any point can lead to a ``black'' point in the RGE simulation. Comparing this to Figure 5(b), we find that across this green subregion the scale of soft SUSY breaking, $m_{\rm susy}$, is roughly within $[1~\text{TeV},10^5~\text{TeV}]$. However, not all points within this green region can produce {\it acceptable} ``black'' points, that is, ``black'' points that can be promoted to ``red'' points. We plot in red those points in the region where {\it at least one} statistical throw of the initial parameters leads to an acceptable ``black'' point. We see that the subregion of possible ``red'' points forms a strip within the green subregion. Note that the ``red'' points are present in the lower part of the green area, in which the range of $m_{\rm susy}$ is approximately  $[1~\text{TeV},10~\text{TeV}]$. Above the red strip, but still within the green region, we do obtain ``black'' points, but none of them are acceptable -- the values of $b/\mu$ at the unification scale are always too large. In the region below this green strip, the scale of the soft SUSY breaking terms becomes too low to produce ``black'' points. The reason for this is that our RGE simulation imposes lower bounds on the superparticle masses. Below the green subregion, at least one of the sparticle masses becomes too light, violating these experimental bounds. By cross-comparison between Figures 5(a) and 5(b), we find that no ``black'' points are produced when $m_{\rm susy} \lesssim 1~\text{TeV}$.
Similarly, in the surface above the green region, the RGE simulation also does not produce ``black'' points. For $m_{\rm susy}$ above $\sim 10^5~\text{TeV}$, we find that the mass of the Higgs boson is always outside current experimental bounds. 

We conclude, therefore, that $m_{\rm susy}$ is a good indicator of how the sizes of the soft SUSY breaking terms vary across the ``viable'' solution space. Specifically, we learn from the red strip of acceptable ``black'' points in Figure 5(a), that for gaugino-induced supersymmetry breaking with a hidden sector bundle built from the line bundle $L=\mathcal{O}_{X}(2,1,3)$, soft supersymmetry breaking is restricted by experimental bounds to lie approximately in the interval $[1~\text{TeV},10~\text{TeV}]$.

\appendix

\subsubsection*{Acknowledgements}

AA is supported by the European Union’s Horizon 2020 research and innovation programme under the Marie
Sk\l{}odowska-Curie grant agreement No.~838776 and acknowledges previous support from the research grant DOE No.~DESC0007901. SD is supported in part by research grant DOE No.~DESC0007901. BO is supported in part by both the research grant DOE No.~DESC0007901 and SAS Account 020-0188-2-010202-6603-0338.

\section{RGE Solutions for \texorpdfstring{$b$}{b} and \texorpdfstring{$\mu$}{mu}}\label{RGE_appendix}

In the observable sector, the simultaneous Wilson lines scenario contains the following scaling regimes. For the ``right-side-up'' hierarchy, that is, when $M_{SUSY} < M_{B-L}$, there are three regimes. From the lowest scale to the highest, these are given by $M_{Z} \rightarrow M_{SUSY}$, $M_{SUSY} \rightarrow M_{B-L}$ and $M_{B-L} \rightarrow \langle M_{U} \rangle$. On the other hand, for the ``upside-down'' hierarchy defined by $M_{B-L}  < M_{SUSY}$, there are three different scaling regimes. These are given by $M_{Z} \rightarrow M_{B-L}$, $M_{B-L} \rightarrow M_{SUSY}$ and $M_{SUSY} \rightarrow \langle M_{U} \rangle$. The RGEs for all parameters in the $B-L$ MSSM, {\it with the exception of $b$ and $\mu$}, are well known for both of these hierarchies and the associated regimes~\cite{Deen:2016vyh,Ovrut:2012wg,Ambroso:2010pe,Ambroso:2009sc,Ambroso:2009jd}. In this appendix, we will present the RGEs for the two remaining parameters, $b$ and $\mu$. 

We begin with the right-side-up hierarchy. In the $M_{Z} \rightarrow M_{SUSY}$ regime, since the sparticles have decoupled, both $b$ and $\mu$ are approximately constant. However, in the $M_{SUSY} \rightarrow M_{B-L}$ interval, we find that the RGEs for $b$ and $\mu$ are given by
\begin{equation}
\label{eq:B_RGE1}
\begin{split}
\frac{d}{dt}b&=\frac{b}{16\pi^2}\left[3y_t^*y_t+3y_b^*y_b+y_\tau^*y_\tau-3g_2^2-\tfrac{3}{5}g_1^2  \right]\\
&\eqspace+\frac{\mu}{16\pi^2}\left[ 6a_ty_t^* +6a_by_b^* +2a_\tau y_\tau^* +6g_2^2M_2+\tfrac{6}{5}g_1^2M_1\right] \ ,\\
%&\approx \frac{b}{16\pi^2}\left[3y_t^*y_t-3g_2^2-\frac{3}{5}g_1^2  \right]+\frac{\mu}{16\pi^2}\left[ 6a_ty_t^* +6g_2^2M_2+\frac{6}{5}g_1^2M_1\right]\ ,\\
\frac{d}{dt}\mu&=\frac{\mu}{16\pi^2}\left[ 3y_t^*y_t+3y_b^*y_b+y_\tau^*y_\tau-3g_2^2-\tfrac{3}{5}g_1^2   \right] \ ,\\
%&\approx \frac{\mu}{16\pi^2}\left[ 3y_t^*y_t-3g_2^2-\frac{3}{5}g_1^2   \right]\ .
\end{split}
\end{equation}
respectively.
In the above, the SM Yukawa couplings, which are $3\times 3$ matrices in flavor space,
were all approximated to be zero except for the three elements which give mass to the third-generation SM fermions.
The experimentally determined values of these three coefficients at the EW scale are
\begin{equation}
\label{eq:Yukawa_Low}
y_t=0.9550\ , \qquad y_b=0.0174\ , \qquad y_\tau=0.0102\ .
\end{equation}
The gaugino masses $(M_1,M_2)$ and the cubic scalar masses from the SUSY breaking Lagrangian 
$(a_t,a_b,a_\tau)$ enter the RG equations above the SUSY breaking scale. 
The RG equations for  $(a_t,a_b,a_\tau)$ and $(M_1,M_2)$, as well as for the gauge and Yukawa couplings in 
the MSSM, can be found in \cite{Ovrut:2015uea,Martin:1997ns}. Together with the RGEs for $b$ and $\mu$ given in \eqref{eq:B_RGE1}, they form a 
complex system of equations that needs to be solved simultaneously to go to higher scales. This system 
can be simplified, however, by observing that at low energies $y_t \gg y_b, y_\tau$. This relation remains accurate 
at higher energy scales and, hence, we can drop the terms containing $y_b$ or $y_\tau$ from the system of equations, 
including the cubic couplings $a_b$ and $a_\tau$ which are directly proportional to the Yukawa couplings. 
Finally, in the last regime $M_{B-L} \rightarrow \langle M_{U} \rangle$, we find the RGEs for $b$ and $\mu$ to be 
\cite{Ovrut:2015uea}
\begin{equation}
\label{eq:B_RGE2}
\begin{split}
\frac{d}{dt}b &=\frac{b}{16\pi^2}\left[3y_t^*y_t+3y_b^*y_b+y_\tau^*y_\tau-3g_2^2-\tfrac{3}{26}g_{BL}^2 -\tfrac{9}{13}g_{R}^2 \right]\\
&+\frac{\mu}{16\pi^2}\left[ 6a_ty_t^* +6a_by_b^* +2a_\tau y_\tau^* \eqspace+6g_2^2M_2+\tfrac{3}{13}g_{BL}^2M_{BL}+\tfrac{18}{13}g_{R}^2M_R\right] \ ,\\
%&\approx \frac{b}{16\pi^2}\left[3y_t^*y_t-3g_2^2-\frac{3}{5}g_1^2  \right]+\frac{\mu}{16\pi^2}\left[ 6a_ty_t^* +6g_2^2M_2+\frac{6}{5}g_1^2M_1\right]\ ,\\
\frac{d}{dt}\mu &=\frac{\mu}{16\pi^2}\left[ 3y_t^*y_t+3y_b^*y_b+y_\tau^*y_\tau-3g_2^2-\tfrac{3}{26}g_{BL}^2 -\tfrac{9}{13}g_{R}^2   \right]\ .\\
%&\approx \frac{\mu}{16\pi^2}\left[ 3y_t^*y_t-3g_2^2-\frac{3}{5}g_1^2   \right]\ .
\end{split}
\end{equation}

The structure of the RGEs in the upside-down hierarchy is somewhat different. In this case, since the sparticles are integrated out, we again find that $b$ and $\mu$ are essentially constant in both the regimes $M_{Z} \rightarrow M_{B-L}$ and $M_{B-L} \rightarrow M_{SUSY}$. However, in the highest interval $M_{SUSY} \rightarrow \langle M_{U} \rangle$, the RGEs given in 
\eqref{eq:B_RGE2} remain valid.

\bibliographystyle{JHEP}
\bibliography{citations}

\end{document}